\documentclass[secnumeq]{FBSart}
\usepackage{amsfonts}
\usepackage{amssymb}
\usepackage{epsfig}

\title{Variational Worldline Approximation for the Relativistic 
Two-Body Bound State in a Scalar Model}

\author{K. Barro-Bergfl\"odt \instnr{1}, R. Rosenfelder \instnr{2}, 
M. Stingl \instnr{3} }

\instlist{Department of Mathematics, ETH Z\"urich, CH-8092 Z\"urich, Switzerland 
\and
Particle Theory Group, Paul Scherrer Institute, CH-5232 Villigen PSI, 
Switzerland 
\and
Institut f. Theoretische Physik, Universit\"at M\"unster,
D-48149 M\"unster, Germany}

\runningauthor{K. Barro-Bergfl\"odt et al.}
\runningtitle{Variational Worldline Approximation for the Relativistic 
Bound State}
\newcommand{\bce}{\begin{center}}
\newcommand{\ece}{\end{center}}
\newcommand{\be}{\begin{equation}}
\newcommand{\ee}{\end{equation}}
\newcommand{\bea}{\begin{eqnarray}}
\newcommand{\eea}{\end{eqnarray}}
\newcommand{\bdes}{\begin{description}}
\newcommand{\edes}{\end{description}}
\newcommand{\E}{\>=\>}
\newcommand{\EA}{&=&}
\newcommand{\EQ}{\> \equiv \>}
\newcommand{\deF}{\> = \, : \>}
\newcommand{\Def}{\> : \, = \>}
\newcommand{\To}{\> \longrightarrow \> }
\newcommand{\non}{\nonumber \\}
\newcommand{\dket}{\rangle \! \! \rangle}
\newcommand{\dbra}{\langle \! \! \langle }
\newcommand{\bfp}{{\bf p}}
\newcommand{\bfpp}{{\bf P}}
\newcommand{\bfq}{{\bf q}}
\newcommand{\bfa}{{\bf a}}
\newcommand{\bfx}{{\bf x}}
\newcommand{\bfy}{{\bf y}}

\newcommand{\Deltaleft}{{\stackrel{\leftarrow}{\Delta}}}
\newcommand{\Deltaright}{{\stackrel{\rightarrow}{\Delta}}}

\begin{document}

\maketitle
\begin{abstract}
We use the worldline representation of field theory together 
with a variational approximation to determine the lowest bound state  
in the scalar Wick-Cutkosky
model where two equal-mass constituents interact via the exchange of mesons. 
Self-energy and vertex corrections are included  approximately in a consistent 
way as well as crossed diagrams. Only vacuum-polarization effects of the heavy 
particles are neglected. In a path integral description of an appropriate 
current-current correlator an effective, retarded action is obtained by 
integrating out the meson field. As in the polaron problem we employ a 
quadratic trial action with variational functions to describe
retardation and binding effects through multiple meson exchange.
The variational equations for these functions are derived, discussed 
qualitatively and solved numerically. We compare our results with the ones from
traditional approaches based on the Bethe-Salpeter equation and find 
an enhanced binding 
contrary to some claims in the literature.
For weak coupling this is worked out analytically and 
compared with results from effective field theories.
However, the well-known instability of the 
model, which usually is ignored, now appears at 
smaller coupling constants than in the one-body case and even when self-energy 
and vertex corrections are turned off. This induced instability
is investigated analytically and the width of the bound state 
above the critical coupling is estimated.
\end{abstract}

\section{Introduction}
\label{sec: intro}

The bound state problem in Quantum Field Theory has a long history
starting with the classic paper by Salpeter and Bethe \cite{SaBe} more than
50 years ago. While the Bethe-Salpeter equation (BSe) is formally exact, its 
building  blocks are only defined as infinite series of increasingly 
higher $n$-point Green functions. This hierarchy of coupled equations 
(called Dyson-Schwinger equations) has to be truncated for practical purposes. 
In this way approximations are introduced which are hard to control and 
sometimes violate symmetries (e.g. gauge invariance). The most common 
approximation is the ladder approximation where bare propagators and vertices
are used. There is a vast literature on the ladder BSe and 
its many deficiencies (e.g. existence of abnormal states, failure in the
large mass limit etc.) \cite{BSreview}. 
Practical methods to include an extended set of diagrams 
(dressed propagators \cite{Saul}
or cross-ladder \cite{crossedBS})
have been developed only recently.

Many other variants have been investigated over the years: 
3-dimensional reductions (also called the ``quasipotential''
approach) \cite{LoTa,BlSu,Gross}, relativistic Hamiltonian models in 
light-front \cite{light} and point form \cite{point} quantization,
Feynman-Schwinger representations \cite{FSR} to name only a few. It is 
close to impossible (and not intended here) to give a comprehensive overview 
over all attempts and theoretical approaches available in the literature.

Apart from the theoretical interest, there
are (at least) three areas of physics in which the relativistic bound state 
problem needs to be addressed: the first one is bound-state Quantum 
Electrodynamics (QED)
where ultra-precise experimental data are available. However, from the 
theoretical point of view this is a special field since one can employ 
perturbation theory in powers of the fine-structure constant and its logarithm
after binding between the constituents has been established. 
Impressive accuracy matching the experimental precision has been achieved in 
recent years \cite{boundQED}.

There is another area for relativistic bound-state calculations
where strong-coupling is required: the electromagnetic structure
of nuclear few-body systems at intermediate and high energies. 
For example, relativistic effects
in the electromagnetic form factor of the deuteron seem to be important 
already at surprisingly low momentum transfers \cite{emdeut,PWD}. Unfortunately,
these effects cannot be clearly separated from those of the phenomenological 
nuclear interaction and of the assumed form of the current operator. 

Finally there is the area of hadronic physics where one has to understand how
(light) relativistic quarks and gluons bind to form the low-energy mesons and 
baryons (possibly also hybrids and exotics). Here, Quantum Chromodynamics (QCD)
at least provides a clear underlying theoretical framework. However, the problems
of confinement, chiral symmetry breaking and strong coupling are formidable
and have not been solved (analytically) at low energies. Lattice Gauge Theory
(see, e.g. ref. \cite{MoMue}) is considered as the prime method to obtain 
gauge-invariant results from first principles, albeit with enormous 
numerical effort and problems of its own. 
The continous progress of lattice calculations not withstanding,
considerable progress has also been made in the last years in solving 
the Dyson-Schwinger equations for Landau-gauge QCD under some simplifications 
\cite{AlSm} and in describing the low-lying hadrons as bound-state of quarks and 
gluons \cite{DS}. While phenomenologically quite succesful and often going 
beyond the ladder BSe, these calculations still have limitations due to 
truncations, gauge dependence and the use of model propagators. 
In a similar way as in QED, remarkable theoretical progress has been achieved 
recently for special systems like heavy quarkonia
by using the techniques of (nonrelativistic) effective field 
theories \cite{NRQCD}.

Given the twisted history of the relativistic bound-state problem, 
it might be useful to consider it from a different angle, namely from the one 
provided by the {\it worldline 
variational} approach. Variational methods are widely used in molecular, 
atomic and  nuclear physics as they offer accurate, non-perturbative results 
for bound and scattering states in a quantum-mechanical framework. In 
field theory their use is more restricted \cite{varfield}
since one cannot handle non-gaussian wave functionals as trial states.
Nevertheless, there have been a number of variational calculations for 
relativistic bound-state systems, most notably by Darewych and collaborators 
\cite{Dar,DarWC}.
In contrast to these works we advocate here the worldline description 
of field theory which entails a huge reduction of the dynamical degrees of 
freedom. 
This is essential for a variational calculation in which only quadratic trial 
actions can be employed for analytical evaluations. The classic example is 
Feynman's treatment of the polaron \cite{Fey} -- a nonrelativistic field 
theory for an electron interacting with the phonons in a crystal -- which 
still stands out among all approaches describing this system 
non-perturbatively for all coupling constants. The key to this success was that,
in a path integral treatment of the problem, he could integrate out  
the phonons exactly leading to a retarded two-time action 
for the electron. For the latter a variational approximation then gave results
unmatched by all other (analytical) methods.
Shortly after this seminal paper, Mano, a student of Feynman's, applied
these results to a scalar relativistic theory \cite{Mano}. However, his 
pioneering work was largely forgotten as was the description of relativistic 
systems by trajectories (worldlines). Recent times have seen a revival of 
worldline methods \cite{worldline} under different names (e.g. Feynman-Schwinger 
representation) and for different purposes, e.g. for an efficient calculation
of diagrams or effective actions. In a series of papers 
(\cite{WC1,WC2,WC3,WC4,WC56,WC7}) we have extended the ``polaron'' variational 
approach to field theory as pioneered by
Mano. This was first done in the context of the scalar  Wick-Cutkosky (WC) 
model \cite{WiCu} and later extended to a more realistic fermionic theory, 
{\it viz.} QED (\cite{QED1,VALE}). As an attempt to catch
non-perturbative features of these field theories, the results have been 
encouraging: for example, the well-known instability of the WC model 
\cite{unstable} was detected by the worldline variational approach, and 
in QED a fully 
gauge-invariant approximate result for the anomalous mass dimension of the 
electron was obtained \cite{QED2}.

Therefore we believe that a fresh look at the relativistic bound-state problem
from the worldline variational perspective may 
be useful even in an unrealistic model like the WC model,
which incidentally was developed just for the
purpose of studying bound states \footnote{Frequently only the exact 
solution of the ladder BSe for massless exchange particles is called 
the ``Wick-Cutkosky model''. Here we use this designation in a more general sense
for a class of scalar models described by Lagrangians of the form 
(\ref{L1 WC},\ref{L2 WC}).}.
In previous applications \cite{WC1} - \cite{WC7} we used a version of 
that scalar model which describes neutral  ``nucleon'' fields $\Phi$ 
interacting with ``meson'' fields $\chi$. It is given by the Lagrangian 
(our metric signature is $+ - - -$ and in general we put $\hbar = c = 1$).
\be
{\cal L}_1 \E \frac{1}{2} \left ( \partial_{\mu} \Phi \right)^2 - \frac{1}{2} 
M_0^2\Phi^2 + \frac{1}{2} \left ( \partial_{\mu} 
\chi \right )^2 - \frac{1}{2} m^2 \chi^2 + g \Phi^2 \chi \> .
\label{L1 WC}
\ee
In order to compare with standard literature (e.g. Chap. 10 in 
Itzykson \& Zuber \cite{ItZu})
it is advisable to  switch to charged particles and to allow for
different masses of the constituents. Then one needs
two different types of nucleon fields and therefore the model which we are 
going to use is described by the following Lagrangian 
\be
{\cal L}_2 \E \sum_{i=1}^2 \> \left [ \, \left | \partial_{\mu} \Phi_i 
\right |^2 - M_i^2 \left |\Phi_i \right |^2 \, \right ] + 
\frac{1}{2} \left ( \partial_{\mu} 
\chi \right )^2 - \frac{1}{2} m^2 \chi^2 + \sum_{i=1}^2 \> g_i  
\left |\Phi_i \right |^2  \chi \> .
\label{L2 WC}
\ee
The coupling constants $g_i$ are now twice as large as 
in the neutral case as can be seen by collecting all pieces 
quadratic in the nucleon fields. Previously we have considered \cite{BRS} a 
bound nucleon-antinucleon system in the neutral model described by the 
Lagrangian (\ref{L1 WC}) while now we will investigate a bound two-particle 
state described by the Lagrangian (\ref{L2 WC}). Both models are, of course,
a far cry from reality and in the present treatment basically equivalent. 
Of course, there are differences from the fact that in nucleon loops
two types of nucleons run around, i.e. that the determinant factor appears twice.
However, since we will work in the ``quenched'' 
approximation in the following this doesn't matter. Note that 
also the meson properties are unchanged in this approximation.
For simplicity the present work will be restricted mostly to the equal-mass case 
\be
M_1 \E  M_2 \EQ M_0 
\label{equal mass}
\ee
where $M_0$ denotes the {\it bare mass} of the nucleon. First results 
from this new approach have been presented in ref. \cite{BRS}.
Here we will give a full account of this formalism, test it in the nonrelativistic
case, present improved numerical calculations  
and explore special cases, in particular when self-energy and vertex effects 
are switched off but crossed-ladder diagrams are retained.

Finally it is worthwhile to recall the historical and conceptual roots of 
the 
worldline variational approach: for a single scalar particle it 
may be considered as a relativistic version of the polaron problem, and hence 
the present work 
is close in spirit (but not in the details) to investigating  a {\it bipolaron}, 
the possible bound state of two electrons in an ionic crystal. 
This occurs when the attractive electron-phonon coupling 
overcomes the repulsive Coulomb interaction and has been the topic of many 
theoretical papers \cite{bipol,VSPD} as a proposed mechanism 
for high-temperature superconductivity.

This paper is organized as follows: in the next Sect. we briefly overview the
well-known nonrelativistic results for the present model. We then study a 
particular current-current correlator (or polarization propagator) 
whose poles will give us the relativistic bound-state energies. Sect. 4 
describes the variational calculation of that object in the worldline formalism,
including the retarded trial action which we use. In the next two Sects.
the variational equations for the retardation functions which enter our trial 
action are derived and numerically solved. The last Sect. contains our 
conclusions and the outlook for further work. Most of the technical details 
such as the calculation of the various averages with the trial action or 
numerical details are collected in several appendices.

\section{Nonrelativistic Results}
\label{sec: nonrel}
A useful guide for the results one may expect for the two-body 
relativistic bound-state problem is provided by the nonrelativistic limit. 
Although this is quite standard we briefly recall the essential facts
and point to the differences in a relativistic treatment. 
It is well-known (and re-derived by 
path integral methods in appendix~\ref{app: nonrel}) that in the nonrelativistic 
limit one obtains a 
Schr\"odinger equation for the two particles interacting via an attractive
Yukawa potential
\be
V \left ( \bfx_1 - \bfx_2 \right ) \E - \frac{\alpha}{|\bfx_1 - \bfx_2|} \, 
e^{- m |\bfx_1 - \bfx_2|} 
\label{Yuk pot}
\ee
where
\be
\alpha \EQ \frac{g_1 g_2}{16 \pi M_1 M_2}
\label{def alpha}
\ee
is the usual (dimensionless) coupling constant \footnote{Sometimes, e.g. in ref. 
\cite{ItZu} , denoted by $\lambda$. Taking into account the different coupling 
terms for the case of neutral particles the present definition is the same 
as in refs. \cite{WC1} - \cite{WC7}.}. One of the 
simplifications of the nonrelativistic treatment is that center-of-mass and 
relative motions decouple.
For massless mesons ($ m = 0$) the Yukawa potential in Eq. 
(\ref{Yuk pot}) becomes an attractive Coulombic potential 
with internal energy levels
\be
\epsilon_n^{\rm nonrelat.} \biggr |_{m=0} \E - 
\frac{\alpha^2}{2 (n+1)^2} M_{\rm red}  \> , \hspace{0.5cm}
n = 0,1,2 \ldots
\ee
where
\be
 M_{\rm red} \E \frac{M_1 M_2}{M_1 + M_2}
\label{def Mred}
\ee
is the reduced mass of the system.
If $m \ne 0$ then first-order perturbation theory gives 
\be
\epsilon_0 \E  - \frac{1}{2} \, \alpha^2 M_{\rm red} \, \left [ \, 1  
- 4 \delta + 6 \delta^2 - 8 \delta^3 + 
{\cal O} \left ( \delta^4 \right ) \, \right ]
\label{E0 Yuk small m} 
\ee
where
\be
\delta \E  \frac{m}{2 \alpha M_{\rm red}} 
\label{def delta}
\ee 
is proportional to the ratio of Bohr radius to meson Compton wavelength. 
This shows that the finite range of the Yukawa potential, as expected, 
{\it decreases} the binding energy. 
Indeed, numerical calculations \cite{Pol} demonstrate that an attractive 
Yukawa potential only develops bound states if
\be
\delta^{-1} \E \alpha  \, \frac{2 M_{\rm red}}{m} \>  > \> 1.67981 \> .
\label{crit}
\ee
Thus for large enough coupling constants and/or small meson mass the WC model 
will most probably - inasmuch as
the nonrelativistic approximation is an useful guideline - have bound 
states. Note that nonrelativistically the masses are always constants which 
are not changed by the interaction whereas relativistically the binding 
energy subtracts from the ``weight'' of the system. If the system has total 
momentum $\bfq$ then its total energy is
\be
M_1 + M_2  + \frac{\bfq^2}{2 (M_1 + M_2) } + \epsilon_0 \> \simeq \> 
\sqrt{ (M_1 + M_2 + \epsilon_0)^2 + \bfq^2} 
\ee
and with $ \epsilon_0 < 0 $ the rest mass of a bound state is less than the sum 
of its constituent masses. Of course, the separation of center-of-mass and 
relative motion is no longer valid in the relativistic case, nor is the concept 
of a reduced mass. However, our approach of calculating the poles of suitable 
Green functions does not require these, and indeed automatically 
gives the correct nonrelativistic center-of-mass energy $\bfq^2/(2(M_1 + M_2))$ 
as shown in appendix~\ref{app: FJ nonrel}.

\section{The Current-Current Correlator}
\label{sec: corr}

``In the relativistic approach, bound states and resonances are identified 
by the occurence of poles in Green functions.'' 
(p. 481 in ref. \cite{ItZu}). Therefore in general one has 
to investigate a 4-point function of the form
\be
G_{4,0}^{(ij)} (x_1,x_2,x_3,x_4) \E \left < 0 \left |\,  {\cal T} \left ( 
\hat \Phi_j(x_1)  \hat \Phi_i(x_2)  \hat \Phi^{\dagger}_j(x_3)  
\hat \Phi^{\dagger}_i(x_4)  
\right ) \, \right | 0 \right > 
\ee
with no external mesons and $i,j = 1,2$. However, for the purpose of 
getting bound-state energies this object contains too much information and
therefore it is advisable to consider the Fourier transform of a 
special 4-point function where arguments are pairwise identical. 
In the context of the model described by the Lagrangian (\ref{L2 WC})
this could be
\be
\Pi_{i j}(q) \Def - i \, \int d^4x \, e^{i q \cdot x } \, 
\left < 0 \left |\,  {\cal T} \left ( 
\hat \Phi_j(x) \hat \Phi_i(x)  \,  \hat \Phi^{\dagger}_i(0)  
\hat \Phi_j^{\dagger}(0) \right ) \, \right | 0 \right >_{\rm connected}  \> .
\label{def Pi(q)}
\ee
Here we already have used translational invariance to set the second 
argument to zero.
Equation (\ref{def Pi(q)}) is a version of the polarization propagator 
in the language of many-body physics \cite{FeWa} or a 2-point correlator for the 
``current'' operator
\be
\hat C_{i j}^{\dagger}(x) \E \hat \Phi^{\dagger}_i(x)  \hat \Phi^{\dagger}_j(x) 
\> .
\label{current}
\ee
It is also (up to a sign and different currents) essentially the quantity 
considered by Shifman 
{\it et al.} in their work on QCD sum rules \cite{SVZ} to obtain hadronic 
masses from quarks and gluons. Note that we have
taken a particular scalar current and that we 
are free to choose different ones: for example, $ \hat \Phi^{\dagger}_i(x)  
\partial_{\mu} \hat \Phi^{\dagger}_j(x) \> $ 
may be used to study bound states with angular momentum $1$.
Due to covariance the present polarization propagator can only
depend on invariants, i.e. $\Pi_{i j}(q) = \Pi_{i j}(q^2) $. 
We will concentrate on the case $ i = 1, \, j = 2 $ in which the operator
(\ref{current}) creates a nucleon of type 1 and one of type 2
out of the vacuum.

To see what the structure of the polarization propagator is we now look at
its spectral representation. This is done in
the standard fashion by inserting a complete set of states with four momenta
$P_n$
\be
1 \E \sum_n \, \left | P_n \left > \right < P_n \right | \> \> \> ; \hspace{1cm} 
P_n^2 = M_n^2
\ee
into Eq. (\ref{def Pi(q)}). Using translational invariance one can perform  
the $x$-integration and obtains
\bea
\Pi_{i j}(q^2) \EA (2 \pi)^3  \sum_n   \frac{1}{q^2 - M_n^2 + i0} 
\Biggl \{  \delta^{(3)} \left ( \bfq - 
\bfpp_n \right ) \left ( q_0 + P_n^0 \right )   
\left | \left < P_n \left | \hat C_{i j}^{\dagger}(0) \right | 0 \right > 
\right |^2 \non
&& \hspace{2.5cm} - \delta^{(3)} \left ( \bfq + \bfpp_n \right ) \left (q_0 - 
P_n^0 \right ) \, 
\left | \left < P_n \left | \hat C_{i j}(0) \right | 0 \right > \right |^2
 \, \Biggr \} 
\label{Pi(q) spectral}
\eea
showing that the polarization propagator indeed develops poles at 
$ q^2 = M_n^2 $. Here $M_n$ is the mass of the intermediate state which 
can be reached from the ground state (the vacuum) by application of either
the current $\hat C^{\dagger}_{i j}(0)$ or by $\hat C_{i j}(0)$.
Using the standard expansion of the field operators in terms of creation and 
annihilation operators for the free-field case,
one easily sees that for it creates a {\it two-particle state} 
(consisting either of two nucleons or two antinucleons
of type  $i$ and $j$) as 
intermediate state when acting on the vacuum.

\subsection{Reduction to a Path integral Over Meson Fields}
\label{subsec: path}
\noindent
With no meson source the generating functional for our model 
is denoted by $Z'$ and reads
\be
Z'[J^*,J] \Def \int \prod_{i=1}^2 {\cal D} \Phi_i  \, {\cal D} \Phi_i^* \, 
{\cal D} \chi \> 
\exp \left \{ i \int d^4x \> \left [ \, {\cal L}_2 + \sum_i \left ( J_i^* 
\Phi_i + \Phi_i^* J_i \right ) \, \right ] \> \right \} \> .
\label{gen func 2}
\ee
One may integrate out the nucleon fields to obtain
\bea
Z'[J^*,J] \EA {\rm const.} \, \int {\cal D} \chi \> e^{i S_0[\chi]} \, 
\prod_{i=1}^2 \Biggl ( \, \frac{1}{{\rm det} \, {\cal O}_i(\chi)} \non
&& \times \, \exp  \left [ \, - i \int d^4x \, d^4y \, J_i^*(x) 
\left < x \left |  
{\cal O}_i^{-1} (\chi) \right | y \right > J_i(y) \, \right ] \, \Biggr )
\label{gen func 3}
\eea
where 
\be
S_0[\chi] \E  \frac{1}{2} \int d^4x \> \left [ \, (\partial \chi)^2 - 
m^2 \chi^2 \, \right ]
\ee
is the free meson action and 
\be
{\cal O}_i(\chi) \E - \partial^2 - M_i^2 + g_i \chi
\label{def O}
\ee
is the operator for the quadratic part of the nucleon action.
Note that charged nucleon fields lead to a power of $-1$ of the determinant 
instead of the usual $-1/2$ for neutral fields. However, in the quenched 
approximation where this determinant is neglected this makes no difference.

The functional differentiation for
\be
\Pi_{i j}(q) \E - i \int d^4x \, e^{i q \cdot x} \, 
\frac{\delta^4 \ln Z'}{\delta 
J_j(x) \,  \delta J_i^*(x) \, \delta J_i(0) \, \delta J_j^*(0)} 
\Biggr |_{J=J^*=0}
\ee
is most easily done by employing a cumulant expansion 
in terms of the sources in Eq. (\ref{gen func 3}). This gives
\bea
\ln Z' \EA  {\cal O} \left ( J^0,J^2 \right ) + \frac{(-i)^2}{2} \, 
\sum_{k,l=1}^2 \, \prod_{i=1}^4 \left ( \int d^4x_i \right ) \> J_k^*(x_1) \,  
J_k(x_2) \, J_l^*(x_3) \,  J_l(x_4') \non
&& \hspace{5cm} \times \,  R_{k l}(x_1,x_2;x_3,x_4) + {\cal O} \left ( J^6 
\right )
\eea
where
\bea
R_{k l}(x,y;x',y') \EA \dbra  \left < x \left |  {\cal O}_k^{-1}(\chi) \right | 
y \right > \left < x' \left |  {\cal O}_l^{-1}(\chi) \right | 
y' \right >   \dket \non
&&  - \dbra   \left < x \left |  {\cal O}_k^{-1}(\chi) \right | 
y \right >   \dket \, \dbra \! \left < x' \left |  {\cal O}_l^{-1}(\chi) 
\right | 
y' \right >   \dket .
\eea
Here we have defined the following average over the meson fields
\bea
\dbra A \dket  &\Def& \left \{  \int {\cal D} \chi \> 
\exp \left  [ \,  i S_0[\chi] \, - \sum_i {\rm Tr} \ln {\cal O}_i(\chi) \, 
\right ] \right \}^{-1}  \non
&&    \hspace{0.5cm} \times \, \int {\cal D} \chi \> A(\chi) 
\, \exp \left  [ \,  i S_0[\chi] \, - \sum_i {\rm Tr} \ln {\cal O}_i(\chi) \, 
\right ] \> .
\label{def av} 
\eea
The functional differentiations can now be performed easily with the result
\bea
\Pi_{i j}(q) \EA  i \int d^4x \, e^{i q \cdot x } \,  
R_{i j}(x,0;x,0) \non
\EA i \int d^4x \, e^{i q \cdot x } \, \Bigl  [ \,
\dbra  \left < x \left |  {\cal O}_i^{-1}(\chi) \right | 
x=0 \right > \, \left < x \left |  {\cal O}_j^{-1}(\chi) \right | 
x=0 \right > \dket \non
&& - \dbra  \left < x \left |  {\cal O}_i^{-1}(\chi) \right | 
x=0 \right >  \dket \, \dbra \left < x \left |  {\cal O}_j^{-1}(\chi) \right | 
x=0 \right >  \dket \> \Bigr ] \> .
\label{pol prop 1}
\eea
This describes the propagation of a nucleon (or an antinucleon) of type $i$ and 
one of type $j$ from the space-time point $0$ to $x$.
In between they emit and absorb  all types of mesons which is
represented by the functional integral over the meson field $\chi$. 

The last term Eq. (\ref{pol prop 1}) subtracts the unconnected pieces and 
consists just of the product of usual propagators. Since this term does not 
contain any poles besides the usual one-particle poles we can drop it when 
we search for {\it additional} bound-state poles in the polarization propagator. 
Taking $ i,j = 1,2 $ we therefore will investigate the simpler form
\be
\Pi(q) \Def 
\E i \int d^4x \, e^{i q \cdot x } \, 
\dbra  \left < x \left |  {\cal O}_1^{-1}(\chi) \right | 
x=0 \right > \, \left < x \left |  {\cal O}_2^{-1}(\chi) \right | 
x=0 \right >  \dket \> .
\label{pol prop 2}
\ee

\subsection{Worldline Description of the Polarization Propagator}
\label{subsec: worldline}
\noindent
We now derive the worldline formulation for $\Pi(q)$ 
given in Eq. (\ref{pol prop 2}) still for unequal masses.
As in the case of the 2-point function this is only 
possible in the {\it quenched approximation}, i.e. by neglecting the determinant 
in the average (\ref{def av})
\bea 
\dbra A \dket \> \longrightarrow \>   \dbra A \dket_{\rm quenched} 
&\Def& \left \{  \int {\cal D} \chi \, 
\exp \Bigl ( \,  i S_0[\chi] \, \Bigr )  \, \right \}^{-1} \non
&& \hspace{0.3cm} \times \, \int {\cal D} \chi \> A(\chi)
\exp \Bigl ( \,  i S_0[\chi] \, \Bigr )  .
\label{quenched av} 
\eea
Only then can one perform the path integral over the meson field $\chi$ after
using the Schwinger representation 
\be
\frac{1}{{\cal O}_i(\chi)} \E \frac{1}{2 i \kappa_0} \, \int_0^{\infty} dT \> 
\exp \left [ \, \frac{iT}{2 \kappa_0} \left ( -\partial^2 - M_i^2 + g_i \chi 
+ i 0 \right ) \, \right ]
\label{Schwinger rep}
\ee
and the quantum-mechanical path-integral\footnote{$\kappa_0 > 0$ reparametrizes 
the proper time and can be considered as ``mass'' of the
equivalent quantum-mechanical particle. In numerical applications 
we will set this parameter (more precisely: its euclidean counterpart) to $1$ in
which case all proper times have mass dimension $ -2 $ as can be seen from Eq. 
(\ref{Schwinger rep}). Another convenient choice is $\kappa_0 = M$ so that
the proper time has the same dimension as the ordinary time to which it reduces
in the nonrelativistic limit \cite{QED1}.} 
\bea
&& \left < y \left |  \, \exp \left [ - i T \left (- 
\frac{\hat p^2}{2 \kappa_0} - \frac{g_i}{2 \kappa_0} \chi(\hat x)  \right ) 
\right ]  \, \right |  x \right > \non
\EA \int_{x(0)=x}^{x(T)=y}  {\cal D} x \> \exp \left \{  \, i \int_0^T dt  \, 
\left [  - \frac{\kappa_0}{2} \dot x^2 + \frac{g_i}{2 \kappa_0} \chi (x)  
\right ]  \, \right \} 
\eea
{\it twice}, i.e. for {\it both} propagators. In this way one obtains
\bea
\Pi(q) \EA i \int d^4x \, e^{i q \cdot x} \int_0^{\infty} 
\frac{dT_1 \, dT_2}{(2 i \kappa_0)^2 } \, 
\exp \left [  -
\frac{i}{2 \kappa_0} \left ( M_1^2 T_1 + M_2^2 T_2 \right )  \right ] \non  
&& \times \, \prod_{i=1}^2 \, \left ( \int_{x_i(0)=0}^{x_i(T_i)=x} {\cal D} x_i 
\right ) \, \exp \left \{ \, i \sum_{i=1}^2 \int_0^{T_i} dt_i 
\left (  - \frac{\kappa_0}{2} \dot x^2_i \right ) \, \right \} \non
&& \times  \dbra \exp \left \{  \, i  \sum_{i=1}^2 \frac{g_i}{2 \kappa_0}  
\int_0^{T_i} dt_i  \> \chi \left ( x_i(t_i) \right ) \, \right \} \dket .
\eea
Writing the argument of the last exponential as $i \int d^4 z \, b(z) \, 
\chi(z) $ with
\be
b(z) \E \frac{1}{2 \kappa_0} \sum_{i=1}^2 \, g_i \, \int_0^{T_i} dt \> 
\delta \left ( \, z - x_i(t) \, \right )
\ee
the required functional average over the meson field can now be performed 
since it is just a gaussian integral
\bea
\int {\cal D} \chi \> e^{i S_0[\chi] + i (b,\chi)} \EA
\int {\cal D} \chi \> \exp \left [ \,\frac{i}{2} \left ( \chi, 
(-\partial^2 - m^2) \chi \right ) + i (b,\chi)  \right ] \non
\EA {\rm const.} \, \exp \left [ \, -\frac{i}{2} \left ( b, \frac{1}{-\partial^2 
- m^2} b \right ) \, \right ] \> .
\eea
Hence
\be
\dbra \exp \left [ i (b,\chi) \right ] \dket  \deF
\exp  \Bigl ( \>  i S_{\rm int}\left [x_1,x_2 \right ] \> \Bigr )
\ee
with
\bea
S_{\rm int}\left [x_1,x_2 \right ] \EA  - \frac{1}{8 \kappa_0^2} \, 
\sum_{i,j=1}^2 \, g_i g_j \, 
\int_0^{T_i} dt \, \int_0^{T_j} dt' \, \left < x_i(t) \left | 
\frac{1}{-\partial^2 
- m^2} \right | x_j(t') \right > \non
&=&
- \sum_{i,j=1}^2   \frac{g_i g_j}{8 \kappa_0^2} 
\int_0^{T_i} \! \! dt  \! \int_0^{T_j} \!\!  dt'  \!\int \frac{d^4 p}{(2\pi)^4} 
\! \frac{\exp \left [-i p \cdot \left ( x_i(t)-x_j(t') \right )  
\right ]}{p^2 - m^2 + i0}  .
\label{Sint 1,2}
\eea
The polarization propagator is therefore given by the double worldline
path integral \cite{FSR}
\bea
\Pi(q) \EA i \int d^4x \, e^{i q \cdot x} \int_0^{\infty} 
\frac{dT_1 \, dT_2}{(2 i \kappa_0)^2}  \, 
\exp \left [\,  - 
\frac{i}{2 \kappa_0} \left ( M_1^2 T_1 + M_2^2 T_2 \right ) \, \right ] \non
&& \times \, \prod_{i=1}^2  \left ( \int_{x_i(0)=0}^{x_i(T_i)=x} {\cal D} x_i 
\right ) \,  \, \exp \left \{ \, i \sum_{i=1}^2 S_0[x_i] + 
i S_{\rm int}[x_1,x_2] \, \right \}
\label{double worldline}
\eea
\noindent
where 
\be
S_0[x_i] \E \int_0^{T_i} dt \> \left (  - \frac{\kappa_0}{2} \dot x^2_i(t)
\right ) \> \>, \> \> i \E 1, 2
\label{Sfree}
\ee
is the standard free action for each particle.
\begin{figure}[hbt]
\bce
\mbox{\epsfxsize=130mm \epsffile{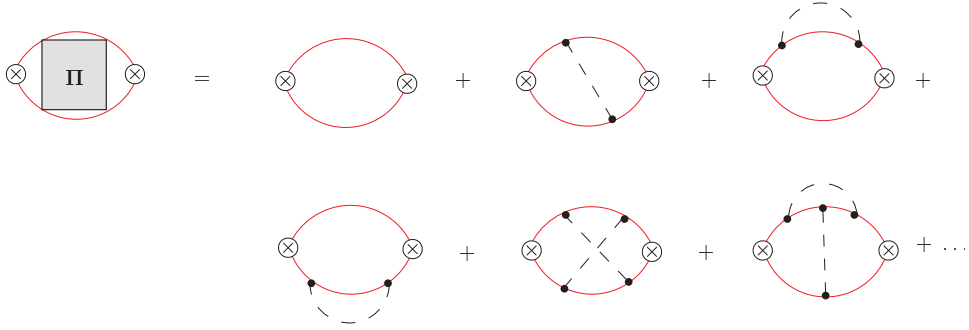}}
\ece
\caption{Perturbative expansion of the correlator in 
the worldline formulation of Eqs. (\ref{Sint 1,2}) 
- (\ref{Sfree}). Full (dashed) lines represent nucleons 
(mesons), filled dots the meson-nucleon vertices and circled dots the spacetime
points $x = 0$ and $ x $ where the nucleon lines start and end, respectively.
}
\label{fig: full corr}

\end{figure}

The perturbative expansion of the polarization propagator is shown in Fig. 
\ref{fig: full corr}. Note that the worldlines of the particles are parametrized 
by their proper times over which one has to integrate; thus crossed and 
ladder-type diagrams are included on an equal footing as are self-energy and 
vertex corrections.

If Eq. (\ref{Sint 1,2}) is split into terms with $ i = j $ and $ i \ne j $ one 
sees that the former generate the self-energies of each particle,
while the latter describe the interaction between the nucleons
by exchange of (any number of) mesons. The vertex corrections come automatically
due to different values of the proper times; for example, if one self-energy 
meson is already ``in the air'' when another one is 

\begin{figure}[thb]
\bce
\mbox{\epsfxsize=130mm \epsffile{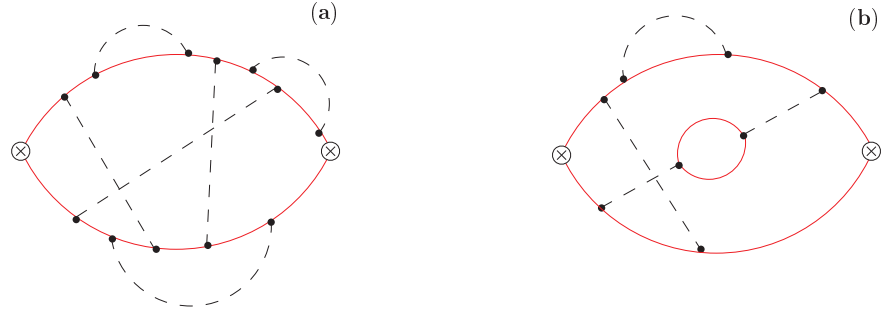}}
\ece
\caption{(a) A typical diagram which is contained, 
(b) a typical diagram which is {\it not} contained  in Eq. 
(\ref{double worldline}).
}
\label{fig: typ diagr}

\end{figure}

\noindent
emitted  
to the other particle. A typical diagram which is contained in the compact 
expression
(\ref{double worldline}) is depicted in Fig. \ref{fig: typ diagr}a . 
A diagram which is omitted due to the quenched approximation (neglect of pair 
production) is shown in Fig. \ref{fig: typ diagr}b.

\noindent
We normalize the path integrals by dividing and multiplying with
\bea
&& i \int d^4x \, e^{i q \cdot x} 
\int_{x_1(0)=0}^{x_1(T_1)=x} {\cal D} x_1 \, \int_{x_2(0)=0}^{x_2(T_2)=x} 
{\cal D} x_2 \, \exp \left \{  i \sum_{i=1}^2 S_0[x_i] \right \} \non
&& \hspace{2.5cm} = \>  \left ( \frac{\kappa_0}{2 \pi (T_1 + T_2)} \right )^2  
\exp \left ( 
i \frac{q^2}{2 \kappa_0} \, \frac{T_1 T_2}{T_1 + T_2} \right )
\eea
which is easily obtained from
\bea
\int_{x(0)=0}^{x(T)=x} {\cal D} x  \> e^{i S_0[x]} \EA
\left < x \left |   \exp \left ( i T \frac{\hat p^2}{2 \kappa_0} \right ) 
\right | x=0 \right > \non
\EA  \int \frac{d^4 p}{(2 \pi)^4} \> \exp \left ( - i p \cdot x + 
i \frac{p^2 T}{2 \kappa_0} \right )  \> .
\eea
Thus
\bea
\Pi(q) \EA - \frac{1}{16 \pi^2} \int_0^{\infty} 
\frac{dT_1 \, dT_2}{(T_1 + T_2)^2} \, \exp \left [ \,  
i \frac{q^2}{2 \kappa_0} \, \frac{T_1 T_2}{T_1 + T_2} - \frac{i}{2 \kappa_0}
 (M_1^2 T_1 + M_2^2 T_2) \, \right ] \non 
&& \hspace{4cm} \times \>  \frac{\int \tilde{\cal D}(x_1,x_2) \, \exp ( i 
\tilde S )}{\int \tilde{\cal D}(x_1,x_2) \, \exp ( i \tilde S_0 )}
\label{Pi worldline}
\eea
where
\bea
\int \tilde{\cal D}(x_1,x_2) &\equiv& \int d^4 x \,  e^{i q \cdot x} \, 
\int_{x_1(0)=0}^{x_1(T_1)=x} {\cal D} x_1 \, \int_{x_2(0)=0}^{x_2(T_2)=x} 
{\cal D} x_2 \\
\tilde S &\equiv& q \cdot x + S \> , \> \> \> \tilde S_0 \EQ  q \cdot x + 
\sum_{i=1}^2 \, S_0[x_i] \> .
\eea
With no interaction the ratio of path integrals in Eq. (\ref{Pi worldline}) is 
unity and one obtains the free polarization propagator. Without
performing the proper-time integrals in Eq. (\ref{Pi worldline})
one sees in this case that the exponential vanishes, i.e. 
there is no damping anymore if the ratio $r = T_2/T_1$ of proper times
fulfills the quadratic equation
\be
r^2 M_2^2 + r \left ( M_2^2 + M_1^2 - q^2 \right ) M_1^2 \E 0 \> .
\ee
This only has positive, real solutions if 
\be
q^2 \> \ge \> q_{\rm th}^2 \E \left ( M_1 + M_2 \right)^2 \> .
\label{begin of cut}
\ee
Hence there are (branch-point) singularities in the free polarization propagator
starting from values of the external momentum $q$ as given in Eq. 
(\ref{begin of cut}): 
this is just the familiar cut in the polarization propagator which describes
the two particles in the continuum. It is a {\it cut} because we integrate over
{\it two} proper times, or equivalently over all the poles generated by 
one proper-time integration (say $T_1$) depending on $r$. In the interacting
polarization propagator this cut also exists although it is modified in strength 
but not in position by the interactions between the unbound nucleon $1$ and $2$.
However, for the bound state problem we are looking for an additional 
{\it pole below} $q_{\rm th}^2$. As in the case of the two-point function 
we expect that such a pole may be generated by the undamped integration of an 
exponential over {\it one} proper time (combination).

\section{Variational Calculation for the Worldline Polarization Propagator}
\label{sec: var}
\setcounter{equation}{0}

\noindent
We now apply the Feynman-Jensen variational principle to the last factor in Eq. 
(\ref{Pi worldline}), i.e. approximate
\bea
\frac{\int \tilde{\cal D} \, \exp ( i \tilde S )}
{\int \tilde{\cal D}\, \exp ( i \tilde S_0 )} \EA 
\frac{\int \tilde{\cal D} \, \exp ( i \tilde S_t )}
{\int \tilde{\cal D}\, \exp ( i \tilde S_0 )} \, \cdot \, 
\frac{\int \tilde{\cal D} \, \exp ( i \tilde S_t ) \, 
\exp (i (\tilde S - \tilde S_t))}
{\int \tilde{\cal D}\, \exp ( i \tilde S_t )} \non
&\simeq& \frac{\int \tilde{\cal D} \, \exp ( i \tilde S_t )}
{\int \tilde{\cal D}\, \exp ( i \tilde S_0 )} \, \cdot \, 
\exp \left [ \, i \left < \tilde S - \tilde S_t \right >_t \, \right ] \> .
\label{Feyn-Jens}
\eea
In the last line the average is defined as
\be
\left < \tilde S - \tilde S_t  \right >_t \E \frac{\int \tilde{\cal D}(x_1,x_2) 
\>  \left ( \tilde S - \tilde S_t \right )  \, \exp ( i \tilde S_t )}
{\int \tilde{\cal D}(x_1,x_2) \> \exp ( i \tilde S_t )} \> .
\label{average}
\ee
It is essential for our approach that the RHS of Eq. (\ref{Feyn-Jens})  
not only yields an approximation
but is {\it stationary} w.r.t. arbitrary variations of the trial action $S_t$. 
Because the path integrals have to be solvable
we are restricted to quadratic, but retarded (i.e. non-local) trial actions. 

\subsection{The Trial Action}
\label{subsec: trial action}
In the following we will use the Fourier path integral method as employed 
originally in refs. \cite{WC1,WC2}~\footnote{In the following these references 
will be denoted by (I) and (II), respectively.}. Although there exist more general 
and elegant treatments 
\cite{WC7}, the expansion in Fourier modes  has the advantage of 
being simple and straightforward.

It is convenient to rescale the 
times in the free and the interacting part of the action
\be
t \E T_i \tau \> , \> \> \> t' \E T_j \tau' 
\ee
with $ \tau, \tau' \in [0,1] $.
The paths fulfilling the required boundary conditions can then be written as
\be
x_i(\tau) \E x \, \tau + \sum_{k=1}^{\infty} \, \frac{\sqrt{2 T_i}}{k \pi} \, 
a_k^{(i)} \, \sin \left ( k \pi \tau \right )
\ee
where, of course, the Fourier coefficients $a_k^{(i)}$ are 4-dimensional 
vectors. The free action becomes
\be
S_0 \E - \frac{\kappa_0}{2} \, \sum_{k=1}^2 \, \left [ \, x^2 \frac{1}{T_i}  +  
 \sum_{i=1}^{\infty} \, a_k^{(i) \> 2} \> \right ] \> , \> \> \> 
\tilde S_0 \E q \cdot x + S_0 \> .
\ee
This serves as a guide for a quadratic trial action which we choose as
\be
\tilde S_t \E  \tilde \lambda q \cdot x - \frac{\kappa_0}{2} \, \sum_{i=1}^2 \, 
\left [ \, A_0 \, x^2 \frac{1}{T_i}  +  
 \sum_{k=1}^{\infty} \, A_k^{(i)} \, a_k^{(i) \> 2} \, \right ] 
+ \kappa_0 \, \sum_{k=1}^{\infty} \, B_k \, a_k^{(1)} \cdot  a_k^{(2)}
\label{tilde St}
\ee
where the last term accounts for the direct coupling of the two worldlines. 
Being diagonal in Fourier space this 
is {\it not} the most general quadratic action but in all known cases (polaron,
WC model, QED) possible non-diagonal pieces do not contribute in the limit
of infinite proper time(s).
However, as shown in ref. \cite{WC7}, additional Lorentz structures 
built from external momenta (here the momentum transfer $q$) can lead to 
considerable improvements. Here we will not include such refinements and 
stick to the simple trial action (\ref{tilde St}). As shown in 
appendix~\ref{app: FJ nonrel} 
for the nonrelativistic case, such a trial action leads to the
correct center-of-mass energy and to an internal ground-state energy of the system
exactly as if evaluated with a variational trial wave function of gaussian type.
It may also be noted that in the relativistic case 
the {\it ansatz} (\ref{tilde St}) generalizes the one for 
a single nucleon, which showed the instability of the WC model in the one-particle 
sector.

\subsection{The Pole of the Variational Polarization Propagator and Mano's 
Equation}
\label{subsec: mano}
Since the trial action is at most quadratic in the dynamical variables,
all path integrals needed for 
Feynman's variational principle can be performed. The calculation of these
averages is outlined in appendix~B and gives the result 
(\ref{app:Pi var 1})
\bea 
\Pi^{\rm var}(q) \EA - \frac{1}{16 \pi^2} \, \int_0^{\infty} \frac{dT_1 \, dT_2} 
{(T_1 + T_2)^2} \, \exp \Biggl \{ \, \frac{i}{2 \kappa_0} \, 
\Biggl [ \, - (M_1^2 T_1 + M_2^2 T_2) \non
&& + \frac{T_1 T_2}{T_1 + T_2} \, q^2  \left ( 
2 \lambda - \lambda^2 \right ) - (T_1 + T_2) \, \left ( \, \Omega_{12}  +  V
\, \right )(T_1,T_2)  \Biggr ]  \Biggr \} 
\label{Pi var 1}
\eea
where $\Omega_{12}$ and $V$ are defined in Eq. (\ref{def Omega12}) and
(\ref{def V}), respectively. The original variational parameter $\tilde \lambda$
has been replaced by
\be
\lambda \E \frac{\tilde \lambda}{A_0} \> .
\label{def lambda}
\ee

As in the case of the 2-point function (or the nonrelativistic 
polarization propagator) the variational expression simplifies 
considerably on the pole. However, now we have {\it two} times $T_1, T_2$ and 
it is not clear {\it a priori} which one, or which combination should go to infinity
in order to produce the (single) pole. For the equal-mass case (\ref{equal mass})
the solution simply is to introduce
\be
T \Def \frac{T_1 + T_2}{2} \> , \hspace{0.5cm} s \Def T_1 - T_2 \> .
\label{def T}
\ee
Since $ T_1 T_2 /(T_1 + T_2) = T/2 - s^2/(8 T) $
we have from Eq. (\ref{Pi var 1})
\be 
\Pi^{\rm var}(q) \E - \frac{1}{64 \pi^2} \, \int_0^{\infty} dT \, 
\exp \Biggl \{ \> \frac{i}{4 \kappa_0} \, T 
\Biggl [ \, - 4 M_0^2   + q^2 \, \left ( 
2 \lambda - \lambda^2 \right ) + \frac{4 \kappa_0}{iT} \ln g(q^2,T) \, \Biggr ] 
\> \Biggr \}
\label{Pi var 2}
\ee
where
\bea
g(q^2,T) \EA \frac{1}{T^2} \int_{-2T}^{+2T} ds \> \exp \Biggl \{ \, 
\frac{i}{2 \kappa_0} \Bigl [ 
- \frac{s^2}{8 T} q^2 (2 \lambda 
- \lambda^2 ) \non
&& \hspace{1cm} - 2 \, T \, \left ( \, \Omega_{12} + V \, \right ) 
\left ( T+ \frac{s}{2}, T - \frac{s}{2} \right ) 
  \, \Bigr ] \, \Biggr \} \> .
\eea
It is obvious that an additional pole develops if $ \ln g(q^2,T) $ has terms linear
in $T$ for large $T$. This happens if both $\Omega_{12}$ and $V$ are 
independent of $T$ and $s$ in the limit $T \to \infty$.  Then 
\be
\frac{4 \kappa_0}{iT} \ln g(q^2,T) \> \stackrel{T \to \infty}{\longrightarrow} \>
- 4 \left ( \, \Omega_{12} + V \, \right ) - \frac{4 \kappa_0}{i} \, 
\frac{\ln T^2}{T} +  {\cal O} \left ( \frac{1}{T} \right )
\ee
because the remaining $s$-integral is convergent.

The proof that $ \Omega_{12}, V$ become constants in the large $T$-limit follows 
along the same lines as in the 2-point case: 
for $T \to \infty$ we may assume that the coefficients $A_k = A(k \pi/T) $ and that 
the sums over Fourier coefficients become integrals over 
$ E = k \pi/T $
\be 
\sum_k f \left ( A_k \right ) \> \longrightarrow \> \int_0^{\infty} dk \> 
f \left ( 
A \left ( \frac{k \pi}{T} \right ) \right ) \E \frac{T}{\pi} \, 
\int_0^{\infty} dE \> f \left ( A(E) \right ) \> .
\label{asympt}
\ee
In this way we obtain from Eq. (\ref{def Omega12})
\bea
\Omega_{12} \EA  \frac{d \kappa_0}{2 i \pi}  \int_0^{\infty} dE  \> \Biggl [ \, 
 \ln \left ( A^{(1)}(E) A^{(2)}(E) - B^2(E) \right ) \non
&& \hspace{3cm} +  
\frac{A^{(1)}(E)+ A^{(2)}(E)}{A^{(1)}(E) A^{(2)}(E) - B^2(E)} 
- 2  \Biggr ]  \> .
\eea
When the coupling function $B(E)$ vanishes we find 
that $ \Omega_{12} |_{B=0} = \Omega_1 + \Omega_2 $, 
i.e. the kinetic term reduces to the sum of usual kinetic terms for the
self-energy of the individual nucleons (see ref. (I)).

To evaluate $V$ from Eq. (\ref{def V}) we go back to the original (unscaled)
times $t_1 = T_1 \tau , t_2 = T_2 \tau'$ and note that for
\be
\sigma \Def t_1 - t_2 \> , \> \> \Sigma \Def \frac{t_1+t_2}{2} \> \> 
\Longrightarrow \> \> 
\int_0^T dt_1 \int_0^T dt_2 \> \ldots \E \int_{-T}^{+T} d\sigma \, 
\int_{|\sigma|/2}^{T-|\sigma|/2} d\Sigma \> \ldots
\ee
The integral over the relative time $\sigma$ remains whereas 
the integral over the total time $ \Sigma$ 
produces the needed factor $T$. Therefore we obtain
\be 
V \E  \sum_{i,j=1}^2 \,  \frac{g_i g_j}{8 \kappa_0}  
\, \int_{-\infty}^{+\infty} \! d\sigma\> \int \frac{d^4 p}{(2 \pi)^4} \> 
\frac{1}{p^2 - m^2 + i0} \, \exp \left \{ \frac{i}{2 \kappa_0} \, 
\left [  p^2 \mu_{i j}^2(\sigma) - \lambda \, p \cdot q \, \sigma  
\right ]  \right \} .
\label{V explicit}
\ee
This only works if the quantity 
$\mu^2_{i j}$ defined in Eq. (\ref{app:mu2(tau,tau')}) becomes 
independent of the total time $\Sigma$ in the limit $T \to \infty$. Inspection of  
Eq. (\ref{app:mu2(tau,tau')}) 
together with the addition theorems of trigonometric 
functions shows that this is indeed the case:
\be
\mu_{i j}^2(t_1,t_2;T_1,T_2) \E
\mu_{i j}^2(\sigma, \Sigma;s,T) 
\> \stackrel{T \to \infty}{\longrightarrow} \>  \mu_{i j}^2(\sigma) 
\ee
with
\bea
\mu_{i j}^2(\sigma) \EA \frac{2}{\pi} \, \int_0^{\infty} \frac{dE}{E^2}
 \>  \frac{1}{A^{(1)}(E) A^{(2)}(E) - B^2(E)} \,
  \Biggl  \{ \, 2 \, \delta_{ij} \, A^{(3-i)}(E)  
\, \sin^2 \left ( \frac{E\sigma}{2} \right ) \non
&& \hspace{1cm} + \left ( 1 - \delta_{ij} \right ) \, \left [ \, 
\frac{ A^{(1)}(E) + A^{(2)}(E)}{2}
- B(E) \, \cos \left ( E \sigma \right ) \, \right ] \, \Biggl \} \> .
\label{mu2 (sigma)}
\eea
This is the generalization of the pseudotime $ \mu^2(\sigma)$ for the 
self-energy of a single nucleon encountered in previous work. 
Even the diagonal $(i = j)$-part now depends on both profile functions 
as well as on the coupling function $B(E)$. If the coupling vanishes then
$ \mu_{i i}^2 $
reduces to the usual pseudotime for the $i^{\rm th}$ particle.
Similarly, the off-diagonal $(i \ne j)$-part also depends 
on both profile functions $A^{(1)}, A^{(2)}$. The mutual dependence  reflects the
change of properties of the one particle in the presence of the other 
(``medium effect'') and the vertex corrections mentioned before. 
Note also the missing factor $2$ in front of the
last term in the exponential of Eq. (\ref{V explicit}) compared to the
single particle (2-point function) case. This means that on the
average each particle in the momentum loop of the polarization propagator has
momentum $q/2$ and -- since $(q/2)^2 < M^2 $ in the bound state  -- is slightly 
off-shell. Finally, it is evident that $ \mu_{i j}^2 =
\mu_{j i}^2$ .

Collecting all terms in the exponent of Eq. (\ref{Pi var 2}) which are linear 
in $T$ we obtain (what we usually call) {\it Mano's equation} 
\be
0 \E - 4 M_0^2 + q^2 \, \left ( 2 \lambda - \lambda^2 \right ) - 4 \, \left (\,  
\Omega_{12} + V \, \right ) \> .
\label{Mano 1}
\ee
determining a possible additional pole in the polarization propagator. 
In view of the previous remark it may also be written as
\be
\left ( \frac{q}{2} \right )^2 \, \left ( 2 \lambda - \lambda^2 \right ) 
\E M_0^2 + \Omega_{12} + V
\label{Mano 2}
\ee
which very much resembles Mano's equation for the 2-point function. The missing 
factor $2$ on the RHS in front of the kinetic and potential terms is 
contained in the definitions (\ref{def Omega12},\ref{def V}). One may split up 
the interaction as
\be
V \deF \sum_{i=1}^2 \, V_{ii} + 2 \, V_{12}
\ee
with 
\be
V_{ij} \E  \frac{g_i g_j}{4 \kappa_0} \, \int_0^{\infty} d\sigma \,  
\int \frac{d^4 p}{(2 \pi)^4} \, 
\frac{1}{p^2 - m^2 + i0} \, \exp \left [ \, \frac{i}{2 \kappa_0} \, 
\left ( \, p^2 \mu_{ij}^2(\sigma) - \lambda \, p \cdot q \, \sigma \, \right ) 
\, \right ] \> .
\label{interaction}
\ee
Here we have used the fact that the pseudotime is even 
to restrict the $\sigma$-integration to the interval $ [0,\infty]$.
One may call $V_{ii}$ the self-energy part and $V_{12}$ the binding part
of the interaction, although these effects are coupled as mentioned 
before.

Let us discuss Mano's Eq. (\ref{Mano 2}) qualitatively: without the kinetic and 
potential terms one would obtain $\lambda =1 $ as a stationary solution 
(as in the nonrelativistic case) and $q_{\star}^2 = q_{\rm th}^2 = 4 M^2$ 
would determine the beginning of the
continuum branch-cut. The effects of the interaction can best be estimated in the 
euclidean formulation ($\kappa_0 = i \kappa_E$). Then for small coupling
\bea
\Omega_{12} &\simeq& \frac{d \kappa_E}{2 \pi} \int_0^{\infty} dE \, \Biggl \{  
\sum_{i=1}^2 
\left [  \, \ln A^{(i)} + \frac{1}{A^{(i)}} - 1  \right ] 
+ \frac{B^2}{A^{(1)} A^{(2)}} \, \left ( \frac{1}{A^{(1)}} + \frac{1}{A^{(2)}}
- 1 \right ) \non
&& \hspace{7.5cm} + \, {\cal O} \left ( B^4 \right )  \Biggr \}
\eea
which is positive 
since $\ln A + 1/A - 1 \ge 0 $ for $ A > 0 $ and the second term 
adds for $ A^{(i)} < 2 $. Thus the kinetic term tends to 
push $q_{\star}^2$ above the threshold value. On the other hand, 
from Eq. (\ref{V explicit}) the potential term can be seen to 
be attractive: $V < 0 $ since $dp_0 = i dp_4$ and $ p^2 - m^2 = -(p_E^2 + m^2) $.
This is the usual competition between kinetic and potential energy and one may
therefore expect a solution of Mano's equation with 
$q_{\star}^2 < 4 M_{\rm phys}^2 $,
i.e. a truly relativistic bound state for sufficient strong coupling. 
Since we know
that the instability of the WC model prevents large coupling,
the nonrelativistic criterion (\ref{crit}) does not apply anymore.

\subsection{Regularization and Renormalization}
\label{subsec: renorm}

\noindent
As usual in quantum field theory divergences show up and have to be absorbed in
physical, i.e. measured quantities ({\it renormalization}). Fortunately, 
the Wick-Cutkosky model
is a super-renormalizable theory (Eq. (\ref{def alpha}) shows that 
the coupling constants $g_i$ have the dimension of a mass) and in the 
quenched approximation only a mass renormalization is necessary. This has been done
already for the 2-point function in refs. (I,II)  and, indeed, the source of the
divergence is again the self-energy of the particles. To be more specific we expect 
that only the diagonal pseudotime vanishes at small relative times:
\be
\mu^2_{ii}(\sigma) \> \stackrel{\sigma \to 0}{\longrightarrow} \> \sigma 
\label{diag mu2 small sigma}
\ee
In contrast, the non-diagonal
part of the pseudotime stays constant (see appendix~C):
\be 
\mu^2_{12}(\sigma) \> \stackrel{\sigma \to 0}{\longrightarrow} \> {\rm const.}
\ee
This means that for $\sigma \to 0, p \to \infty $ the 
$p$-integrand for $V_{ii}$ (but not for $V_{12}$ !) loses its
exponential suppression which leads to a (large $p$ or small
$\sigma$) UV-divergence. The solution is straightforward
and simple (compared to a {\it real}, i.e. renormalizable theory with
a dimensionless coupling constant): regularize the divergent
expression, isolate the divergence and combine it with the (squared) bare mass
of the particle. In ref. (I) we simply subtracted a particular term with 
the meson mass as a scale.

Another possibility for regularization is a cut-off 
at small proper time ({\it proper-time regularization}), i.e.
\be
\int_0^{\infty} d\sigma \ldots \> \longrightarrow \> 
\int_{1/\Lambda^2}^{\infty} d\sigma \ldots
\label{prop time reg}
\ee
It is understood that at the end of the calculation we let the cut-off $\Lambda$
(with dimension mass squared) tend to infinity. We then write the regularized
expression for $V_{ii}$ as
\bea
V_{ii} \EA \frac{g_i^2}{4 \kappa_0} \,  
\int_{1/\Lambda^2}^{\infty} d\sigma\> \int \frac{d^4 p}{(2 \pi)^4} \> 
\frac{1}{p^2 - m^2 + i0} \, \exp \left [ \, \frac{i}{2 \kappa_0} \, p^2 \sigma
\right ] \non
&& 
+ \frac{g_i^2}{4 \kappa_0}  
\int_{1/\Lambda^2}^{\infty} \!  d\sigma \int \frac{d^4 p}{(2 \pi)^4} 
\frac{1}{p^2 - m^2 + i0} \, \Biggl \{  
 \exp \left [  \frac{i}{2 \kappa_0} 
\left ( p^2 \mu_{ii}^2(\sigma) - \lambda  p \cdot q  \sigma 
\right )  \right ] \non 
&& \hspace{4cm} - \exp \left [  \frac{i}{2 \kappa_0}  p^2 \sigma
\right ]  \Biggr \} \> \deF \> V_{ii}^{\rm sing} + V_{ii}^{\rm reg} \> . 
\label{Vsing + Vreg}
\eea
The singular part can be evaluated easily:
\bea
V_{ii}^{\rm sing} \EA  \frac{g_i^2}{4 \kappa_0} 
\, \int_{1/\Lambda^2}^{\infty} d\sigma 
\int_0^{\infty} \frac{du'}{2 \kappa_0 i}  
\> \int \frac{d^4 p}{(2 \pi)^4} \,
\exp \left \{   \frac{i}{2 \kappa_0} \, \left [ (p^2 - m^2) u' + p^2 \sigma
\right ]  \right \} \non
\EA - \frac{ g_i^2}{32 \pi^2}   \int_0^{\infty} du'
\, \frac{1}{u' + 1/\Lambda^2} \, \exp \left [ - \frac{i}{2 \kappa_0} 
m^2 u' \right ] = - \frac{ g_i^2}{32 \pi^2} \, e^z  E_1(z) 
\label{V1 sing}
\eea
where $ z = i m^2/(2 \kappa_0  \Lambda^2) $ and 
$E_1(z)$ is the exponential integral (ref. \cite{Handbook}, Chap. 5). Using its 
expansion for small arguments (i.e. large cut-off $\Lambda$) one obtains
\be
V_{ii}^{\rm sing} \Bigr |_{\rm prop. \, time \, reg.} \> 
\stackrel{\Lambda\to \infty}{\longrightarrow} \> - 
\frac{ g_i^2}{32 \pi^2} \, \ln \left ( 
\frac{\Lambda^2}{m^2} \right ) + C
\label{V prop time reg}
\ee
where 
\be
C  \Bigr |_{\rm prop. \, time \, reg.} \E  \frac{ g_i^2}{32 \pi^2} \, \left [ 
\, \gamma_E + \ln \left ( \frac{i}{2 \kappa_0} \right ) \, \right ]
\label{constant}
\ee
is a constant specific for this regularization scheme. The
divergent part plus the constant (or part of it) can be absorbed in the bare mass 
of each particle by defining the finite (intermediate) mass
\be
\bar M^2_i \E  M_0^2 - \frac{ g_i^2}{16 \pi^2} 
\, \ln \left ( \frac{\Lambda^2}{m^2} \right ) + 2 \, C \,  \> .
\label{bar M}
\ee
For $ g_1 = g_2 = 2 g, C = 0 $ this is exactly the quantity 
\footnote{Note that in these
references the intermediate mass has been called $M_1$. 
To avoid confusion with the {\it bare} mass of particle $1$ we use 
$\bar M_1$, etc. in the present work.}
which has been calculated in refs. (I,II) 
by requiring a pole of the 2-point function at the physical mass $M$. The 
remaining, regularized interaction just provided a finite mass
shift to the observed mass of the particle. Note that in $V_1^{\rm reg}$ 
we now can send the cut-off to infinity 
since the integrals are convergent by construction. In this way all traces of 
the cut-off have disappeared, or better are hidden in the masses $ \bar M_i $. 
As can be seen in Eq. (\ref{constant}) proper-time and other
regularization methods with a cut-off should only be used in euclidean time 
($\kappa_0 = i \kappa_E$) to make everything real.
Other regularization schemes do not need that restriction:
for example, after performing the $\sigma$-integration, one has
in $d = 4 - 2 \epsilon $ dimensions
\be
V_{ii}^{\rm sing} \Bigr |_{\rm dim} \E \frac{g_i^2}{4 \kappa_0} \, \nu^{2 \epsilon}
\> \int \frac{d^d p}{(2 \pi)^d} \> \frac{1}{p^2 - m^2 + i0} \, 
\frac{2 i \kappa_0}{p^2}
\ee
where $\nu$ is a mass parameter introduced for keeping the mass 
dimension of the coupling constant fixed in arbitrary dimensions. The $p$-integral
is a standard one (see, e.g. appendix~B in ref. \cite{Ynd}) and 
one obtains
\bea
V_{ii}^{\rm sing}  \Bigr |_{\rm dim} \EA - \frac{g_i^2}{32 \pi^2} \, 
\, \left (\frac{4 \pi \nu^2}{m^2} \right )^{\epsilon} \, 
\frac{\Gamma(\epsilon)}{ 1 - \epsilon} \non
&& \stackrel{\epsilon \to 0}{\longrightarrow} \>  - \frac{g_i^2}{32 \pi^2}
\, \left [ \, \frac{1}{\epsilon} + 1 - \gamma_E + \ln (4 \pi) + 
\ln \left ( \frac{\nu^2}{m^2} 
\right ) + {\cal O}(\epsilon) \, \right ] 
\label{V1 sing dim reg}
\eea
which is independent of $\kappa_0$ and real. As usual one may combine the bare mass
with the $1/\epsilon$-divergence leading to the minimal subtraction (MS) scheme or 
include some of the constants ($\overline{\rm MS}$).
In the Pauli-Villars regularization
one stays in $d=4$ dimensions but modifies the meson propagator
\be
\frac{1}{p^2 - m^2 + i0} \> \longrightarrow \> \frac{1}{p^2 - m^2 + i0} - 
\frac{1}{p^2 - \Lambda^2 + i0} \> .
\ee
This just amounts subtracting from Eq. (\ref{V1 sing dim reg}) the same 
expression with $m^2 \to \Lambda^2$. Hence one obtains the same result 
as in Eq. (\ref{V prop time reg}) with no additional constant.

Whatever regularization one chooses, the divergent part of $V_{ii}$ 
can be combined with the (squared) bare masses of the nucleons. 
Of course, there is the ambiguity of adding possible constants to the singular part 
of the interaction and subtracting them from the regular part,
leading to different renormalization schemes.
The scheme used previously did not have
a constant term in Eq. (\ref{V prop time reg}) and we will also choose this
convention. However, it should be stressed that this only affects the relation 
between the unobservable bare mass and the intermediate mass and all physical 
results are independent of this choice.

\subsection{Special Cases}
\label{subsec: special}

\noindent
Let us now consider the case where the two particles not only have the same mass
but also the same coupling constant
\be
g_1 \E  g_2 \equiv g' \> .
\ee
Their dynamics being the same we can assume
\be
A^{(1)} (E) \E A^{(2)} (E) \EQ A(E)
\ee
and Mano's Eq. reads
\be
\bar M^2 \E \left ( \frac{q}{2} \right )^2 \, ( 2 \lambda - \lambda^2 ) - 
\Omega_{12} - 2 \left ( \, V_{11}^{\rm reg} + V_{12} \, \right ) \> .
\label{Mano 3}
\ee
Here
\be
\Omega_{12} \E \frac{d \kappa_0}{2 \pi i} \, \int_0^{\infty} dE \> \left \{ \,
\ln \left [  A^2(E) - B^2(E) \right ] + \frac{2 A(E)}{A^2(E) - B^2(E)} - 2 \, 
\right \}
\ee
and
\bea
V_{1j} \EA \frac{g'^2}{4 \kappa_0} \, \int_0^{\infty} d\sigma \int 
\frac{d^4 p}{(2 \pi)^4} \> \frac{1}{p^2 - m^2 + i0} \,  
\Biggl \{ \, \exp \left [ \frac{i}{2 \kappa_0} \left (
p^2 \mu^2_{1j}(\sigma) - \lambda p \cdot q \sigma \right ) \right ] \non
&& \hspace{6.5cm} - \delta_{1j} \,  \exp \left [ \frac{i}{2 \kappa_0} \, 
p^2 \sigma  \right ]  \, \Biggr \} 
\label{V1alpha}
\eea
for $j = 1,2$ . The pseudotimes are related to the profile functions by
\bea 
\mu_{11}^2(\sigma) \EA \frac{4}{\pi}  \int_0^{\infty} dE \>  \frac{1}{E^2} \, 
\frac{A(E)}{ A^2(E) - B^2(E)} \,  \sin^2 \left ( 
\frac{E \sigma}{2} \right ) \\
\mu_{12}^2(\sigma) \EA \frac{2}{\pi}  \int_0^{\infty} dE \,  \frac{1}{E^2} \, 
\frac{1}{ A^2(E) - B^2(E)} \, \Bigl [ \, A(E) - B(E) 
\cos (E \sigma) \, \Bigr ] .
\eea
It is very useful to introduce the combinations
\be
A_{\pm}(E) \Def A(E) \pm B(E)
\label{def Aplusminus}
\ee
because then the kinetic term separates in two distinct pieces
\be
\Omega_{12} \E \Omega[A_-] + \Omega[A_+] \> .
\label{total Omega}
\ee
Here 
\be 
\Omega[A] \E  \frac{d \kappa_0}{2 i \pi} \, 
\int_0^{\infty} dE \> \left [ 
\, \ln A(E)  + \frac{1}{A(E)} - 1 \, \right ]  
\label{usual Omega}
\ee
is just the usual kinetic term encountered in the self-energy of 
a single nucleon (see (I)). The pseudotimes now become
\bea 
\mu_{11}^2(\sigma) \EA \frac{2}{\pi} \, \int_0^{\infty} dE \>  \frac{1}{E^2} \, 
\left [ \frac{1}{A_-(E)} + \frac{1}{A_+(E)} \right ] \, \sin^2 \left ( 
\frac{E \sigma}{2} \right ) 
\label{mu11 by Aplusminus}\\
\mu_{12}^2(\sigma) \EA \frac{2}{\pi} \, \int_0^{\infty} dE \>  \frac{1}{E^2} \, 
\left [ \frac{\sin^2(E \sigma/2)}{A_-(E)} + \frac{\cos^2(E\sigma/2)}{A_+(E)} 
\right ] \> .
\label{mu12 by Aplusminus}
\eea
Note the appearance of the cosine function in the `interaction pseudotime'
$\mu^2_{12}$ which leads to 
a finite value at $\sigma = 0$, in contrast to the 'self-energy pseudotime'
$\mu^2_{11}$ which vanishes at that point. As explained before this requires
a subtraction in $V_{11}$ but not in the interaction part $V_{12}$. 
We may perform the $p$-integration in the usual way by exponentiating the
meson propagator (cf. Eq. (\ref{V1 sing})). Then we obtain explicitly
in euclidean time ($ \kappa_0 = i \kappa_E$ ) 
\bea
V_{11}^{\rm reg} \EA - \frac{\alpha}{2 \pi} M^2 \, \int_0^{\infty} d\sigma \> 
\int_0^1 du \> 
\Biggl [ \, \frac{1}{\mu^2_{11}(\sigma)} \, e \left ( m \, \mu_{11}(\sigma), 
\frac{\sigma \lambda q/2}{\mu_{11}(\sigma)},u \right ) \non
&& \hspace{5cm} - \frac{1}{\sigma} \, 
e \left ( m \, \sqrt{\sigma},0,u \right ) \, \Biggr ] 
\label{V11 explicit}\\
V_{12} \EA - \frac{Z \alpha}{2 \pi} M^2 \, \int_0^{\infty} d\sigma \> 
 \frac{1}{\mu^2_{12}(\sigma)} \, 
\int_0^1 du \> e \left ( m \, \mu_{12}(\sigma), 
\frac{\sigma \lambda q/2}{\mu_{12}(\sigma)},u \right ) 
\label{V12 explicit}
\eea
where
\bea
e \left ( m \, \mu_{1j}(\sigma), 
\frac{\sigma \lambda q/2}{\mu_{1j}(\sigma)},u \right ) \EA \exp \left \{ \, - 
\frac{1}{2 \kappa_E} \left [ m^2 \mu^2_{1j}(\sigma)  \frac{1-u}{u}
+ \frac{(\lambda q/2)^2 \sigma^2}{ \mu^2_{1j}(\sigma) } u \right ] \,  
\right \} \non
&& \deF \>  e_j(u,\sigma) \> .
\label{def e(u,sigma)}
\eea
is the function used in refs. (I,II). Here $q/2$ is a  shorthand notation for 
$\sqrt{q^2}/2 $ and the dimensionless coupling constant as defined in Eq. 
(\ref{def alpha}) is given by
\be
\alpha \E \frac{g'^2}{16 \pi M^2} \> .
\label{def alpha equal mass}
\ee
In Eq. (\ref{V11 explicit}) we now explicitly
use the renormalization scheme of ref. (I) , i.e. subtract a simple
exponential in the proper-time integral which has the meson mass $m$ as a scale.
Subtraction at a different scale $\nu$ (which, e.g. is needed for  $m = 0 $) 
would add the term 
\bea
&& \frac{\alpha}{2 \pi} M^2 \int_0^{\infty} d\sigma \> \frac{1}{\sigma} \, 
\int_0^1 du \> 
\left [ \, \exp \left ( - \frac{\nu^2}{2 \kappa_E} \frac{1-u}{u} \, \sigma \right )
-   \exp \left ( -\frac{m^2}{2 \kappa_E} \frac{1-u}{u} \, \sigma \right) \, 
\right ] \non
\EA  \frac{\alpha}{2 \pi} M^2 \, \int_0^1 du \> \ln \frac{m^2}{\nu^2} \E 
\frac{\alpha}{2 \pi} M^2 \, \ln \frac{m^2}{\nu^2} \> \stackrel{!}{=} \> C_{\nu}
\eea
on the RHS of Eq. (\ref{V11 explicit}). This is exactly the same term as obtained
in Eq. (\ref{V1 sing dim reg}) directly from dimensional regularization and amounts to 
chosing another constant $C$ in Eq. (\ref{bar M}). Therefore one obtains a new
intermediate mass
\be
\bar M^2(\nu) \E M_0^2 - \frac{\alpha}{\pi} M^2 \, \ln \left ( 
\frac{\Lambda^2}{\nu^2} \right )
\ee
which is independent of $m$ but $\nu$-dependent.
Obviously, the scale $\nu$ is arbitrary and the physical 
observables do not depend on it. Note also that $ \Omega_{12}$ and $V_{1j}$ are 
independent of the reparametrization parameter $\kappa_E$. This is because 
profile functions and pseudotimes scale as
\be
A_{\pm}(\kappa_E,E) \> \equiv \> A_{\pm}(\kappa_E E) \> \> 
\Longrightarrow \>  \mu_{1j}^2(\kappa_E,\sigma) \> \equiv \> \kappa_E \, 
\mu_{1j}^2 \left ( \sigma/\kappa_E \right )  
\label{rep behav}
\ee
where the last relation follows from Eqs. 
(\ref{mu11 by Aplusminus}, \ref{mu12 by Aplusminus}).
Using the explicit forms of kinetic terms and potentials it is easily seen that
these quantities do not depend on the parameter $\kappa_E$.
Mano's equation then implies that also the variational parameter 
$ \lambda $ is reparametrization-invariant.

\noindent
In addition, in Eq. (\ref{V12 explicit}) we have
introduced an artificial strength factor $Z$ which in the end has to be set to one
but allows one to distinguish  between the binding part and the radiative corrections.
This is standard practice in QED bound-state calculations for atomic systems 
including hydrogen. In the present case, however, the factor $Z$ has no 
physical meaning since it does not reflect the different coupling constant of 
one of the particles; otherwise we would have to include it in the corresponding 
self-energy of this particle.
With this artificial parameter to switch on and off one now
can study several special cases:

\bdes
\item{(i) $Z = 0 \> $ : } Since there is no dependence on $\mu^2_{12}$ and 
$\mu^2_{11}$ is
symmetric in $A_-,A_+$ (see Eq. (\ref{mu11 by Aplusminus})) both profile functions
give the same contribution and Mano's Eq. (\ref{Mano 3}) for the 2-body case 
becomes identical to the one for the self-energy of a single nucleon.

\item{(ii) $\alpha = 0, Z\alpha \ne 0 \> $ :} This corresponds to the case studied 
by practically all relativistic bound-state approaches (with the exception
of atomic systems where one does perturbative QED calculations): neglect 
self-energy and vertex corrections. Naively one then expects no sign of an 
instability in the WC model but we will see that this is {\it not} the case.
Since now $\bar M = M $ Mano's Eq. simplifies to
\be
M^2 \E \left ( \frac{q}{2} \right )^2 \, ( 2 \lambda - \lambda^2 ) - 
\Omega[A_-] - \Omega[A_+] - 2 \, V_{12} \> .
\label{Mano alpha = 0}
\ee
Still, when varying this equation one includes all crossed exchange diagrams 
(in variational approximation with our quadratic trial action ...), not just
iteration of the ladder diagrams. It is unclear how one
can simulate the ladder Bethe-Salpeter approximation in the present 
worldline approach as it
naturally contains all orderings of internal lines. Similar remarks apply to
approximations where a 3-dimensional reduction has been performed \cite{BlSu})
or one particle has been put on the mass-shell \cite{Gross}. 

\item{(iii) $ c \to \infty \> $ : } For the nonrelativistic limit
one could re-introduce the velocity of light $c$ and let it tend to infinity
as in appendix~\ref{app: nonrel}.
A simpler method is to perform the limit $M \to \infty$
in the unsubtracted expression (\ref{interaction}) keeping the coupling constant
$\alpha$ fixed. Setting $ q = (2 M + \epsilon_{\bfq}, \bfq) $ one sees that the 
term $ p \cdot q$ in the exponential forces $ p_0 \to 0 $ in all other terms.
The $p_0$-integration then produces a factor
\be
2 \pi \delta \left ( \frac{\lambda q_0 \sigma}{2 \kappa_0 } \right ) \E 
\frac{2 \kappa_0 }{ \lambda q_0} \, 2 \pi \delta (\sigma) \> ,
\ee
i.e. an {\it instantaneous} pion exchange. Therefore
\be
V_{1j} \> \stackrel{c \to \infty}{\longrightarrow} \> - 
\frac{4 \pi Z^{j-1} \alpha M^2}{\lambda q_0} 
\, \int \frac{d^3 p}{(2 \pi)^3} \> \frac{1}{{\bf p}^2 + m^2} \,  
\exp \left [ \, - \frac{i}{2 \kappa_0} {\bf p}^2 \, \mu^2_{1j}(0) \, \right ]
\> .
\ee
Because $ \mu^2_{11}(0) = 0 $ 
the unregularized self-energy interaction $V_{11}$ now is just a (divergent) 
constant which changes $M_0^2 \to M^2 $ but has no effect on the dynamics of the
particles (we anticipate $ \lambda \to 1$ for $ M \to \infty$).
Writing $ \, \mu^2_{12}(0) \deF 1/\omega \, $
Mano's Eq. (\ref{Mano 3}) then reads in the nonrelativistic limit
\be
M^2 \E \left ( \, M^2 + \epsilon_{\bfq} M - \bfq^2/4 \,  \right )^2 \, ( 2 \lambda - 
\lambda^2 ) - \Omega_- - \Omega_+ - 2 \, V_{12} (\omega)
\ee 
from which we immediately deduce by variation that $A_-(E) = 1$ 
as $V_{12}$ does not depend on it anymore. Hence $\Omega_- = 0 $ . 
Assuming $ \, V_{12}(\omega) \ll M^2 \, $ one finds 
$ \, \lambda \to 1 \,$ as expected
and therefore for the total nonrelativistic energy
\be
\epsilon_{\bfq}  \E \frac{\bfq^2}{4 M} + \frac{1}{M} 
\Bigl [ \, \Omega_+ + 2 \, V_{12} (\omega) \, \Bigr ]   \> .
\ee
Variation w.r.t. $A_+(E)$ shows that this profile function has the 
familiar (euclidean) form $ \, A_+(E) = 1 + \omega^2/E^2 \, $,
in agreement with the nonrelativistic variational calculation in 
appendix~\ref{app: FJ nonrel} (see Eq. (\ref{solution Ai,B}) for $M_1 = M_2$). 
Consequently, Mano's equation for the internal energy $\epsilon_0$  
(with $\kappa_0 = i \kappa_E $)
\be
\epsilon_0  \E \frac{d}{4 M} \, \kappa_E \omega
- 4 \pi Z \alpha \int \frac{d^3 p}{(2 \pi)^3} 
\> \frac{1}{\bfp^2 + m^2} \, \exp \left [ - 
\frac{\bfp^2}{2 \kappa_E \, \omega} \right ]
\ee
is the correct nonrelativistic variational equation (\ref{Mano omega}) for the 
Yukawa potential if we put $ d = 3 $.
For the Coulomb case ($m = 0$) 
where all calculations can be done analytically this gives the results
\bea
\kappa_E \, \omega \EA \frac{8}{d^2 \pi} \, (Z \alpha)^2 M^2 
\label{om nonrel d} \\
\epsilon_0 \EA - \frac{4}{d \pi} \, (Z \alpha)^2 \, \frac{M}{2} \> .
\label{Coul nonrel d}
\eea
However, there is no compelling reason to reduce the dimensionality to $d = 3$ 
by {\it fiat} and therefore we conclude that the nonrelativistic limit of the
variational calculation is different from the variational calculation in the 
nonrelativistic limit which starts from the beginning with 
$d = 3 $. This phenomenon can be traced back to the ansatz (\ref{tilde St}) for the 
trial action which is covariant but too rigid to allow a different treatment of 
time and space-like dynamical variables needed for the correct
nonrelativistic limit. An obvious remedy is to give the trial action more freedom, 
for example by allowing more general Lorentz structures for the profile
functions
$ A(E) \to A^{\mu \nu} (E) =  A_L(E) \, q^{\mu} q^{\nu}/q^2 + A_T(E) \, 
\left ( g^{\mu \nu} -  q^{\mu} q^{\nu}/q^2 \right ) $ 
and similarly for $ B(E) $. Indeed, 
the kinetic term then splits into $ \Omega_L/d + (d-1) \Omega_T/d \> $ (see Eq. (19) 
in ref. \cite{WC7}) which would provide the necessary dimensional reduction for
the nonrelativistic limit without destroying the covariance of the whole approach.
Actually, as shown in appendix~C of ref. \cite{WC7}, the variational principle
{\it demands} such a general form if the profile functions are left 
completely free.

\item{(iv) $\alpha \to 0 \> $ : } Although our main interest is in the 
strong-coupling case a detailed investigation of the weak-coupling limit
is worthwhile, for example to see whether the logarithmic terms of 
the ladder Bethe-Salpeter approximation (see, e.g. Eq. (10-78) in ref. \cite{ItZu}) 
also appear in the present approach. Since the nonrelativistic limit coincides
with the weak-coupling limit we can expect that Eq. (\ref{Coul nonrel d}) with 
$d = 4 $ is the leading term for $ m = 0$ . Indeed, as shown in appendix~
\ref{app: weak var bind} the binding energy in the worldline variational approach
has the expansion
\be
\frac{\epsilon_0}{M/2}  \E - \frac{(Z \alpha)^2}{\pi} \, \left [ \, 1 + 
\frac{7}{2} \, \frac{\alpha}{\pi} + \ldots \, 
\right ] \, - \frac{(Z \alpha)^4}{\pi^2} \, \Bigl [ \, 1 + \ldots \, \Bigr ] -  \ldots 
\> .
\label{perturb binding}
\ee
This shows an {\it increased} binding caused both by radiative effects 
(indicated by the $\alpha/\pi$-terms) and by
relativistic corrections (indicated by the $(Z \alpha)^4$-term). Further 
discussion is postponed until Sect. \ref{subsec: comp}.

\edes

\section{Variational Equations}
\label{sec: var eqs}
\setcounter{equation}{0}

\noindent
Here we derive the variational equations for the $\lambda-$parameter and profile
functions $A_{\pm}(E)$ from Mano's equation (\ref{Mano 3}).
As this is straightforward we just quote the results:  
assuming that $\kappa_E$ does not depend on $\lambda$ 
variation w.r.t. $\lambda$ gives
\footnote{One can choose a 
$\lambda$-dependent reparametrization parameter to simplify the potential 
energy term $V$ at the price of making the kinetic term $\Omega_{12}$ 
$\lambda$-dependent. Due to the virial theorem (see below) nothing new
is obtained by such a procedure.}
\be
\lambda \E 1 - \frac{4}{q^2} \, \frac{\partial}{\partial \lambda} \left ( V_{11} + 
V_{12} \right)
\label{var eq for lambda 1}
\ee
with
\be
\frac{\partial  V_{1j} }{\partial \lambda} \E Z^{j-1} \frac{\alpha}{2 \pi} M^2 \, 
\lambda \, \frac{q^2}{ 4 \kappa_E} \, \int_0^{\infty} d\sigma \> 
\frac{\sigma^2}{\mu_{1j}^4(\sigma)} \int_0^1 du \> u \, e_j(u,\sigma)
\> . 
\ee
The explicit $\lambda-$factor in the derivative can be combined with the LHS of 
Eq. (\ref{var eq for lambda 1}) to obtain the variational Eq. for  $\lambda$ in the
form (cf. Eq. (139) in ref. [I]).
\be
\lambda \E  \left \{ \, 1 + \frac{\alpha}{2 \pi \kappa_E} M^2 \, 
\int_0^{\infty} d\sigma \> \sigma^2 \, \int_0^1 du \> u \, \sum_{j=1}^2
\frac{Z^{j-1}}{\mu_{1j}^4(\sigma)} \, e_j(u,\sigma)
\, \right \}^{-1} \> .
\label{var eq for lambda 2}
\ee
This is still a nonlinear equation but is expected to better converge 
under iteration
as part of the interaction is already included. In addition, Eq. 
(\ref{var eq for lambda 2}) explicitly shows that $ 0 < \lambda \le 1$ since the 
integrand is positive.

Variation w.r.t. the profile functions $A_{\pm}(E)$ gives
\bea
A_-(E) \EA 1 + \frac{2}{\kappa_E \, E^2} \,  \int_0^{\infty} d\sigma
\>  \sum_{j = 1}^2 \, \frac{\delta V_{1j}}{\delta \mu^2_{1j}(\sigma)} 
\, \sin^2 \left ( \frac{E \sigma}{2} \right ) 
\label{var eq for Aminus}\\
A_+(E) \EA \! 1 + \frac{2}{\kappa_E \, E^2} \int_0^{\infty} \! \! d\sigma
  \left [  \frac{\delta V_{11}}{\delta \mu^2_{11}(\sigma)} 
 \sin^2 \left ( \frac{E \sigma}{2} \right )  + 
\frac{\delta V_{12}}{\delta \mu^2_{12}(\sigma)} 
 \cos^2 \left ( \frac{E \sigma}{2} \right )  \right ] .
\label{var eq for Aplus}
\eea
The functional derivative of $V_{1j}$ w.r.t. the pseudotimes is easily found to be
\be
\frac{\delta V_{1j}}{\delta \mu^2_{1j}(\sigma)} \E   
\frac{Z^{j-1} \alpha}{2 \pi} \frac{M^2}{\mu^4_{1j}(\sigma)} \, \int_0^1 du \, 
\left [  1 + \frac{m^2}{2 \kappa_E}  \mu^2_{1j}(\sigma) \, \frac{1-u}{u}
- \frac{(\lambda q/2)^2 \sigma^2}{2 \kappa_E  \, \mu^2_{1j}(\sigma)} u  
\right ] \, e_j(u,\sigma)  
\label{delta V1j 1}
\ee
which should be compared with Eq. (140) in ref. [I].
With a suitable integration by parts the result can also be written as
\be
\frac{\delta V_{1j}}{\delta \mu^2_{1j}(\sigma)} \E   
Z^{j-1} \frac{\alpha}{2 \pi} M^2 \frac{1}{\mu^4_{1j}(\sigma)} \, \int_0^1 du \> u \, 
 \left [ \, 2 - \frac{(\lambda q/2)^2 \sigma^2}{2 \kappa_E  \, 
\mu^2_{1j}(\sigma)} u \, \right ] \,  e_j(u,\sigma) 
\label{delta V1j 2}
\ee
and is particularly simple for massless pions:
\be
\frac{\delta V_{1j}}{\delta \mu^2_{1j}(\sigma)} \Biggr |_{m=0} \E   
Z^{j-1} \frac{\alpha}{2 \pi} M^2 \, \frac{1}{\mu^4_{1j}(\sigma)} \, 
\exp \left [ -\frac{(\lambda q/2)^2 \sigma^2}{2 \kappa_E  \, 
\mu^2_{1j}(\sigma)} \, \right ]  \> .
\label{delta V1j m=0}
\ee
Note that we have to deal with one-dimensional but highly non-linear 
integral equations. However, even without
solving them they can be used to deduce
the behaviour of $A_{\pm}(E), \mu_{1j}^2(\sigma) $ for small and large 
values of its arguments.
This is derived in appendix~C. Here we just summarize the results: 
at large $E$ both profile functions approach unity in the same way: 
\be
A_{\pm}(E) \> \stackrel{E \to \infty}{\longrightarrow} \> 
1 + \frac{\alpha}{4} \, \frac{M^2}{\kappa_E \, E} + \ldots \> .
\ee
The same holds for the pseudotimes in the limit $\sigma \to \infty$ 
\be
\mu^2_{1j}(\sigma) \> \stackrel{\sigma \to \infty}{\longrightarrow} \> 
\frac{\sigma}{2A_-(0)} + 
\frac{1}{\pi} \, \int_0^{\infty} dE \,\frac{1}{E^2} \, \left [ \, \frac{1}{A_-(E)}
- \frac{1}{A_-(0)} + \frac{1}{A_+(E)} \, \right ] + \ldots \> .
\ee
For $E, \sigma \to 0 $ there is an essential difference between the self-energy
and the interaction part: whereas $A_-(E)$ 
approaches a constant value $A_-(0)$ , $A_+(E)$ diverges for small $E$:
\be
A_+(E) \> \stackrel{E \to 0}{\longrightarrow} \> \frac{\omega_{\rm var}^2}{E^2} + 
{\rm const.} + \ldots
\> \> , \hspace{1cm} \omega_{\rm var}^2 \Def  \frac{2}{\kappa_E} \, \int_0^{\infty} 
d\sigma \> \frac{\delta V_{12}}{\delta \mu^2_{1 2}(\sigma)} \> .
\label{A+ for small E}
\ee
Therefore this term cannot be considered as ``small'' but has to be kept 
consistently. This is, of course, what we expect for a bound state 
which cannot be reached by perturbation theory from the free solution.
Similarly, the pseudotimes behave very differently for small $\sigma$: 
$\mu_{11}^2(\sigma) \to \sigma$ as usual, but
\be
\mu^2_{12}(\sigma) 
\> \stackrel{\sigma \to 0}{\longrightarrow} \>  
\frac{2}{\pi} \, \int_0^{\infty} dE \, \frac{1}{E^2 A_+(E)} \> \equiv \> 
\mu_{12}^2(0) \> .
\label{mu12(0)}
\ee
tends to a constant. This different behaviour in the UV-region was crucial
for the mass renormalization in the WC model.

Given the solution of the variational equation it is possible to express
the kinetic term $\Omega_{12} = \Omega_- + \Omega_+$ not only from 
its definition in 
terms of $A_{\pm}(E)$ but also from a ``virial theorem'' in which case it is given 
in terms of the pseudotimes. Indeed, following the derivation
given in appendix~E of ref. \cite{WC7} one can derive straightforwardly that 
\be
\Omega_{12}^{\rm vir} \E 2 \int_0^{\infty} d\sigma \> \sum_{j=1}^2 \, 
\frac{\delta V_{1j}}{\delta \mu^2_{1j}(\sigma)} \, \left [ \,  \mu^2_{1j}(\sigma) 
- \sigma \, \frac{\partial \mu^2_{1j}(\sigma)}{\partial \sigma} \, \right ] \> .
\label{virial theorem}
\ee
Note that this relation only holds {\it after} variation and for the sum of the
two kinetic terms. It is also possible to derive expressions for
the individual terms but they are more involved and will not be 
considered here.
The factor of ``$2$'' in front of Eq. (\ref{virial theorem}) is due to our 
definition of $\mu^2_{1j}$ in Eqs. 
(\ref{mu11 by Aplusminus},\ref{mu12 by Aplusminus}),
whose normalization is reduced by one half (compared to the usual definition of 
the pseudotime) but where one has to sum over two profile functions. Indeed, if 
either $V_{11} = 0$ or $V_{12} = 0$ one obtains the standard virial theorem 
in the polaron approach (see ref. (II)). 

Another useful relation  is obtained by 
differentiating Mano's equation with respect to some parameter, say 
the artificial coupling strength $Z$, and using the 
variational equations to greatly simplify the result. For this purpose 
we first note that the
interaction potentials depend on the combination
\be
x \Def \lambda \, \frac{q}{2} \> \equiv \> \lambda \, \frac{\sqrt{q^2}}{2} 
\label{def x}
\ee
and, of course, functionally on the pseudotimes, i.e. on the profile functions.
Introducing this combination as a variational parameter instead of $\lambda $, 
Mano's equation therefore reads
\be
\bar M^2 \E  q  x - x^2  -  \left ( \, \Omega_+
+ \Omega_- \, \right ) - 2 \left ( \, V_{11}^{\rm reg} + V_{12} \, \right ) \> .
\label{Mano x}
\ee
Note that $ \Omega_{\pm} $ is a functional of the profile functions which,
after solving the variational equations, are
complicated functions of $Z$ as are the parameter $x$ and the momentum $q$. 
Thus by differentiation
the RHS of Eq. (\ref{Mano x}) w.r.t. $Z$ we obtain
\bea
0 \EA \frac{\partial q}{\partial Z} \, x  - 2 \, \frac{V_{12}}{Z} 
+ \left \{  \, q  - 2 x -  2 \frac{\partial }{\partial x} \left (\, 
V_{11}^{\rm reg} + V_{12} \, \right ) \, \right \}  \, 
\frac{\partial x}{\partial Z} \non
&& - \sum_{j = \pm} \int_0^{\infty} dE \> 
\left \{ \, \frac{\delta}{\delta A_j(E)} \,  \Bigl [ 
\, \Omega_+ + \Omega_- \,  + 2 \left ( \, V_{11}^{\rm reg} + V_{12} \, \right ) \, 
\Bigr ] \,  \right \} \, \frac{\partial A_j(E)}{\partial Z} \> .
\eea
At first sight this looks rather complicated, but
due to the variational equations the terms in the curly brackets
vanish identically and after reintroducing the parameter $\lambda$ 
we obtain the simple result
\be
 \frac{\partial }{\partial Z} \, \left ( \frac{q}{2} \right )^2 \E 
\frac{2}{\lambda} \,  \frac{V_{12}}{Z} \> .
\label{Hell-Fey}
\ee 
This is our version of the ``Feynman-Hellmann theorem'' \cite{Fey-Hell} 
for the worldline approach. Note that the RHS originates from the explicit 
$Z$-dependence of $V_{12}$ and that the dependence of the variational
solutions on that strength only enters implicitly. We use the Feynman-Hellmann 
theorem in appendix~\ref{app: weak var bind} to derive the first terms in the 
weak-coupling expansion of the binding energy.

\newpage
\section{Results and Discussion}
\label{sec: results}
\subsection{Numerical Solutions and Binding Energies}
\label{subsec: num result}
The system of coupled integral equations (\ref{var eq for lambda 2}) - 
(\ref{var eq for Aplus}), together with Eqs. (\ref{delta V1j 2}),
(\ref{mu11 by Aplusminus}) and (\ref{mu12 by Aplusminus}) can be solved 
by iteration using similar methods as in ref. (II). The basic idea 
is to use a grid of gaussian integration points $E_i, \sigma_j$ on which
profile functions and pseudotimes are evaluated and used in the corresponding
$E, \sigma$-integrals. This is similar to the mesh methods employed
in ref. \cite{mesh} for quantum-mechanical eigenvalue problems.
The necessary modifications for the worldline bound-state case 
are discussed in appendix~D. To obtain stable and reliable results 
it was crucial to understand and match the analytical solutions at small and large
values of the variables (see appendix~C) to the numerical outcome.

After considerable effort we have developed a program which 
solves the variational bound-state equations numerically with sufficient 
accuracy. Setting $V_{12} = 0$ reproduced the 
one-body results -- a necessary but unfortunately rather weak check of its 
correctness. More stringent is the ``virial check''~, i.e. comparing the total
kinetic term (\ref{total Omega}) evaluated from the profile functions 
with the one in Eq. (\ref{virial theorem}) obtained from the pseudotimes.
To find a bound-state solution one has to search for values $q^2 < 4 M^2$ 
where the LHS minus the RHS of Mano's equation 
(\ref{Mano 3}) changes sign. This was done by applying the {\it regula falsi} once
the possible solution had been bracketed.
The values of the intermediate renormalized mass $\bar M$ have been 
recalculated at the
beginning of the program by setting $Z \alpha = 0 $ and $ \sqrt{q^2} = 2 M $.

\begin{figure}[hbt]
\vspace{0.5cm}
\bce
\mbox{\epsfxsize=7cm \epsffile{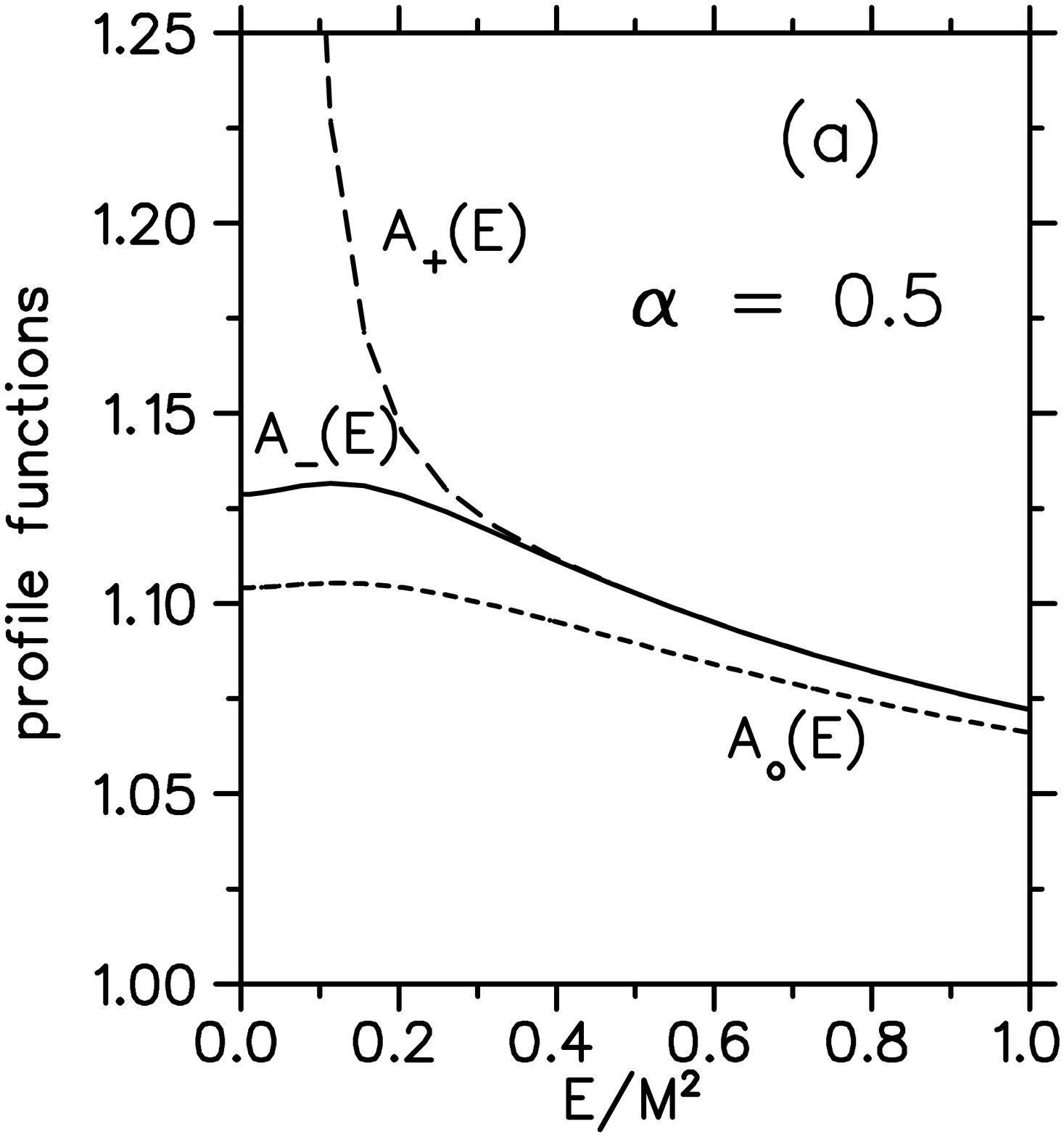}}
\vspace{0.5cm}

\mbox{\epsfysize=7cm \epsffile{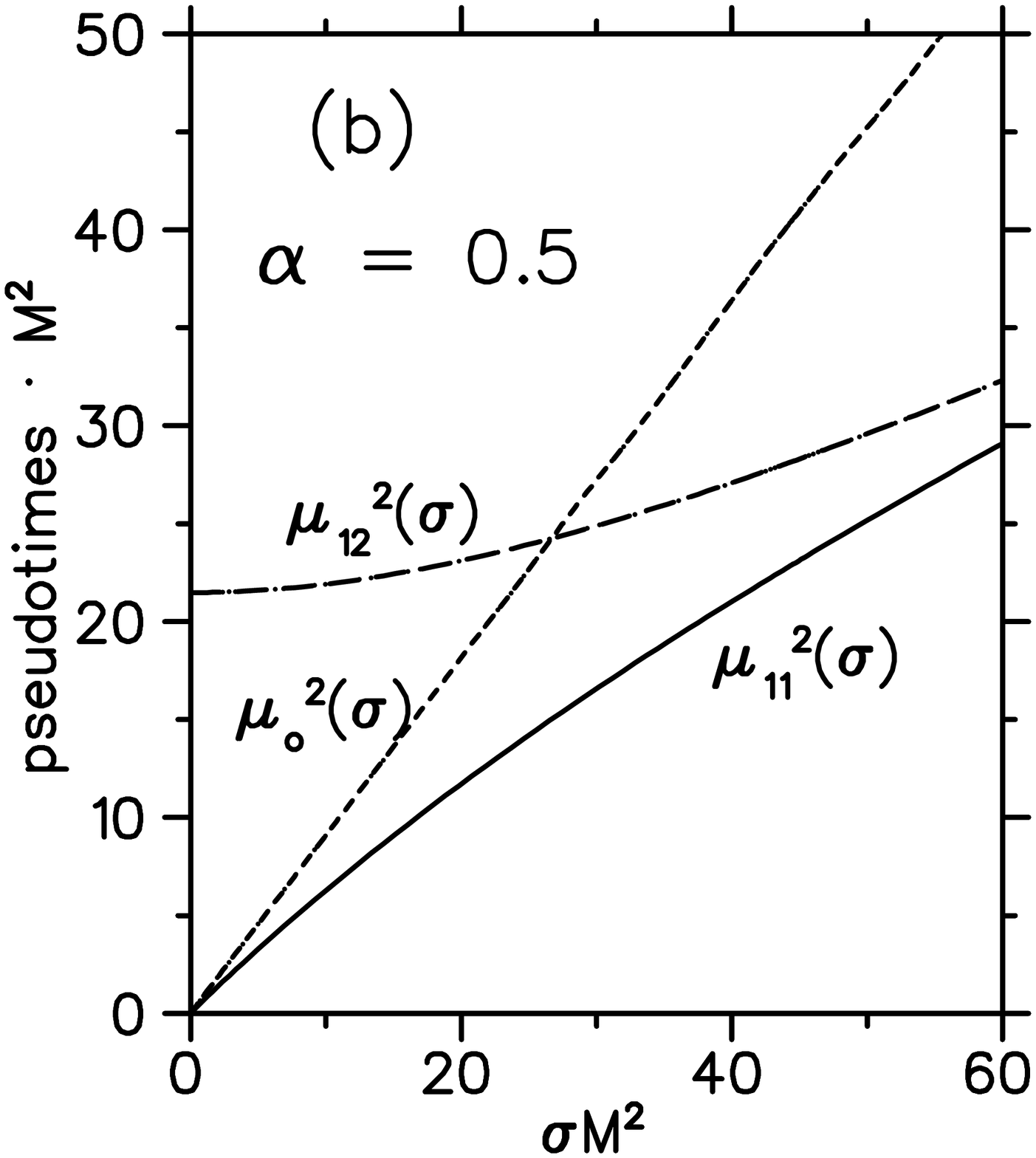}}
\ece
\caption{(a) The profile functions $A_-(E)$ and $A_+(E)$, (b) the 
pseudotimes $ \mu_{12}^2(\sigma) $ and $ \mu_{11}^2(\sigma)$  
for the bound-state solution at $\alpha = 0.5$ with 
reparametrization parameter $ \kappa_E = 1$.
For comparison the one-body results (subscript $o$) are also shown.
}
\label{fig: profile+pseudo}

\end{figure}

\begin{table}[htb]
\vspace{0.5cm}
\begin{center}
\begin{tabular}{lcrlcl} 
\firsthline 
\footnotesize      
 \quad $\alpha$  &  $\> \bar M/M$  & $  \Delta_{\rm min} \, $  &  
vir. check & $\> \epsilon/M$ & $ \> \Gamma/M$ \\
\normalsize 
\midhline
0.44    &  0.75757     &  $< 10^{-3}$         & - 0.0019    & \quad - 0.00063  & \quad 0 \\
0.45    &  0.75156     &  $< 10^{-3}$         & - 0.0001    & \quad - 0.00138  & \quad 0 \\
0.46    &  0.74553     &  $< 10^{-3}$         & - 0.0015    & \quad - 0.00232  & \quad 0 \\
0.47    &  0.73947     &  $< 10^{-3}$         & - 0.0001    & \quad - 0.00338  & \quad 0 \\
0.48    &  0.73339     &  $< 10^{-3}$         &  ~0.0001    & \quad - 0.00462  & \quad 0 \\
0.49    &  0.72728     &  $< 10^{-3}$         &  ~0.0001    & \quad - 0.00610  & \quad 0 \\
0.50    &  0.72114     &  $< 10^{-3}$         &  ~0.0002    & \quad - 0.00782  & \quad 0 \\
0.51    &  0.71498     &  $< 10^{-3}$         &  ~0.0002    & \quad - 0.00985  & \quad 0 \\
0.52    &  0.70879     &  $< 10^{-3}$         &  ~0.0003    & \quad - 0.01226  & \quad 0 \\
0.53    &  0.70258     &  $< 10^{-3}$         &  ~0.0005    & \quad - 0.01516  & \quad 0 \\
0.54    &  0.69634     &  $< 10^{-3}$         &  ~0.0004    & \quad - 0.01900  & \quad 0 \\
0.541   &  0.69572     &  $< 10^{-3}$         &  ~0.0008    & \quad - 0.01945  & \quad 0 \\
0.542   &  0.69509     &  $< 10^{-3}$         &  ~0.0008    & \quad  -0.01998  & \quad 0 \\
0.543   &  0.69446     &  1.8 $\cdot 10^{-3}$ &  ~0.0033    & \quad  -0.02048  & \quad 
0.00012 \\ 
0.545   &  0.69321     &  4.4 $\cdot 10^{-3}$ &  ~0.0083    & $\sim$ -0.0212   & \quad 
0.00060 \\
0.55    &  0.69007     &  1.1 $\cdot 10^{-2}$ &  ~0.021~    & $\sim$ -0.0233   & \quad 
0.0026  \\
0.56    &  0.68378     &  1.7 $\cdot 10^{-2}$ &  ~0.048~    & $\sim$ -0.0248   & \quad 
0.0088  \\
0.57    &  0.67747     &  2.3 $\cdot 10^{-2}$ &  ~0.069~    & $\sim$ -0.0288   & \quad 
0.0170  \\
0.58    &  0.67112     &  3.1 $\cdot 10^{-2}$ &  ~0.091~    & $\sim$ -0.0331   & \quad 
0.0269 \\
0.59    &  0.66475     &  4.0 $\cdot 10^{-2}$ &  ~0.118~    & $\sim$ -0.0391   & \quad 
0.0382 \\
0.60    &  0.65836     &  4.6 $\cdot 10^{-2}$ &  ~0.143~    & $\sim$ -0.0443   & \quad 
0.0508 \\ 
\lasthline

\end{tabular}

\end{center}
\caption{Results for the binding energy $ \epsilon = \sqrt{q_{\star}^2} - 2 M $ 
from the solution of the variational bound-state equations for $m/M = 0.15 $
and different coupling constants. 
The value of the intermediate mass $\bar M$ of the single nucleon (renormalized
at $\nu = m$) is also given.
Calculations were performed with  $3 \times 72$
gaussian points and the iteration was stopped when the relative deviation $\Delta$ 
(defined in Eq. (\ref{global dev}))  was less than $10^{-3}$ or when it increased 
twice consecutively. In this case $\Delta_{\rm min}$ is given. The column 
``vir. check '' gives the relative deviation 
$(\Omega_{12}^{\rm vir} - \Omega_{12})/\Omega_{12} $. The width of the bound 
state above the critical coupling $\alpha_{\rm crit}^{(2)} = 0.542 $ 
is estimated from Eq. (\ref{Gamma near crit 2}).}
\label{tab:varbind}
\end{table}
Figure \ref{fig:  profile+pseudo}a shows the behaviour of the profile 
functions $A_{\pm}(E)$ as functions of the variable $E$ after a bound-state solution 
had been found in this way
at $\sqrt{q_{\star} ^2} = 1.99218 M $ for $\alpha = 0.5$. As expected from our analytic
results the two profile functions have a very different behaviour for small $E$ but it
is also clearly seen that $A_-(E)$ differs 
appreciably from the one-body profile function obtained at the same 
coupling constant. This illustrates how binding affects the nucleon self-energies
or -- in other language -- the in-medium effects. Similar changes can also be observed
in  Fig. \ref{fig:  profile+pseudo}b for the pseudotimes.

\noindent
Table \ref{tab:varbind} gives the result for the binding energy
\be
\epsilon \E \sqrt{q_{\star}^2} - 2 M
\ee
obtained for the mass ratio \footnote{Our previous results \cite{BRS} were 
obtained with masses $ M = 0.939 $ GeV and $ m =  0.14 $ GeV, i.e.
$ m/M = 0.149095 $. In order to facilitate comparison with results in the literature 
we have repeated the calculations for this standard mass ratio.} 
\be
\frac{m}{M} \E 0.15
\label{mass ratio}
\ee
and a range of coupling constants $\alpha$. 
The lowest value comes from the nonrelativistic criterion
(\ref{nonrel criterion}) which for the chosen parameters says that 
$\alpha > 0.4067 $ in order to get a bound state in the Yukawa potential 
(alternatively no solution of Mano's equation with 
$q_{\star}^2 < 4 M^2$ can be found for too small $\alpha$). 
An upper value is determined by the critical value 
\be
\alpha_{\rm crit}^{(1)} \E  0.817
\ee
above which no real solutions 
of Mano's equation in the one-body case are found~\footnote{This value is slightly 
larger than the one obtained in ref. (II) because of the slightly higher 
mass ratio. Similar values have been found with truncated Dyson-Schwinger 
equations \cite{AhAl,Saul}.}; this is the variational
sign of the instability of the WC model when the
quantum corrections are taken into account.

However, already for $\alpha > 0.542 $ one observes that the
iteration only converges for a certain number of iterations and 
then the global measure of deviation defined in Eq. (\ref{global dev})
starts to rise again .
In the one-body case this was a signal for the 
instability of the WC model and here also we can interpret this phenomenon as
the impossibility of finding {\it real} solutions of the variational 
equations for coupling constants larger than 
\be
\alpha_{\rm crit}^{(2)} \E 0.542 \> \> .
\label{alpha crit 2body}
\ee 
Consequently the virial check deteriorates rapidly.
We also observe that this
(infamous) instability now occurs at {\it lower} values of the coupling constant.
This is not unreasonable as, 
in general (e.g. in a laser), transitions may be 
induced by the presence of other particles \cite{induced}: the higher the number 
of particles in a system the faster it decays. 
In ref. (II) we showed that above the critical coupling 
only {\it complex} solutions of Mano's equation were possible where 
the imaginary parts determined the width of the particle. Similarly, 
in Sect. \ref{subsec: instab} we will give a rough, analytical estimate of 
$\alpha_{\rm crit}^{(2)}$ and 
determine the width of the bound state above this critical coupling.
\vspace{-0.2cm}

\subsection{Comparison with Other Work}
\label{subsec: comp}
Of course, not only the possible width of the bound state due to the instability
of the WC model is of interest but also the magnitude of the binding. Here our 
approach should be compared with the nonrelativistic variational method 
(using a non-retarded quadratic trial action, i.e. gaussian wavefunctions) and 
with the exact  nonrelativistic results from solving the s-wave Schr\"odinger equation 
for a Yukawa potential in the reduced sytem. The results are 
listed in Table \ref{tab:bind comp}. 

In addition, we also give binding energies from 
the ladder Bethe-Salpeter equation (BSe). To be more 
precise we display the coupling constant belonging to a given $ \sqrt{q^2} < 2 M $ 
calculated from Efimov's variational approximation \cite{Efi1} to the ladder BSe. 
Comparison with some exact numerical solutions available in the literature 
(e.g. ref. \cite{NiTj2})
shows that this approximation with 2 variational parameters is surprisingly good.
For example, for a binding energy of $ 0.01 M $ ref. \cite{KaCa} reports a necessary
coupling constant $ \alpha = 0.5716 $ whereas  Efimov's approximation requires
 $ \alpha = 0.5730$
\footnote{Actually, upon repeating the calculations listed in Table 3 of Efimov's 
paper for $m/M = 0.1$, better values were found in a few cases because 
the author had missed 
the true minimum \cite{Efi2}.}. The 2-dimensional integrals needed
in Eqs. (56), (58) of Efimov's paper were evaluated by standard numerical 
Gauss-Legendre integration and the minimum was found easily with the help of the 
CERN routine MINUIT.

\begin{figure}[ht]
\vspace{0.5cm}
\begin{center}
\mbox{\epsfysize=8.5cm\epsffile{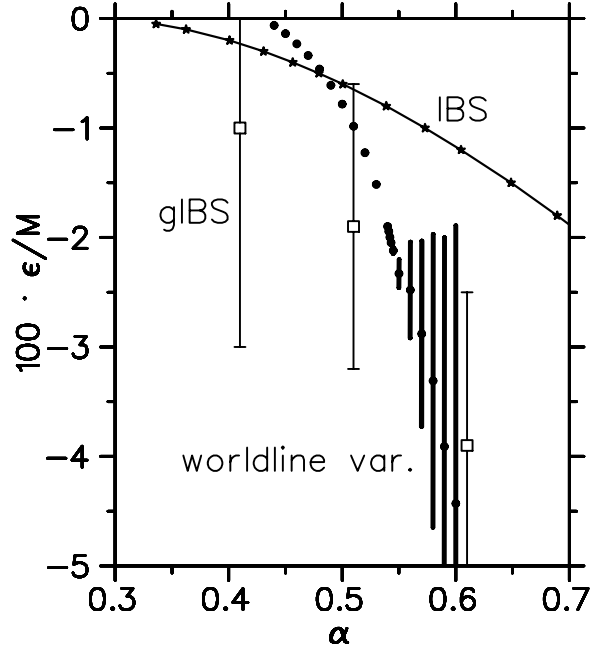}}
\end{center}
\caption{The binding energy  $\epsilon/M $ of the two-body bound state for the 
equal-mass case with $m/M = 0.15 $  as a function of the dimensionless
coupling constant defined in Eq. (\ref{def alpha equal mass}).
The results from the worldline variational approach including self-energy and 
vertex corrections are compared with those from Efimov's variational approximation 
to the ladder 
Bethe-Salpeter (lBS) equation \cite{Efi1}. Also shown are the Monte-Carlo 
results (with errors) from ref. \cite{NiTj1} in 
the ``generalized ladder approximation'' to the 
Bethe-Salpeter equation (glBS). The thick bars for the worldline 
results represent the width of the bound state above the critical coupling 
estimated from Eq. (\ref{Gamma near crit 2}). 
}
\label{fig: binding}

\end{figure}

As can be seen in Fig. \ref{fig: binding} the binding energies of the full 
variational approach including the 
self-energy and vertex corrections are much larger in magnitude than the results 
from the other approaches 
provided the coupling constant $\alpha$ is appreciably larger than  
the threshold value needed to support a bound state. Of course, 
in the present variational approximation this threshold value is 
too large due to the insufficient approximation of a Yukawa potential by a harmonic
oscillator one. 
Also the comparison of nonrelativistic variational results with 
the exact values from the solution of the Schr\"odinger equation shows that
quantitatively the variational approximation is not a very good one 
(for reasons that are discussed in appendix~\ref{app: FJ nonrel} 
after eq. \ref{var energy}).
Nevertheless
the field theoretical effects are clearly visible and produce much more binding. 
That this must be due to self-energy/vertex corrections can be inferred by
comparison with Nieuwenhuis \& Tjon's 
Monte-Carlo calculation  \cite{NiTj1} which gives
the two-body binding energy beyond the ladder approximation 
but still without self-energy and vertex corrections
\footnote{Unfortunately, the results there are only given in the form of Fig. 1 
from which values have to be read off in an approximate way: 
$ \sqrt{q^2}/M \simeq 1.990 \pm 0.020, 1.981 \pm 0.013, 1.961 \pm 0.014, 
1.921 \pm 0.018, 1.851 \pm 0.020, 1.770 \pm 0.016 $ for 
$\alpha = 0.4, 0.5, 0.6, 0.7, 0.8, 0.9 $ (the abscissa in Fig. 1 is $4 \alpha$).
The BSe curve is in good agreement with the values obtained 
from Efimov's approximation.}. 
\begin{table}[htb]
\begin{center}
\begin{tabular}{lllllc} 
\firsthline 
\footnotesize    
$\alpha$    &  worldline var. & NR var.     &  NR Schr\"od.  & \quad lBSe &
gen. lBSe  \\
\normalsize
\midhline
0.30        &                 &               & -0.000463    &           &  \\
0.33605     &                 &               &              &  -0.00050 &  \\
0.35        &                 &               & -0.001975    &           &  \\
0.36254     &                 &               &              &  -0.00100 &  \\
0.40        &                 &               & -0.004592    &           & -0.010 (20) \\
0.40076     &                 &               &              &  -0.00200 &  \\
0.43072     &                 &               &              &  -0.00300 &   \\
0.44        & \quad -0.00063  &  -0.001278    &              &           &     \\
0.45        & \quad -0.00138  &  -0.001754    & -0.008348    &           &   \\
0.46        & \quad -0.00232  &  -0.002271    &              &           &   \\
0.47        & \quad -0.00338  &  -0.002830    &              &           &   \\
0.47943     &                 &               &              &  -0.00500 &   \\
0.48        & \quad -0.00462  &  -0.003430    &              &           &   \\
0.49        & \quad -0.00610  &  -0.004072    &              &           &   \\
0.50        & \quad -0.00782  &  -0.004754    & -0.013267    &           & -0.019 (13) \\
0.51        & \quad -0.00985  &  -0.005478    &              &           &    \\
0.52        & \quad -0.01226  &  -0.006243    &              &           &    \\
0.53        & \quad -0.01516  &  -0.007049    &              &           &   \\
0.54        & \quad -0.01900  &  -0.007896    &              &           &   \\
0.55        & $\sim$ -0.0233  &  -0.008783    & -0.019367    &           &   \\
0.56        & $\sim$ -0.0248  &  -0.009712    &              &           &   \\
0.57        & $\sim$ -0.0288  &  -0.010682    &              &           &   \\
0.57304     &                 &               &              &  -0.01000 &   \\
0.58        & $\sim$ -0.0331  &  -0.011693    &              &           &   \\
0.59        & $\sim$ -0.0391  &  -0.012745    &              &           &  \\
0.60        & $\sim$ -0.0443  &  -0.013838    & -0.026661    &           & -0.039 (14) \\
0.64863     &                 &               &              &  -0.01500 &   \\
0.65        &                 &  -0.019920    & -0.035159    &           &   \\
0.70        &                 &  -0.027033    & -0.045867    &           & -0.079 (18)\\
0.71497     &                 &               &              &  -0.02000 &   \\
\lasthline
\end{tabular}

\end{center}
\caption{The binding energies  $\epsilon/M $
from the worldline variational approach  with $ m/M = 0.15 $
compared with those from the 3-dimensional nonrelativistic (NR) variational method, 
the exact solution of the nonrelativistic Schr\"odinger equation and 
(Efimov's variational 
approximation to) the ladder Bethe-Salpeter equation (lBSe). The last column gives the
Monte-Carlo results (estimated errors of the last digits in parenthesis) 
of ref. \cite{NiTj1} in which ladder and crossed-ladder graphs 
have been included 
(``generalized lBSe'') but self-energy and vertex corrections were still 
neglected.}
\label{tab:bind comp} 
\end{table}
\clearpage
 
Having introduced the artificial strength parameter $Z$ for the direct interaction
it is, of course, possible to check the self-energy effects on the binding energy by
considering the limit
\be
\alpha \To 0 \> , \hspace{1cm} Z \alpha \To \alpha' 
\label{no self}
\ee
in which the self-interaction $V_{11}$ vanishes. For the single nucleon we then have
$\bar M = M$ and the
variational equations (\ref{var eq for Aminus}, \ref{var eq for Aplus}) simplify
slightly because
\be
A_-(E) + A_+(E) \Bigr |_{\alpha=0} \E 2 + \frac{\omega_{\rm var}^2}{E^2} \> .
\ee
Also the asymptotic behaviour of the profile functions changes to 
$A_{\pm}(E) \to 1 + \omega_{\rm var}^2/(2 E^2)$.

\begin{figure}[hbt]

\begin{center}
\mbox{\epsfysize=8cm\epsffile{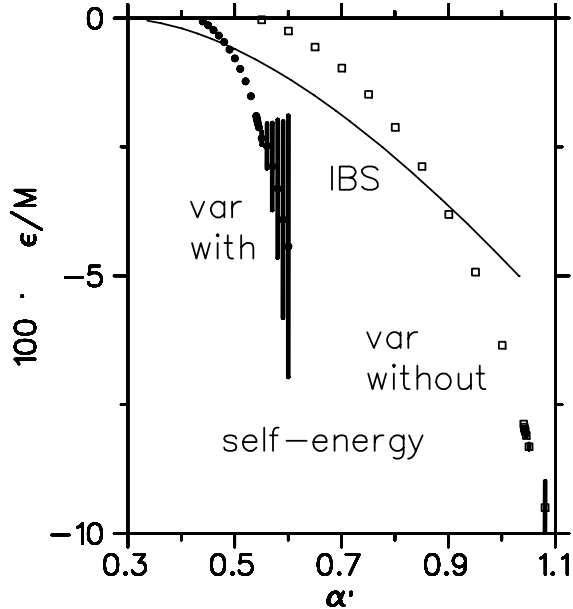}}
\end{center}
\caption{Variational binding energies with and without 
inclusion of self-energy effects for $m/M = 0.15$. For comparison 
the results from the ladder Bethe-Salpeter equation (denoted by ``lBS'') are also 
shown.}
\label{fig:no self binding}

\end{figure}

Fig. \ref{fig:no self binding} shows a comparison of variational binding energies with 
and without self-energy effects as functions of the coupling constant $\alpha'$.
It is seen that the 
binding energies without self-energy effects are much smaller especially near the
threshold where the interaction is strong enough for a bound state. 
In this case the particles move slowly, and without 
self-energy and vertex corrections the minimal coupling is therefore the same as in the
nonrelativistic case. However, we have to correct for the fact that our trial 
action is too restricted to reduce to the correct 3-dimensional kinetic term 
(as discussed above)
and therefore we have to use Eq. (\ref{nonrel criterion}) with $d  = 4 $.
This gives
\be
\alpha'_{\rm min} \Bigr|_{\alpha=0} \E 0.54207 \> .
\label{alp min for noself}
\ee
Our numerical results obtained by solving the variational equations and 
collected in Table \ref{tab:varbind noself} are consistent with this minimal value,
which is an additional, nice check on the correctness and stability of our
numerical procedures. Note that the minimal coupling (\ref{alp min for noself}) 
for a bound state {\it without} self-energy effects 
numerically coincides with the value (\ref{alpha crit 2body}) found for the 
instability of the bound system {\it with} self-energy and vertex corrections.
This points to an important role of the direct binding as a 
new ``doorway'' for the decay of the system.

As noted before, 
the variational calculation ``without'' still contains (approximately) all crossed 
diagrams and therefore leads to more binding than the ladder Bethe-Salpeter equation 
for sufficiently large $\alpha'$. Note that up to $\alpha' = 1 $ no sign of an 
instability is found. However, for slightly higher coupling constants the 
variational calculation does not converge anymore and we find again a 
critical coupling
\be
\alpha'_{\rm crit} \Bigr|_{\alpha=0} \E 1.041 \> \>.
\ee

\begin{table}[htb]
\begin{center}
\begin{tabular}{lrlcl} 
\firsthline 
\footnotesize    
 \quad $\alpha'$ & $ \Delta_{\rm min} \> $  & vir. check  & $\epsilon/M$ & $ \> \Gamma/M$ \\
\normalsize 
\midhline
0.55       &  $< 10^{-3}$ & - 0.0013             & \quad - 0.00031  & \quad 0 \\
0.60       &  $< 10^{-3}$ & - 0.0013             & \quad  -0.00252  & \quad 0 \\
0.65       &  $< 10^{-3}$ & - 0.0012             & \quad - 0.00564  & \quad 0 \\
0.70       &  $< 10^{-3}$ & - 0.0012             & \quad - 0.00973  & \quad 0 \\
0.75       &  $< 10^{-3}$ & - 0.0007             & \quad - 0.01485  & \quad 0 \\
0.80       &  $< 10^{-3}$ & - 0.0011             & \quad - 0.02116  & \quad 0 \\
0.85       &  $< 10^{-3}$ & - 0.0008             & \quad - 0.02879  & \quad 0 \\
0.90       &  $< 10^{-3}$ & - 0.0011             & \quad - 0.03807  & \quad 0 \\
0.95       &  $< 10^{-3}$ &  ~0.0003             & \quad - 0.04930  & \quad 0 \\
1.00       &  $< 10^{-3}$ &  ~0.0005             & \quad - 0.06353  & \quad 0 \\
1.04       &  $< 10^{-3}$ &  ~0.0013             & \quad - 0.07888  & \quad 0 \\
1.041      &  $< 10^{-3}$ &  ~0.0012             & \quad - 0.07942  & \quad 0 \\
1.042      &  1.7 $\cdot 10^{-3}$ &  ~0.0028     & $\sim$ - 0.0798  & \quad 0.00004 \\
1.043      &  2.7 $\cdot 10^{-3}$ &  ~0.0043     & $\sim$ - 0.0802  & \quad 0.00012 \\ 
1.045      &  4.4 $\cdot 10^{-3}$ &  ~0.0072     & $\sim$ - 0.0810  & \quad 0.00034 \\
1.05       &  8.9 $\cdot 10^{-3}$ &  ~0.015~     & $\sim$ - 0.0832  & \quad 0.00114 \\
1.08       &  3.4 $\cdot 10^{-2}$ &  ~0.055~     & $\sim$ - 0.095~  & \quad 0.0103~ \\
1.10       &  5.1 $\cdot 10^{-2}$ &  ~0.087~     & $\sim$ - 0.104~  & \quad 0.0191~  \\
\lasthline

\end{tabular}

\end{center}
\caption{Same as in Table \ref{tab:varbind} but for $\alpha = 0 , Z\alpha = \alpha'$,
i.e. without self-energy and vertex corrections which implies $ \bar M = M $. The width
of the unstable state above the critical coupling $\alpha'_{\rm crit} = 1.041 $ 
is estimated from Eq. (\ref{Gamma near crit noself}).}
\label{tab:varbind noself}
\end{table}
At first sight this is surprising as the instability of the WC model has been 
attributed to the self-energy effects which are absent in this limit and one would 
suspect a numerical instability. However, Table \ref{tab:varbind noself} shows that 
the virial test (checking how well the variational equations are fulfilled 
numerically) is excellent up to the critical
value. In addition, the very same analysis which explains the induced instability 
in the full model (with self-energy effects) leads to the prediction 
$ \alpha' \sim 0.814 $ for zero meson mass (see next Sect.). We are thus led to the 
conclusion that the WC model is ``unstable in nearly any case'' except for rather
special approximations like the ladder approximation. The Monte-Carlo calculations by
Nieuwenhuis \& Tjon \cite{NiTj1} may easily have missed this instability 
given their large error bars, systematic errors and the fact that they only 
consider coupling constants not larger than $\alpha' = 0.9$.

The increased binding due to the self-energy/vertex corrections is also evident in 
the weak-binding case  
where (for $ m = 0$ ) we have found in appendix \ref{app: weak var bind}
\be
\epsilon_0 \E - \frac{M}{2} \, \left [ \, \frac{(Z \alpha)^2}{\pi} \, K_{\rm var}
+ \frac{(Z \alpha)^4}{\pi^2}  + \ldots  \, \right ] \> .
\label{weak bind}
\ee
with an enhancement factor
\be
K_{\rm var} \E 1 + \frac{7}{2} \, \frac{\alpha}{\pi} + \ldots 
\label{enhanc var}
\ee
\begin{figure}[hbt]

\begin{center}
\mbox{\epsfxsize=9cm\epsffile{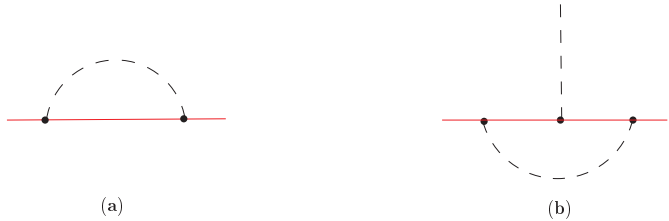}}
\end{center}
\caption{One-loop corrections: (a) self-energy,
(b) vertex.}
\label{fig: one_loop}

\end{figure}

\noindent
for the leading nonrelativistic term. That this 
is not an artefact of the variational approximation but also occurs in the 
full theory can be seen as follows: the one-loop vertex correction and 
the wavefunction renormalization 
for a {\it free nucleon} depicted in fig. \ref{fig: one_loop}  give rise 
to an effective coupling constant
\be 
g' \> \longrightarrow \> g'_{\rm eff} \E g' \, \left ( \, 1 + \frac{\alpha}{\pi} + 
\ldots \, \right )
\label{geff}
\ee
as obtained in Eq. (58) of ref. \cite{WC4}. As derived in appendix 
\ref{app: perturb bind}
this is also the exact one-loop result since  
the first-order variational calculation reduces to that in the weak-coupling limit.
Thus we have $ \alpha \to \alpha (1 + \alpha/\pi + \ldots)^2 $ and we therefore 
expect that the exact
bound-state energy of the equal-mass WC model has the weak-binding expansion
\be
\epsilon_n \Bigr |_{\rm exact} \E - \frac{M}{2} \, \frac{(Z \alpha)^2}{2 (n+1)^2} \, 
K_{\rm exact} + \ldots
\ee
with an enhancement factor
\be
K_{\rm exact} \E \left ( \, 1 + \frac{\alpha}{\pi} + \ldots \, \right )^4 \E 
 1 + 4 \, \frac{\alpha}{\pi} + \ldots \> \> \> .
\label{enhanc exact}
\ee
The result (\ref{enhanc var}) from the variational worldline
approach is in good agreement with this except that the numerical factor 
is $7/2$ and not $4$. Since the variational
calculation underestimates the leading term -- the 
nonrelativistic binding  in units of the reduced mass is $ (Z \alpha)^2/\pi$ instead 
of the correct Coulomb value $ (Z \alpha)^2/2 $ -- such a small quantitative 
discrepancy is of no concern. Note that the procedure used to determine the exact 
enhancement factor (\ref{enhanc exact}) is equivalent to the matching procedure 
used in the application of effective field theories to bound-state 
problems \cite{EFT}. Indeed, Eq. (2.2) in ref. \cite{Gries} is exactly the 
nonrelativistic effective field theory for the massless WC model and in that 
language Eq. (\ref{geff}) determines the Wilson coefficient $c_1$ 
to one-loop order.

However, the enhancement observed above is at variance with Ji's claim \cite{Ji} that
inclusion of self-energy terms leads to {\it less} binding.
This result was obtained in the light-cone formalism whose boosted states
contain many equal-time Fock components. Therefore
a direct comparison with equal-time methods is difficult although the physical
results should be the same ... 

The origin of the disagreement is not clear to us: it
could be due to a missing nucleon mass renormalization \footnote{For a recent 
discussion of renormalization in light-front dynamics see ref. \cite{Math}.}
or a shortcoming of Ji's restricted Tamm-Dancoff approximation which only retains
(light-cone) states with at most one meson.
In view of the fact that both variational and
effective field theory predict {\it more} binding, serious doubts remain whether 
the light-cone calculation is correct and/or complete. 
This assessment is also supported by the results from the dressed ladder 
approximation which give more  binding compared to the calculations 
with bare propagators (see Fig. 8 in ref. \cite{Saul}).
The observed enhancement of
the effective coupling in the WC model is similar to
the enhancement of the electron-photon 
coupling in QED: however, in the latter case only the magnetic coupling (the famous 
anomalous magnetic moment of the electron) is affected and not the electric charge 
which is conserved in a gauge theory.

The $ {\cal O}( (Z \alpha)^4 )$-terms in Eq. (\ref{weak bind}) may be compared with 
the corresponding result from the Todorov equation, one of the 
relativistic quasi-potential equations which describe spinless binaries bound 
by a Coulomb potential. However, what is found
in the textbooks (e.g. ref. \cite{Pil}, Eqs. (4.197) and (4.206)) corresponds 
to scalar QED, not to our coupling of mesons (``scalar photons'') to scalar 
particles. 
The latter case was evaluated by Brezin, Itzykson and Zinn-Justin 
who anticipated Todorov's equation \cite{Tod} in a study of 
the relativistic eikonal approximation. 
This approach includes relativistic recoil and 
``an approximate summation of the crossed-ladder Feynman diagrams'' 
but ``does not include radiative corrections of the self-energy type'' \cite{BIZ}.
Using the (unnumbered) equation following Eq. (20) in that paper or Eq. (3.8) in
ref. \cite{Tod} one obtains in the equal-mass case 
\bea
\epsilon_{n}  \Bigr |_{\rm Todorov} \EA  \left \{ \, 2 M^2 \, \left [ 1 + 
\sqrt{1-(Z \alpha/(n+1))^2} \right ] \, \right \}^{1/2} - 2 M \non
\EA - \frac{M}{2} \, \left [ \, \frac{1}{2} \left ( \frac{Z \alpha}{n+1} \right )^2 
+ \frac{5}{32} \,  \left ( \frac{Z \alpha}{n+1} \right )^4 + \ldots \, \right ] \> .
\eea
For $ n = 0 $ this relativistic Balmer formula 
gives a coefficient $ 5/32 = 0.156 $ for the  $(Z \alpha)^4 $-term
whereas the variational result (\ref{weak bind}) leads to $ 1/\pi^2 = 0.101 $. 
Thus both approaches predict an enhanced binding of the lowest-lying state
also from relativistic effects. Although the variational calculation gives a smaller 
coefficient  -- as it does for the leading 
nonrelativistic  $ {\cal O}( (Z \alpha)^2 )$-term -- it has the clear advantage of 
also containing self-energy and vertex corrections. 

Comparison with the weak-binding expansions shows that in the variational approach 
{\it all} calculated expansion coefficients have the same sign but are smaller in 
magnitude than in the exact case. Therefore it is tempting to deduce
that the true binding energy is always below the variational 
bound-state energy as in the standard Rayleigh-Ritz variational principle 
of quantum mechanics. However, we have not succeeded in proving  this conjecture 
rigorously. The main obstacle is that 
renormalization requires the intermediate mass $\bar M$ of a single nucleon 
as input in Mano's equation (\ref{Mano 3}) and this quantity is the result of 
a variational calculation itself. Actually,
as shown in Eq. (85) in ref. [I],  $\bar M^2$ is bounded from below 
\footnote{In the 1-body case the value of 
$\bar M$ served as a measure of the goodness of different parametrizations for the
profile function.} and one may expect that a similar inequality also exists for 
the 2-body case. But this does not lead to a bound for the  binding energy 
because this quantity is determined {\it relative} to the single-nucleon result.
Note that this argument does not depend on whether the renormalization 
is finite or infinite: the minimum property of the variational calculation is 
also lost in the finite bipolaron problem \cite{VSPD}.

\subsection{Induced Instability}
\label{subsec: instab}

In order to understand the numerical results it is also useful to have some 
rough analytical insight into the solutions of the variational equations. 
In particular, we would like to understand the mechanism of the induced 
instability. Such an insight is provided by the {\it ansatz} 
\be
A_-(E) \E 1
\ee
with $\lambda$ as the only free variational parameter. This was used in ref. 
(II) to determine the critical coupling beyond which no real solution 
of the variational equations exist anymore.
However, in the two-body case
the binding certainly has to be taken into account and the simplest choice is a 
harmonic oscillator-like form ($\kappa_E = 1$ )
\be
A_+(E) \E 1 + \frac{\omega^2}{E^2} \> .
\label{A+ supersimp}
\ee
This is also the nonrelativistic solution and has the correct form 
for $E \to 0 $ (see Eq. (\ref{A+ for small E})).
Hence there are {\it two} variational parameters to be determined from varying 
Mano's Eq. To allow an analytical treatment we furthermore assume $m = 0$ and
weak binding
\be
\frac{\omega}{(q/2)^2} \> \simeq \> \frac{\omega}{M^2} \> \ll \> 1
\ee
together with the property that the relevant proper times are ${\cal O}(1/M^2)$ , 
i.e. $  \> \omega \sigma  \> \ll \> 1  \> $.
As shown in appendix~E the critical coupling can then be worked out 
analytically and is given by
\be
\alpha_{\rm crit} \> \simeq \>  \frac{\pi}{8} \, \frac{(1 + \sqrt{1 + 3 z})^3}{(1 + z + 
\sqrt{1 + 3 z})^2} 
\> \> , \hspace{0.5cm} z \Def 2 \pi Z^2 \> .
\label{alpha crit}
\ee
For $ Z = 0 $, i.e. for the one-body case, this reduces to the value 
$ \alpha_{\rm crit} = \pi/4 = 0. 7854$  obtained in ref. (II).
As a function of $Z$ the critical coupling has the behaviour
\bea
\alpha_{\rm crit} & \stackrel{Z \to 0}{\longrightarrow} &  \frac{\pi}{4} \, 
\left [ \, 1 - \frac{\pi}{2} Z^2 + \frac{3 \pi^2}{4} Z^4 + \ldots \, \right] 
\label{alpha crit z small}\\
\alpha_{\rm crit} &  \stackrel{Z \to \infty}{\longrightarrow} &  
\sqrt{ \frac{27 \pi}{128}} \, \frac{1}{Z} + \ldots 
\label{alpha crit z big}
\eea
which shows that the critical coupling for the bound-state system 
is {\it reduced} compared to the one-body case.
\begin{figure}[htb]
\vspace{0.5cm}
\mbox{\epsfysize=70mm \epsffile{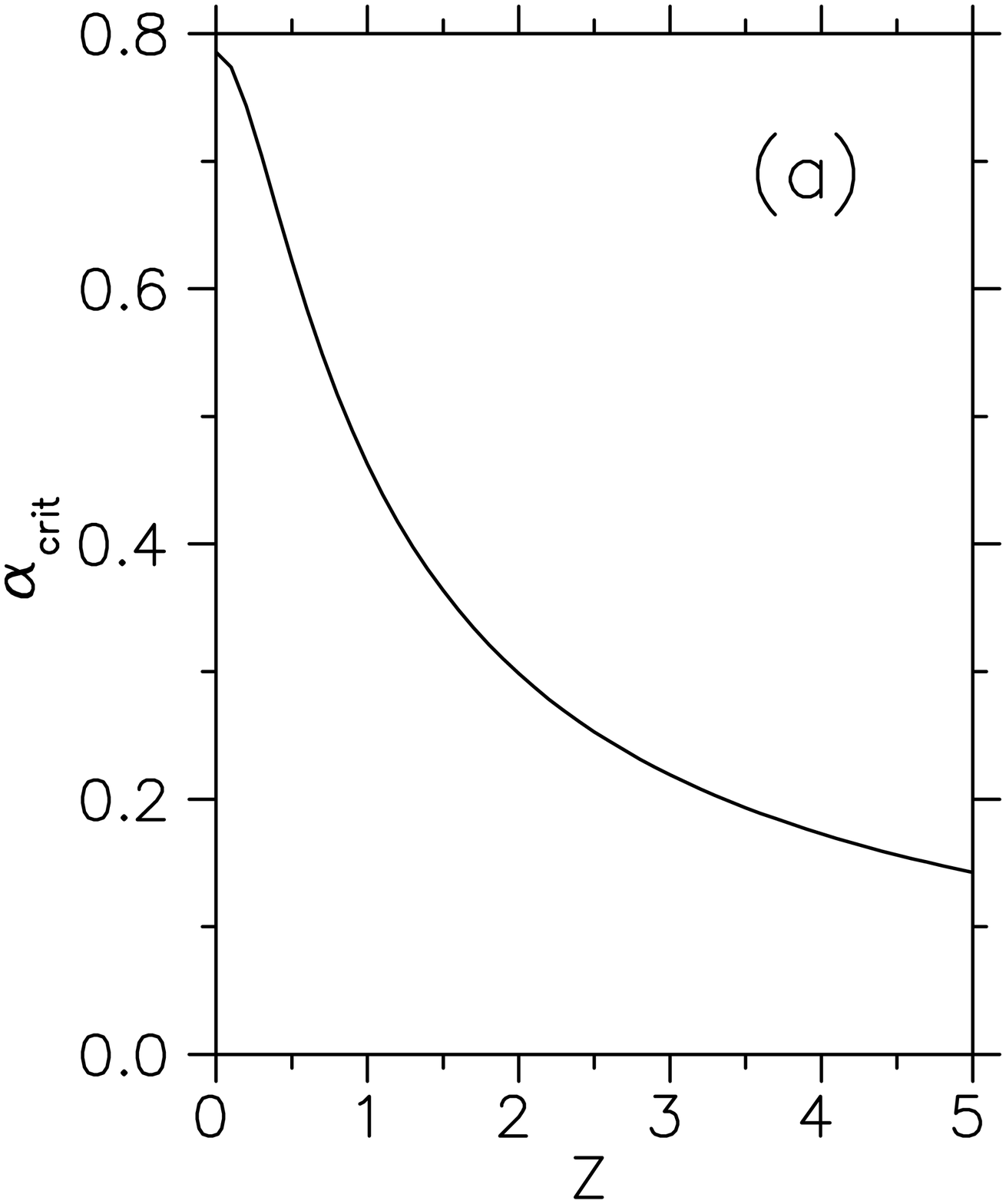}}
\vspace*{-70mm}
\hspace*{10mm}
\mbox{\epsfysize=70mm \epsffile{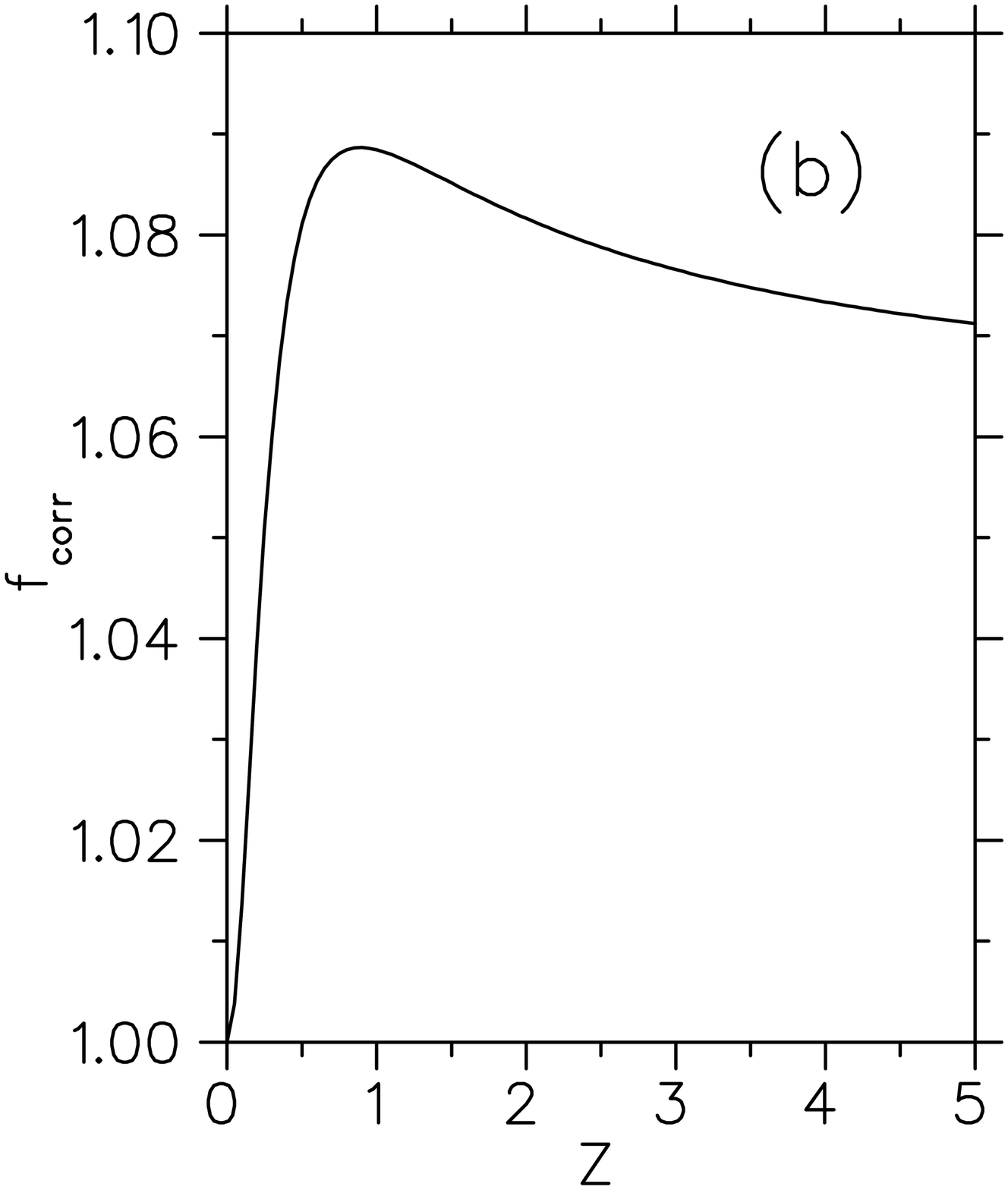}}
\vspace*{70mm}
\caption{(a) The critical coupling $\alpha_{\rm crit}$ and (b) 
the correction factor $f_{\rm corr}$ as function of the 
artificial strength $Z$ of the binding interaction in the simplified 
treatment of the induced instability.}
\label{fig: crit/corr}

\end{figure}

Indeed ~fig.~\ref{fig: crit/corr}a ~demonstrates that the reduction is monotonous 
with increasing $Z$. For $Z = 1 $ one obtains
\be
\alpha_{\rm crit} \, \biggr |_{Z = 1} \> \simeq \>  0.4627
\ee
which is in reasonable agreement with the value $ \alpha_{\rm crit} = 0.542 $ 
obtained from the numerical solution of the variational equations.
It is also interesting that 
the low-$Z$ expansion (\ref{alpha crit z small}) does not converge for the value
$Z = 1 $, a phenomenon which has also been observed in bound-state QED calculations
\cite{Jen}.

It is also possible to determine approximately the width of the bound-state for 
$ \alpha > \alpha_{\rm crit} $ following the treatment in Sect. IV. A of 
ref. (II).
To be specific we write \footnote{In relativistic calculations it is more 
customary to denote the square of the mass of an unstable particle as 
$M^{\star \, 2} - i M^{\star} \Gamma$ but this does not make any significant 
difference for the rough analytical estimate which we will derive.}
\be
q^2 \E \left ( \, M^{\star} - i \frac{\Gamma}{2} \, \right )^2 \> \> , 
\hspace{0.5cm} M^{\star} \> \simeq \> 2 M 
\ee
and introduce
\be
\zeta \Def \lambda \, \frac{q}{2} \> \simeq \>  \lambda  \left ( \, M - 
i \frac{\Gamma}{4} \, \right ) \deF \zeta_0 \, e^{-i \chi} \> .
\label{def zeta}
\ee
A nonvanishing phase $\chi$ signals a complex (approximate) solution of 
the variational equations for coupling constants above the critical one.
Near the critical coupling $\alpha_{\rm crit}$ given in Eq. (\ref{alpha crit}) 
the phase $\chi(\alpha)$ is determined by the implicit equation (see appendix~
E)
\be
\alpha \E \alpha_{\rm crit} \, \left [ 1 + \alpha_2 \, \chi^2 + {\cal O}(\chi^4) 
\, \right ]
\label{alpha by chi}
\ee
where $\alpha_2$ is given in Eq. (\ref{alpha2}).
This can be used to evaluate the width near the critical coupling for which 
one obtains
\be
\Gamma  \stackrel{\alpha \to \alpha_{\rm crit}}{\longrightarrow} \> \frac{2}{3} \, 
2 M \, \left ( \frac{\alpha - \alpha_{\rm crit}}{\alpha_{\rm crit}} \right )^{3/2} 
\, f_{\rm corr} (Z)
\label{Gamma near crit 2}
\ee
with a correction factor 
\be
f_{\rm corr}(Z) \E \frac{\sqrt{1+3z} (1 + \sqrt{1+3z}) (1 + z +\sqrt{1+3z})^{1/2}}{(1 
+ 2 z + \sqrt{1+3z})^{3/2}} \> , \> z \equiv 2 \pi Z^2 \> .
\label{fcorr}
\ee 
The limiting cases are
\be
f_{\rm corr}(Z) \> \stackrel{Z \to 0}{\longrightarrow} \> 1 + \frac{\pi}{2} Z - 
\frac{19 \pi^2}{8} Z^2 + \ldots \> \> , \hspace{0.5cm} f_{\rm corr}(Z) \> 
\stackrel{Z \to \infty}{\longrightarrow} \> \frac{3}{4} \sqrt{2} + 
\frac{1}{16} \sqrt{ \frac{3}{\pi Z}} + \ldots \>
\ee
and ~fig.~\ref{fig: crit/corr}b ~shows that this function is rising rapidly from 
$f_{\rm corr}(0)= 1$ to its asymptotic value
$f_{\rm corr}(\infty) = 1.0607$. At $Z = 1 $ its value is 
$ f_{\rm corr}(1) = 1.08845$. However, this enhancement of the width pales
(and may be not even significant in the light of the 
drastic approximations which have been employed) in comparison with the 
factor $2$ which
distinguishes Eq. (\ref{Gamma near crit 2}) from the one-body result. This factor 
$2$ simply comes from the fact that the bound state has mass $M^{\star} \simeq 2 M$.
Thus, the {\it induced instability} by the presence of the second particle 
not only shows up in the lower critical coupling constant but also in the 
width which is roughly doubled if one exceeds this coupling 
by the same relative amount as in the single-particle case.

What happens in the limit (\ref{no self}), i.e. for the case where the self-energy and 
vertex corrections are neglected ? This can be easily analyzed by writing 
$Z = \alpha'/\alpha $ which goes to infinity when $\alpha$ tends to zero
and $\alpha'$ is kept fixed. 
With the help of Eq. (\ref{alpha crit z big}) we immediately obtain
\be
\alpha_{\rm crit} \To \frac{9}{8} \sqrt{\frac{\pi}{6}} \, 
\frac{\alpha_{\rm crit}}{\alpha'_{\rm crit}} \> \> 
> \Longrightarrow \> \> \> \alpha'_{\rm crit} \> \simeq \> \sqrt{\frac{27 \pi}{128}} 
\E 0.8141
\ee
which is close to the numerically observed value. Similarly above the critical coupling
the estimate (\ref{Gamma near crit 2}) for the width becomes 
\be
\Gamma \Bigr |_{\alpha = 0} \> \simeq \> \sqrt{2} M \, 
\left ( \frac{\alpha' - \alpha'_{\rm crit}}{\alpha'_{\rm crit}} \right )^{3/2} \> .
\label{Gamma near crit noself}
\ee
Therefore the induced instability also shows up 
if self-energy and vertex corrections are switched off.

\section{Summary and Outlook}
\label{sec: summ}
We have investigated the relativistic binding problem of two particles with equal mass
in the scalar Wick-Cutkosky model employing worldline variational methods. Compared to
the standard field-theoretical description the worldline (``particle'') representation
of the system entails a huge reduction in the degrees of freedom,
leading to a quantum-mechanical path integral over the trajectories of the heavy 
particles (``nucleons'') like in the famous polaron problem. 
While previous work in this approach concentrated on single-particle properties and 
processes this is the first application of the method to the inherently 
non-perturbative binding of two relativistic particles and can therefore be 
considered as a relativistic analogue of the bipolaron problem: although the 
delicate balance between repulsive two-body interaction and attractive self-energy 
is missing, there is the additional problem of non-perturbative renormalization. 
Fortunately, the WC model is a super-renormalizable theory so that -- in the quenched 
approximation -- only an identical mass renormalization both in the 1-nucleon and 
the 2-nucleon sector is necessary.

Whereas most bipolaron studies employed parametrized {\it ans\"atze} for the 
variational functions we used a quadratic trial action with free retardation 
functions which are determined by the variational principle itself. This variational 
principle takes the form of a specific equation (``Mano's equation'' for the 
two-body case) which is stationary under 
unrestricted variation of the variational functions and determines the lowest 
bound state in a fully covariant manner. No definition of center-of-mass coordinates, 
factorization of wave functions or choice of special frames is necessary. 
This was checked in the nonrelativistic case where the correct center-of-mass and 
internal energies were obtained from the variational calculation. In the 
relativistic case we have derived the (non-linear) variational equations, 
solved them numerically and studied the asymptotic behaviour of the variational 
functions for small and large values of their arguments. 

In this approach the binding energy is not determined by an eigenvalue equation but 
by the specific value of the external variable $ q_{\star}^2 < 4 M^2 $ where 
both sides of
Mano's two-body equation are equal. We have obtained this value analytically for 
the special case of small coupling constant and zero mass of the 
exchanged particle (``pion'') and numerically without any restriction below a 
critical coupling. In both cases we found more binding due to 
the effects of self-energy and vertex corrections which are included approximately 
in our calculation. 
This is in contrast to results from light-cone calculations \cite{Ji} which claim that 
self-energy effects are repulsive. However, a variety of results obtained by different 
techniques -- dressed-ladder Bethe-Salpeter equation \cite{Saul}, 
effective  field theories, our variational approach -- all point to the opposite 
conclusion.
We have identified
the source of this increased binding as mainly due to the pion-nucleon coupling 
enhanced by vertex corrections, in a similar way as the anomalous magnetic moment of 
the electron or the muon is enhanced by radiative corrections.
Additional binding also comes from crossed-ladder contributions as shown in ref. 
\cite{NiTj1}.
This was confirmed  
by artificially switching off the 
radiative effects in our program and retaining only
ladder and crossed-ladder contributions to the binding. 

However, whereas Quantum Electrodynamics has a stable ground state for not too strong 
fields the scalar Wick-Cutkosky model has not -- a  deficiency which 
too often is forgotten or neglected in bound-state calculations employing this model.
Indeed, whereas previous worldline calculations in the one-particle sector already
yielded a critical coupling above which no real solutions of the variational equations 
exist anymore, the two-particle bound state ceased to exist for even smaller 
coupling constants indicating an induced instability. Surprisingly this instability 
also shows up when the self-energy and vertex corrections are neglected and only 
ladder and crossed-ladder diagrams are included.
Although the variational equations 
in our approach are highly non-linear integral equations we were able to explain this 
phenomenon analytically by simple {\it ans\"atze} which lead to solvable, algebraic
equations. Actually, the ability to combine numerical results with simple analytical 
insights seems to be a feature which the worldline variational approach shares with 
ordinary Quantum Mechanics: the virial theorem which was used to control the 
accuracy of the numerical calculations and the Feynman-Hellmann theorem employed 
for the weak-coupling expansion of the binding energy are other examples.

One of the main weaknesses of the present approach certainly is the use of a 
quadratic, oscillator-like trial action also for the direct interaction of the 
two particles which in the nonrelativistic limit reduces to a Yukawa potential. 
This limits the accuracy, in particular near the threshold where the coupling 
constant is strong enough to support
a bound state. With the present (isotropic) trial action this deficiency is even worse
since in the nonrelativistic (weak coupling) limit the kinetic term reduces
to a 4-dimensional and not to a 3-dimensional theory. This could 
(and will) 
be easily improved by 
allowing anisotropic terms as was done in the one-particle sector 
in ref. \cite{WC7} thus giving the trial action more flexibility. 
The necessary modifications of Mano's equation are straightforward with a modest 
increase in numerical complexity due to the appearance of longitudinal and transverse
components w.r.t. to the external momentum $q$.
Second-order 
corrections to the variational result can also be calculated rather straightforwardly
\cite{cum2} and typically reduce deviations from known exact results by 
a factor of 3 -- 4.
Still, one would like to obtain the correct quantum-mechanical binding energy 
for weak coupling also in this approach which could be possible following the 
work of Luttinger and Lu in the polaron case \cite{LuLu}. At the expense of applying
Jensen's inequality twice these authors have introduced a (variational) 
potential into the trial action and were able to reproduce the polaron energy for
strong coupling where Feynman's quadratic trial action leads to the largest deviations.

Further extensions may include the unequal-mass case, binding of more than two 
particles similar as in ref. \cite{AlDi} and inclusion of vacuum polarization 
effects, i.e. going beyond the quenched approximation (see ref. \cite{RoPANIC} 
for attempts in this direction). 
Calculating  bound states in QED would be another step towards approaching a realistic
theory although the present-day perturbative 
techniques are sufficient for the small electromagnetic coupling constant.
Spin degrees of freedom and gauge invariance are no obstacles for a worldline
description \cite{QED1}.
As we were able to extract the anomalous mass dimension of the electron, i.e. to 
determine the singularity structure of renormalization constants in the worldline 
variational approach \cite{QED2}, the necessary non-perturbative renormalization 
for a renormalizable theory like QED requires more work but should be feasible.
It is clear, however, that the ultimate goal and challenge for
a non-perturbative approach lies in the dynamics of quarks and gluons at low energies.
Whereas color degrees of freedom can be included in a worldline formalism  
\cite{HoGa} it is unclear, at present, how to treat the non-abelian -- 
and therefore nonquadratic -- gluon action in a gauge-invariant way. 
Auxiliary-field methods in which three- and four-gluon self-interactions are
eliminated \cite{Kondo} may be a possible way to proceed as can be shown  
for the simpler case of a $\Phi^4$-theory \cite{Rophi4}.
We think that the results obtained so far in the worldline variational approach, 
in particular the novel
way of treating the relativistic bound-state problem presented in this work, 
are encouraging enough to make an effort in this direction.

\begin{acknowledge}
All diagrams were drawn by using JaxoDraw \cite{Jaxo}.
\end{acknowledge}

\vspace{1cm}
\appendix

\setcounter{equation}{0}

\section{Nonrelativistic Binding}
\subsection{The Nonrelativistic Limit}
\label{app: nonrel}

It has long been known that the exchange of a scalar particle 
(a meson with mass $m$) is approximately equivalent 
to an {\it attractive} Yukawa potential between the heavy nucleons 
with mass $M$.
The usual derivation starts from the one-meson-exchange scattering 
amplitude in the static limit and equates it with the Born approximation 
of potential scattering (see e.g. ref. \cite{ErWe}, appendix~10).
A more consistent way is to integrate out the meson field in 
the generating functional
\be
Z[J^*,J;j] \E \int \prod_{i=1}^2 {\cal D} \Phi_i  \, {\cal D} \Phi_i^* \> 
{\cal D} \chi \> \exp \left \{ \,  i \int d^4x \, \left [ {\cal L}_2
+ \sum_i \left ( J_i^* \Phi_i + \Phi_i^* J_i\right ) + j \chi \, \right ] \, 
\right \} 
\label{gen func 1}
\ee
for vanishing sources, keeping the velocity of 
light $c$ (and for completeness also 
Planck's constant $\hbar$) and then letting $c \to \infty $ in order to obtain
the nonrelativistic limit \cite{Efi3}.
The velocity of light (re)appears in the following places
\bea
x_0  \EA c t \> \> \Longrightarrow \> \> 
\partial^2 \E \Box \E \frac{1}{c^2} \frac{\partial^2}{\partial t^2} - \Delta 
 \hspace{1cm} ({\rm d'Alembert \> operator}) \\
M_i  & \to & \frac{M_i c}{\hbar} \hspace{2.5cm} ({\rm inverse \> Compton \> 
wavelength \> of \> nucleons}) \nonumber
\eea
and it should be remembered that the action is the time integral over the
Lagrange function
\be
S \E \int dt \int d^3x \> {\cal L} \E \frac{1}{c} \int d^4x \>  {\cal L}
\ee
entering as $\exp ( i S/\hbar) $ in the path integral.
The standard shifted gaussian integral 
then leads to an effective action for the nucleons 
\bea
S_{\rm eff}[\Phi_i^*,\Phi_i] \EA \frac{1}{c} \sum_i \, 
\int d^4x \> \Phi_i^*(x) 
\left ( -\Box -  \frac{M_i^2 c^2}{\hbar^2}  \right ) \Phi_i (x) 
- \frac{1}{c} \sum_{i,j} \frac{g_i g_j}{2} \non  
&& \times \, \int d^4 x \, d^4 y \> 
\Phi_i^*(x) \Phi_i (x) \left < x \left |  
\left ( -\Box - \frac{m^2 c^2}{\hbar^2} \right )^{-1} \right | y \right > 
\Phi_j^*(y) \Phi_j (y)
\eea
which is a non-local $\Phi^4$-theory. 
Making the ansatz \cite{BeFu}
\be
\Phi_i \E \frac{ \hbar \phi_i}{\sqrt{2 M_i}} \, e^{-i M_i c^2 t/\hbar} \>  
\Longrightarrow \> 
\frac{1}{c^2}\partial_t^2 \Phi_i \E \left [ \,  - \frac{M_i^2 c^2}{\hbar^2}  
\phi_i - 
\frac{2 i M_i}{\hbar} \, \partial_t \phi_i + \frac{1}{c^2} \partial_t^2 \phi_i 
\, \right ]  \frac{\hbar}{\sqrt{2 M_i}} \, e^{-i M_i c^2 t/\hbar} \> .
\ee 
and sending $c \to \infty$
\footnote{Since the meson mass $m$ may be small or zero 
we do not perform the limit in the meson propagator.} gives
\bea
S_{\rm eff}[\phi_i^*,\phi_i] & \longrightarrow & \frac{1}{c} 
\sum_i \, \int d^4x \phi_i^*(x) 
\left (i \hbar \partial_t + 
\frac{\hbar^2}{2M_i} \Delta \right  ) \phi(x)  
- \frac{1}{c}\sum_{i,j} \, 
\frac{\hbar^4 g_i g_j}{8 M_i M_j} \non
&& \times \int d^4 x \, d^4 y \> 
\phi_i^*(x) \phi_i(x)  \, \left < x \left | (-\Box - m^2 c^2/\hbar^2)^{-1} 
\right | y \right > \, 
\phi_j^*(y) \phi_j(y) \> .
\eea
The first term is the kinetic term of a system of nonrelativistic
 particles (of two types) whereas the 
last one describes their interactions. In the limit $ c \to \infty $ 
the retardation of meson exchange (i.e. the dependence on $p_0$) is suppressed 
and one obtains an {\it instantanous} interaction
\bea
\left < x \left | \frac{1}{-\Box - m^2 c^2/\hbar^2 +i0} \right | y \right > \EA 
\int \frac{d^4 p}{(2 \pi \hbar)^4} \, 
\frac{\exp [ - i p \cdot (x-y)/\hbar]}{p_0^2/(c^2 
\hbar^2)  - {\bf p}^2/\hbar^2 - m^2 c^2/\hbar^2 + i0} \non
&\stackrel{c \to \infty}{\longrightarrow}&  - \hbar^2 \delta 
\left ( x_0-y_0 \right ) \, 
\int \frac{d^3 q}{(2 \pi \hbar)^3} \, \frac{\exp [  i {\bf p} \cdot ( {\bf x}-
{\bf y} )/\hbar ]}{{\bf p^2} + m^2 c^2} \non
\EA - \delta \left ( x_0-y_0 \right ) \,
\frac{1}{4 \pi |{\bf x} - {\bf y}| } \, 
\exp \left ( - \frac{m \hbar}{c}  |{\bf x} - 
{\bf y}| \right ) \> .
\eea
ence, if the effective, nonrelativistic action is written in the standard 
many-body form for two-body potentials 
\bea
S_{\rm eff}[\phi_i^*,\phi_i] \EA \int dt \, \Biggl \{ \, \int d^3x \, \sum_i \, 
\phi_i^*({\bf x},t) \left ( i \hbar \partial_t + \frac{ \hbar^2 \Delta}{2M_i} 
\right ) \phi_i({\bf x},t)  \non
&& - \frac{1}{2} \int d^3x d^3y \, \sum_{i,j} \phi_i^*({\bf x},t)  
\phi_j^*({\bf y},t)  
V_{ij} ( {\bf x} - {\bf y} )  \phi_j ({\bf y},t) \phi_i({\bf x},t) \, \Biggr \}
\eea
one finds that an attractive Yukawa potential 
\be
V_{ij} \left ({\bf x} - {\bf y} \right ) \E - 
\frac{\hbar^4 g_i g_j}{16 \pi M_i M_j} 
\frac{\exp \left ( - m c|{\bf x} - {\bf y}|/\hbar
\right )}{|{\bf x} - {\bf y}| }
\label{Yukawa pot} 
\ee
with range $ \hbar/(m c) $, the Compton wavelength of the exchanged meson, 
acts between the different types of nucleons.

\subsection{Variational Approximation for the Nonrelativistic 
Polarization Propagator}
\label{app: FJ nonrel}

Since we make a variational approximation of the ground state 
energy of the relativistic system it is appropriate to assess the accuracy of our 
calculation in the nonrelativistic limit to which it should reduce for 
$ c \to \infty $. 
The standard approach in Quantum Mechanics is, of course, 
the Rayleigh-Ritz variational principle 
which involves a trial {\it wave function}. Although it is obvious that 
a quadratic trial {\it action} corresponds to a gaussian trial wave function 
we will derive the corresponding approximation in close analogy to the 
field-theoretic (worldline) case. In particular, we do not use the simplification
that the nonrelativistic Hamiltonian can be split up into a center-of-mass part
and a relative part. 
Instead we will deal with the translationally invariant full system whose
Hamiltonian reads
\be 
\hat H \E \frac{\bfp_1^2}{2M_1} + \frac{\bfp_2^2}{2M_2} + V \left ( \bfx_1 
- \bfx_2 \right )
\ee
and investigate a quantity where one projects on a fixed (total) momentum 
$\bfq$. This is just the nonrelativistic 
polarization propagator in $d = 3$ dimensions 
\bea
\Pi ( \bfq, E ) \EA \int d^3 x \, e^{- i \bfq \cdot \bfx} 
\, \left < \bfx_1 =   \bfx_2 =  \bfx \left | \, \frac{1}{E - \hat H + i0} \, 
 \right | \, \bfx_1 =  \bfx_2 = 0 \right > \non
\EA \! \! - i  \int d^3 x   e^{- i \bfq \cdot \bfx} 
 \int_{0}^{ \infty} \! \! dT \left < \bfx_1 =   \bfx_2 =  \bfx 
\left |  \exp\left[ i T  \left(E - \hat H \right) \right]
\right | \bfx_1 =  \bfx_2 = 0 \right >  
\eea
which has poles at the total energy of the system
\be
\Pi (\bfq, E ) \E \sum_n \, |\psi_n(0)|^2 \, \left \{ \, 
E - \left [ \frac{\bfq^2}{ 2(M_1+M_2)} + \epsilon_n \right ] \, 
\right \}^{-1} \> .
\ee
The time-evolution operator has the path integral representation
\be
\left < \bfx_1 =   \bfx_2 =  \bfx \left | \, \exp \left[ - i T \left(  
\hat H \, - i0 \right) \right] \, \right | \, \bfx_1 =  \bfx_2 = 0 \right > 
\E
\int_{\bfx_i (0)= 0}^{\bfx_i (T) = \bfx } {\cal D}^3 x_1 {\cal D}^3 x_2 \> \exp  
\left \{ \, i S \left[ \bfx_1 ,\bfx_2 \right] \, \right \} 
\ee
where the action is
\be
S \left[ \bfx_1 ,\bfx_2 \right ] \E \int_0^T dt \> \left [ \, \
\frac{M_1}{2} \dot{\bfx}_1^2 + 
 \frac{M_2}{2} \dot{\bfx}_2^2 - V\left ( \bfx_1 - \bfx_2 \right ) \, \right ]
\> .
\ee
Note that both particles have one common time - in contrast to the relativistic
case. This facilitates the limit $ T \to \infty $ even for the
case of unequal masses $ M_1 \neq M_2 $.
For vanishing interaction the free polarization propagator
can be calculated easily
\be 
\Pi^{(0)}(\bfq,E) \E \frac{1}{(2 \pi)3} \left(\frac{ \pi M_{\rm red}}{i} 
\right)^{3/2}\int_0^{\infty} \frac{dT}{T^{3/2}} \, 
\exp \left [ \,  iT E - iT \frac{q^2}{2 (M_1 + M_2)}  \, \right ] \> ,
\ee
where $ M_{\rm red} $ is the reduced mass of the system defined in Eq. 
(\ref{def Mred}). This is used for the normalization of the path integral 
in the general case. Thus we have
\be
\Pi(\bfq,E) \E \left(\frac{M_{\rm red}}{4 \pi i} \right)^{3/2}
\int_0^{\infty} \frac{dT}{T^{3/2}}  \exp \left [   i E T  - 
i \frac{\bfq^2 T}{2 (M_1 + M_2)}   \right ] \,  
\cdot \frac{\int \tilde{\cal D}(x_1,x_2) \, \exp ( i \tilde S )}
{\int \tilde{\cal D}(x_1,x_2) \, \exp ( i \tilde S_0 )}
\label{Pi worldline-nonrel}
\ee
where
\bea
\int \tilde{\cal D}(x_1,x_2) &\equiv& \int d^3 x \,  e^{ - i \bfq \cdot \bfx} \, 
\int_{\bfx_1(0)=0}^{\bfx_1(T)=\bfx} {\cal D}^3 x_1 \, 
\int_{\bfx_2(0)=0}^{\bfx_2(T)=\bfx} {\cal D}^3 x_2 \\
\tilde S &\equiv& - \bfq \cdot \bfx + S[\bfx_1,\bfx_2] \> , \> \> \> \tilde S_0 
\EQ  - \bfq \cdot \bfx + \sum_{i=1}^2 \, S_0[\bfx_i] \\
S_0[\bfx_i] &\equiv& \int_0^{T} dt \> \frac{M_i}{2} \, \dot{\bfx}_i^2 \> .
\eea
We now apply the Feynman-Jensen variational principle (\ref{Feyn-Jens})
with a suitable (and manageable) trial action. 
For a proper choice we first 
look at the free action: with the Fourier expansion of the paths
\be
\bfx_i(t) \E \bfx \, \frac{t}{T} + \sum_{k=1}^{\infty} \, 
\frac{2 \sqrt{ T}}{k \pi \sqrt{M_i}} \, 
\bfa_k^{(i)} \, \sin \left ( \frac{k \pi t}{T} \right )
\ee
the free action becomes
\be
\tilde S_0 \E  -  \bfq \cdot \bfx + \sum_{i=1}^2 \, \left [ \, 
\bfx^2 \frac{M_i}{2T}  +  
 \sum_{k=1}^{\infty} \, \bfa_k^{(i) \> 2} \> \right ] \> .
\ee
In the one-body case a rule of thumb is to ``decorate'' the free action 
with variational parameters to obtain a quadratic trial action 
which reduces to the free one if the interaction
is switched off. In the two-body case we need an additional coupling term 
which should account for the interaction of the particles.
Therefore we take as trial action
\be
\tilde S_t \E  - \tilde \lambda \bfq \cdot \bfx + \sum_{i=1}^2 \, 
\left [ \, A_0 \, \bfx^2 \frac{M_i}{2T}  +  
 \sum_{k=1}^{\infty} \, A_k^{(i)} \, \bfa_k^{(i) \> 2} \, \right ] 
- \, \sum_{k=1}^{\infty} \, 2 B_k \> \bfa_k^{(1)} \cdot  \bfa_k^{(2)} \> .
\label{tilde St-nonrel}
\ee
The Fourier coefficients $A_0, A_k^{(i)}, B_k $ as well as the
parameter $\tilde \lambda$ are variational parameters. The relative minus sign
between the $A$ and $B$ terms has been chosen for consistency with the relativistic
ansatz (\ref{tilde St}).
Since the trial action is at most quadratic the averages can be calculated
analytically: 
\bea
\frac{\int \tilde {\cal D} \, \exp (i \tilde S_t)}{\int \tilde {\cal D} \, 
\exp (i \tilde S_0)} \EA \exp \Biggl \{  \, 
i \Biggl [ \, \bfq^2 \frac{T}{2\left(M_1 + M_2\right)} 
\left ( 1 - \frac{\tilde \lambda^2}{A_0} \right ) \non
&& \hspace{1cm} + i \frac{3}{2} \left ( \ln A_0 + \sum_{k=1}^{\infty} 
\ln \left (A_k^{(1)} A_k^{(2)}  - B_k^2 \right ) 
\right ) \,  \Biggr ] \, \Biggr \}  \\
\left <  \tilde S_0 - \tilde S_t  \right >_t
\EA i \frac{3}{2} \left[ \frac{1-A_0}{A_0} + \sum_{k=1}^{\infty} 
\frac{2B_k^2}{A_k^{(1)}A_k^{(2)}-B_k^2} + \sum_{k=1}^{\infty} \frac{A_k^{(1)} 
+ A_k^{(2)} -2A_k^{(1)}A_k^{(2)}}{A_k^{(1)}A_k^{(2)}-B_k^2} \right] \non
&&  \hspace{2cm} + \bfq^2 \frac{T}{2\left(M_1 + M_2 \right)} 
\left( \frac{\tilde \lambda^2}{A_0^2} 
+ \frac{\tilde \lambda^2}{A_0} - \frac{2 \tilde \lambda}{A_0} \right) \\
\left <  S_1\right >_t 
\EA - \int_{0}^{T}dt \int \frac{d^3p}{(2 \pi)^3} \> 
\tilde V (\bfp) \, 
\left < e^{i \bfp \cdot \left( \bfx_1(t) - \bfx_2(t) \right)} 
\right >_t \non
\EA  - \int_{0}^{T}dt \int \frac{d^3p}{(2 \pi)^3} \
\tilde V (\bfp) \,  e^{ -i \bfp^2 \nu(T,t)} \deF - T \, V(T) \> .
\eea
Here
\be 
\tilde V (\bfp) \E \int d^3 x \> V(\bfx) \, e ^{i \bfp \cdot \bfx}
\ee
is the Fourier transform of the potential between the two particles and the
abbreviation
\be
\nu(T,t) \E \frac{T}{M_1M_2} \sum_{k=1}^{\infty}\frac{\sin^2 \left (
k \pi t/T \right )}{k^2 \pi^2 }   \, 
 \frac{ M_1 A_k^{(1)} + M_2 A_k^{(2)} - 2 \sqrt{M_1 M_2} B_k}{
A_k^{(1)} A_k^{(2)}  - B_k^2} 
\ee
has been used. Thus we find
\bea
\Pi(\bfq,E) &\simeq &\frac{1}{(2 \pi)^3} \left(\frac{ \pi M_{red}}{i} 
\right)^{3/2}
\int_0^{\infty} \frac{dT}{T^{3/2}} \, \exp \Biggl [ \,  i T \Bigl ( \, E - 
 \frac{\bfq^2}{2 (M_1 + M_2)} \non
&& \hspace{2cm} + \frac{\bfq^2}{2 (M_1 + M_2)} \left( 1 - 
\frac{\tilde \lambda  }{A_0}\right)^2 - \Omega(T)  -  V (T) \, \Bigr ) 
\, \Biggr ]  
\label{Pi nonrel}
\eea
where the kinetic term is defined as
\bea
\Omega (T) \EA   \frac{3}{2iT} \, \Biggl [ \, \ln A_0 + 
\frac{1 - A_0}{A_0} + \sum_{k=1}^{\infty} \ln \left( A_k^{(1)}A_k^{(2)}-B_k^2 
\right) \non
&& \hspace{1cm} + \sum_{k=1}^{\infty} \frac{A_k^{(1)} + A_k^{(2)} -2A_k^{(1)}
A_k^{(2)}}{A_k^{(1)}A_k^{(2)}-B_k^2} + 2 \sum_{k=1}^{\infty} 
\frac{B_k^2}{A_k^{(1)}A_k^{(2)}-B_k^2} \, \Biggr ]  \> .
\eea
As usual the pole of the polarization propagator is determined by the
limit $T \to \infty$ in which case the sums over Fourier coefficients become 
integrals over $ E = k \pi/T $ as in Eq. (\ref{asympt}). In addition, we may use 
the fact that only those terms in the exponent which develop a linear 
$T$-dependence contribute to the pole position. This means that the 
$A_0$-terms can be dropped and that we can replace\footnote{This can be 
seen, for example, by scaling $ t = T \, x$ and using well-known results
for the asymptotic limit of Fourier cosine transforms \cite{Light}.}
\be
\sin^2 (ET) \E \frac{1}{2} \left [ \, 1 - \cos ( \, 2 E t \, ) \, \right ] 
\> \longrightarrow \> \frac{1}{2}
\ee
in the argument of the function $\nu(T,t)$ making it a {\it constant}:
\be
\nu(T,t) \> \stackrel{T \to \infty}{\longrightarrow} \> \nu \E \frac{1}{2 \pi} 
\frac{1}{M_1 M_2}  \, \int_0^{\infty} dE \> \frac{1}{E^2} 
\frac{M_1 A^{(1)}(E) 
+  M_2 A^{(2)}(E) - 2 \sqrt{M_1 M_2} B(E)}{A^{(1)}(E) A^{(2)}(E) - B^2(E)} \> .
\label{nu asy}
\ee
However, this constant is still a functional of the variational functions
$A^{(i)}(E), B(E) $. Thus we obtain
\bea
\Omega \EA \! \frac{3 }{2 \pi i}  \int_0^{\infty} \! \! dE  \left [ 
\, \ln \left ( A^{(1)}(E) A^{(2)}(E) - B^2(E) \right ) + 
\frac{A^{(1)}(E)+ A^{(2)}(E)}{A^{(1)}(E) A^{(2)}(E) - B^2(E)} - 2  \right ]  
\label{Om nonrel} \\
V \EA \int\frac{d^3p}{(2 \pi)^3} \> \tilde V (\bfp) \, e^{- i \bfp^2 \, \nu} \> .
\eea
Collecting the terms linear in $T$ in Eq. (\ref{Pi nonrel}) and setting
\be
\lambda \equiv \frac{\tilde \lambda}{A_0}
\ee
Mano's equation reads 
\be
E_0 \E \frac{\bfq^2}{2(M_1 + M_2)} \left( 2\lambda - \lambda^2 \right) + 
\left ( \, \Omega + V  \, \right )\, \deF E_{\rm CM} + \epsilon_0 \> .
\label{Mano nonrel}
\ee
Due to our metric ($q^2 = q_0^2 - \bfq^2$) there is now
a different sign between the center-of-mass (CM) and internal parts 
compared with the relativistic case. Variation with respect to $\lambda $ 
yields 
\be
\lambda \E 1
\ee
since $\Omega$ and $V$ are independent of $\lambda$. This is crucial both for 
obtaining the correct center-of-mass energy
\be
E_{\rm CM} \E \frac{\bfq^2}{2(M_1 + M_2)}
\ee
and for guaranteeing that the internal energy $\epsilon_0$ is independent of the 
total momentum of the system.

To determine the optimal variational functions we now vary
Mano's equation (\ref{Mano nonrel}) with respect to $A^{(1)}(E), A^{(2)}(E)$ and
$B(E)$ 
\be
\frac{\delta}{\delta A^{(i)} (E)}\left(\Omega + V\right) \E 0 \> , \> \> 
\frac{\delta}{\delta B (E)}\left(\Omega + V\right) \E 0 \> .
\ee
The variation of the individual terms gives
\bea
\frac{\delta \Omega}{\delta A^{(i)}} \EA 
c_1\left(\frac{A^{(3 - i)}}{A^{(1)}A^{(2)} 
- B^2} - \frac{B^2 + A^{(3 - i)2}}{\left(A^{(1)} A^{(2)} - B^2\right)^2} 
\right) \> \> , \> \> i = 1,2\\
\frac{\delta \Omega}{\delta B} \EA c_1\left(\frac{- 2 B}{A^{(1)} A^{(2)} - B^2} 
+ \frac{2\left(A^{(1)} + A^{(2)}\right)B}{\left(A^{(1)} A^{(2)} - B^2\right)^2} 
\right)\\
\frac{\delta V}{\delta A^{(i)}} \! \EA \!   \frac{c_2}{E^2} 
\frac{ \sqrt{\frac{M_i}{ M_{3-i}}} 
\left(A^{(1)} A^{(2)} - B^2\right) - \left( \sqrt{\frac{M_1}{ M_2}}  A^{(1)} + 
\sqrt{\frac{M_2}{ M_1}} A^{(2)} + 2B\right) A^{(3 - i)}}{\left(A^{(1)} A^{(2)} - 
B^2\right)^2}   \\ 
\frac{\delta V}{\delta B} \EA   \frac{c_2}{E^2} \frac{- 2 \left(A^{(1)} A^{(2)} 
- B^2\right) + 2 \left( \sqrt{\frac{M_1}{M_2}} A^{(1)} + \sqrt{\frac{M_2}{M_1} } A^{(2)} -
2B \right) B}{\left(A^{(1)} A^{(2)} - B^2\right)^2} 
\eea
where we have omitted the argument $E$ in $A^{(i)}(E)$ and $B(E)$; 
$c_1 = 3/(2 \pi i)$ is a numerical constant whereas the constant
\be
c_2 \E \frac{1}{2 \pi i} \frac{1}{\sqrt{M_1 M_2}} \, \int \frac{d^3 p}{(2 \pi)^3} 
\> \bfp^2 \, \tilde V(\bfp) \, e^{- i \bfp^2 \, \nu}
\label{def c2}
\ee
depends on $\nu$, i.e. on the variational functions. 
We observe the relations
\bea
2B \frac{\delta \Omega}{\delta A^{(i)}} + A^{(3-i)} \frac{\delta \Omega}{\delta B}
\EA c_1 \frac{2B}{A^{(1)}A^{(2)} - B^2} \\
2B \frac{\delta V}{\delta A^{(i)}} + A^{(3-i)} \frac{\delta V}{\delta B} \EA c_2 
\frac{1}{E^2} \frac{2\left(\sqrt{M_i / M_{3-i} }B - A^{(3-i)}\right)}{A^{(1)}A^{(2)} -
 B^2} 
\eea
which (by adding suitable combinations of the variational equations) gives us
\be
A^{(1)}(E)  \E B(E) \, \left ( \, \sqrt{\frac{M_{3-i}}{M_i}} + 
\frac{c_1}{c_2} E^2 \, \right ) \> .
\ee
If this is inserted  into one of the variational equations one finds after 
some algebra the simple solution
\be
B(E) \E  \frac{c_2}{c_1} \,  \frac{1}{E^2} 
\ee
and hence 
\be
A^{(i)}(E) \E 1 + \sqrt{\frac{M_{3-i}}{M_i}} \,  \frac{c_2}{c_1} \frac{1}{E^2} \> .
\ee
Defining 
\be
\omega^2 \Def - \frac{M_1 + M_2}{\sqrt{M_1 M_2}} \> \frac{c_2}{c_1} 
\label{def omega}
\ee
we finally have as solutions of the variational equations in real time
\be
A^{(i)}(E) \E 1 - \frac{1}{1 + \frac{M_i}{M_{3-i}}} \, \frac{\omega^2}{E^2} 
\> \> , \hspace{0.5cm} 
B(E) \E - \frac{\sqrt{M_1M_2}}{M_1 + M_2} \frac{\omega^2}{E^2} \> .
\label{solution Ai,B}
\ee
Note that for equal masses 
\be
A^{(1)}(E) \Bigr |_{M_1 = M_2} \E   A^{(2)}(E) \deF A(E) \E
1 - \frac{1}{2} \, \frac{\omega^2}{E^2} \> \> , \hspace{0.5cm}
B(E) \Bigr |_{M_1 = M_2} \E  - \frac{1}{2} \, \frac{\omega^2}{E^2}
\ee
so that $ A(E) - B(E) = 1$ and $ A(E) + B(E) = 1 - \omega^2/E^2 $.
With Eq. (\ref{def c2}) the definition (\ref{def omega}) for the quantity $\omega$
reads
\be
\omega^2 \E - \frac{1}{3} \frac{M_1 + M_2}{M_1 M_2} \, \int \frac{d^3 p}{(2 \pi)^3} 
\> \bfp^2 \, \tilde V(\bfp) \, e^{- i \bfp^2 \, \nu} \> .
\label{omega eq}
\ee
This is an implicit equation for $\omega$ since $\nu$ depends on $\omega$. 
Indeed substituting the solutions (\ref{solution Ai,B}) into Eq. (\ref{nu asy})
one obtains
\be
\nu \E \frac{1}{2 \pi} \frac{1}{\sqrt{M_1 M_2}} \, \int_0^{\infty} dE \> 
\frac{1}{E^2} \, \frac{\sqrt{M_1/M_2} + \sqrt{M_2/M_1}}{1 - \omega^2/E^2 + i0} 
\E - \frac{i}{4 \omega} \left ( \, \frac{1}{M_1} + \frac{1}{M_2} \, \right ) \> .
\label{nu from omega}
\ee
Here we had to specify how the singularity at $ E = \omega$ 
has to be treated. This can be inferred from the original real-time 
path integral by noting that the profile functions need an infinitesimal
positive imaginary part for convergence of the integral. This is equivalent
to $\omega^2 \to \omega^2 - i0$, Feynman's prescription for the causal propagator.
Note that the RHS of Eq. (\ref{omega eq}) is positive for an attractive potential
as anticipated. The implicit equation becomes more familiar if
we consider Mano's equation {\it before} variation w.r.t. $\omega$.
The kinetic term (\ref{Om nonrel}) is
\be
\Omega \E \frac{3}{2 \pi i} \, \int_0^{\infty} dE \, \left [ \, \ln \left ( 1 - 
\frac{\omega^2}{E^2} + i 0 \right ) + \frac{2 - \omega^2/E^2}{1 - \omega^2/E^2 
+ i0} - 2 \, \right ] \E \frac{3}{4} \omega
\label{Omega nonrel}
\ee
and therefore the internal energy in Mano's equation (\ref{Mano nonrel}) 
is determined from 
\be
\epsilon_0 \E \frac{3}{4}\omega + 
\int \frac{d^3 p}{(2 \pi)^3} \> \tilde V (\bfp) \, \exp \left (
\frac{-\bfp^2}{4M_{\rm red} \,  \omega } \right) 
\label{Mano omega}
\ee
which is {\it exactly} the Rayleigh-Ritz variational principle 
with the normalized gaussian trial wave function
\be
\psi_t(\bfx) \E \left ( \frac{M_{\rm red} \,  \omega}{\pi} \right )^{3/4} \, 
\exp \left ( - \frac{M_{\rm red} \,  \omega}{2} \bfx^2 \right ) \> .  
\ee
Indeed variation of Eq. (\ref{Mano omega}) w.r.t. $\omega$ 
leads to Eq. (\ref{omega eq}) if Eq. (\ref{nu from omega}) is taken into account.

Thus -- as expected -- in the nonrelativistic limit the Feynman-Jensen variational 
principle with a quadratic trial action is equivalent 
to the Rayleigh-Ritz variational principle with a gaussian trial wavefunction 
\footnote{As we have worked in real time
where the Feynman-Jensen variational principle only ensures stationarity, we have 
missed the minimum property of the method.  However, this
is easily established in euclidean time where we can use the Feynman-Jensen
{\it inequality}.}.
Apart from the mass-dependent factors the 
profile functions (\ref{solution Ai,B}) are recognized as the standard ones 
for a harmonic oscillator with frequency parameter $\omega$ . It is this 
variational parameter which has to be optimized for the specific potential 
$V(\bfx)$ in the quantum mechanical problem.

For the Yukawa potential $V(\bfx) = - \alpha \exp(- m x)/x \> , x = |\bfx| $ ,
a simple calculation gives 
\be
\epsilon_0 \> \le \> \epsilon_{var}(y) \E M_{\rm red} \, \alpha^2 \, \left [ \, 
\frac{d}{4} \frac{1}{y^2} - \frac{2}{\sqrt{\pi}} \frac{1}{y} + 
2 \delta \, e^{\delta^2 y^2}
\, {\rm erfc} \left ( \delta y \right ) \, \right ]
\label{var energy}
\ee
where $ y = \alpha \sqrt{M_{\rm red} \, /\omega} $ is the appropriate dimensionless 
variational parameter and $\delta = m/(2 \alpha M_{\rm red}) $ was 
already defined in Eq. (\ref{def delta}). erfc($x$) = 1 -- erf($x$)  denotes 
the complementary error function and we have 
explicitly introduced the dimension $ d (= 3) $
in the kinetic-energy term. $ d = 4 $ then also covers the weak-coupling limit
of the relativistic variational calculation where -- as discussed -- the rigid, 
four-dimensional trial action leads to an enhanced kinetic term and therefore to 
less binding in the nonrelativistic limit.
For small $\delta$ one obtains 
\be
\epsilon_{var} \E  M_{\rm red} \, \alpha^2 \, \left [ \, - 
\frac{4}{d \pi} + 2 \delta - d \delta^2 + \frac{d^2 \pi}{8} \delta^3 + 
\ldots \, \right ]
\ee
which means that the exact Coulomb result in Eq. (\ref{E0 Yuk small m}) is
missed by $15 \% $ for $ d = 3 $ and by $ 36 \% $ for $ d = 4 $.
However, the first-order (for $ d = 4 $) and (for $ d = 3 $) 
also the second-order correction due to the finite meson mass
come out correctly. At first sight this looks not very 
impressive for the variational calculation unless one realizes that a 
Yukawa potential is quite different from a trial harmonic oscillator 
potential.

\noindent
The variational energy becomes zero or positive when 
\be
\delta \> \ge \> \delta_{\rm crit} = \frac{4}{d \sqrt{\pi} } \,  
\frac{s_0}{1 +  s_0^2  }
\ee
where $ s_0 \equiv \delta y $ fulfills the transcendental equation
\be
\sqrt{\pi} \, s_0 \left ( 1 +  s_0^2 \right ) \,  \exp \left ( s_0^2\right )  
{\rm erfc}\left (s_0 \right ) - s_0^2 \E \frac{1}{2}
\> .
\ee
Numerically one finds $ s_0 = 0.821324 $ and therefore
$\delta_{\rm crit} = 1.10687/d $. This means that the variational calculation 
predicts bound states for the Yukawa potential if
\be
\alpha \, \frac{2 M_{\rm red}}{m} \> \ge \left \{ \begin{array}{l@{\quad:\quad}l}
                                                2.71035 & d = 3\\
                                                3.61380 & d = 4
                                                \end{array} \right.
\label{nonrel criterion}
\ee
Due to the minimum property of the variational calculation this is always larger 
than the exact number (\ref{crit}).

\section{Calculation of Averages}
\label{app: aver}
\setcounter{equation}{0}

To perform path integrals involving the trial action (\ref{tilde St}) we need 
the ``master'' integral in $d$ dimensions
\be
I_M  \E \int d^dx \, \int {\cal D}^d a^{(1)} {\cal D}^d a^{(2)} \, 
e^{i \tilde S_M} \deF \int \tilde {\cal D} \, e^{i \tilde S_M}
\ee
where
\be
\tilde S_M \E p \cdot x - \frac{\kappa_0}{2} A_x x^2 - 
\frac{\kappa_0}{2} \sum_{i=1}^2 \, 
\sum_{k=1}^{\infty} \, A_k^{(i)} \, a_k^{(i) \> 2} + \kappa_0 \sum_{k=1}^{\infty} 
\, 
B_k \, a_k^{(1)} \cdot a_k^{(2)} + \sum_{i=1}^2  \sum_{k=1}^{\infty} \, f_k^{(i)} 
\cdot a_k^{(i)} \> .
\ee
This is practically the trial action with sources 
$f_k^{(i)}$ coupled to the modes which will be needed to evaluate 
the average of the interaction.
With the (ubiquitous) gaussian integral
\be
\int d^da \, \exp \left [ \, - i \frac{\kappa_0}{2} A \, a^2 + i b \cdot a \, 
\right ] 
\E \frac{\rm const.}{A^{d/2}} \, \exp \left [ \, \frac{i}{2 \kappa_0} \, \frac{b^2
}{A} 
\, \right ]
\ee
we get by completing the square
\bea
I_M \EA {\rm const.} \, A_x^{-d/2}  \, \prod_{k=1}^{\infty} \left ( A_k^{(1)} 
A_k^{(2)} - B_k^2 \right )^{-d/2} \, \exp \Biggl \{ \> \frac{i}{2 \kappa_0} \, 
\frac {p^2}{A_x} 
+  \frac{i}{2 \kappa_0} \, \sum_{k=1}^{\infty} \, \frac{1}{A_k^{(1)} A_k^{(2)} - 
B_k^2} \non
&& \cdot \left [ \, A_k^{(2)} f_k^{(1) \> 2} + A_k^{(1)} f_k^{(2) \> 2} + 
2 B_k f_k^{(1)} \cdot f_k^{(2)}
\, \right ] \> \Biggr \} \deF {\rm const} \, \cdot \, \exp \left ( \, 
\frac{i}{2 \kappa_0} {\cal F}_M \, \right ) \> ,
\label{master int}
\eea
where $ {\cal F}_M $ is a function of $ p, A_x, A_k^{(i)}, B_k $ and $f_k^{(i)}$.

\noindent
We are now in a position to calculate the averages required in the Feynman-Jensen
variational principle (\ref{Feyn-Jens}). First
\be
\frac{\int \tilde {\cal D} \, \exp (i \tilde S_t)}{\int \tilde {\cal D} \, 
\exp (i \tilde S_0)} \E \exp \left [ \, 
\frac{i}{2 \kappa_0} \left ( {\cal F}_t - {\cal F}_0 \right )  \, \right ] \> . 
\ee
To evaluate the numerator (i.e. ${\cal F}_t$) one simply has to put
$p = \tilde \lambda q, A_x = A_0 (1/T_1 + 1/T_2)$ in Eq. (\ref{master int}); 
for the denominator (i.e.  ${\cal F}_0$) one sets
$p = q,  A_x = 1/T_1 + 1/T_2$ and in both cases the sources are zero: 
$f_k^{(1)} = f_k^{(2)} = 0$. This gives
\bea
&& \frac{\int \tilde {\cal D}  \, \exp (i \tilde S_t)}{\int \tilde {\cal D} 
\, \exp (i \tilde S_0)} \E \non
&& \exp \left \{    
\frac{i}{2 \kappa_0} \left [ q^2 \frac{T_1 T_2}{T_1 + T_2} 
\left ( \frac{\tilde \lambda^2}{A_0} - 1 \right ) + 
i d \kappa_0  \left ( \ln A_0 + \sum_{k=1}^{\infty} \ln \left (
A_k^{(1)} A_k^{(2)}  - B_k^2 \right ) 
\right )  \right ]  \right \} . 
\eea
$d = 4$ is understood but kept general for pieces which contribute to the 
``kinetic'' term in the variational functional.
Next we obtain the average of $\tilde S_0$ with respect to the trial action
by differentiation
\bea
\left < \tilde S_0 \right >_t \! \EA \! \frac{1}{i} \left [  
\frac{\partial}{\partial \tilde \lambda} + \frac{\partial}{\partial A_0}
+ \sum_{i=1}^2 \sum_{k=1}^{\infty}  \frac{\partial}{\partial A_k^{(i)}}  
\right ]  \ln \int \! \tilde {\cal D} e^{i \tilde S_t} 
= \left [   \frac{\partial}{\partial \tilde \lambda} 
+ \frac{\partial}{\partial A_0}
+ \sum_{i=1}^2 \sum_{k=1}^{\infty}  \frac{\partial}{\partial A_k^{(i)}}  
\right ] \, \frac{{\cal F}_t}{2 \kappa_0} \non
\EA  \frac{1}{2 \kappa_0} \, \left [ \, \frac{q^2 T_1 T_2}{T_1 + T_2} \left ( 
\frac{2 \tilde \lambda}{A_0} - \frac{\tilde \lambda^2}{A_0^2} \right )
+ i \kappa_0 d \left ( \frac{1}{A_0} + \sum_{k=1}^{\infty} \,
\frac{A_k^{(1)} + A_k^{(2)}}{A_k^{(1)} A_k^{(2)}  - B_k^2} \right ) \, \right ]
\> .
\eea
Similarly we may obtain 
\bea
\left < \tilde S_t  \right >_t \EA  
\frac{1}{i} \frac{\partial}{\partial r}   \ln \int  \tilde {\cal D} \exp \left \{
  i \tilde S_t
\left [ \tilde \lambda \to r \tilde \lambda, A_0 \to r A_0,
A_k^{(i)} \to r  A_k^{(i)}, B_k \to r B_k \right ]  \right \}
\Biggr|_{r=1} \non
\EA  \! \frac{1}{2 \kappa_0}  \frac{\partial}{\partial r}  {\cal F}_t [r \tilde 
\lambda, r A_0, r A_k^{(i)}, r B_k] \Biggr|_{r=1} 
=  \frac{1}{2 \kappa_0} \left [ \frac{ q^2 T_1 T_2}{T_1 + T_2} 
\frac{\tilde \lambda^2}{A_0}  + i \kappa_0 d \left ( 1 + \sum_{k=1}^{\infty} 2 
\right )  \right ]  .
\eea
Finally we calculate the average of the interaction part (\ref{Sint 1,2})
\be
\left < S_{\rm int}  \right >_t \E 
- \sum_{\alpha,\beta=1}^2 \,  
\frac{g_{\alpha}' g_{\beta}'T_{\alpha} T_{\beta}}{ 8 \kappa_0^2}
\int_0^1 d\tau \, d\tau'  \int \frac{d^4 p}{(2 \pi)^4} \,
\frac{1}{p^2 - m^2 + i0} \, 
\left < \exp \left [  - i p \cdot \left (  x_{\alpha}(\tau) -  
x_{\beta}(\tau') \right ) \right ] \right > .
\label{Sint av 1}
\ee
Since
\be 
- p \cdot \left (  x_{\alpha}(\tau) -  x_{\beta}(\tau') \right ) \E 
- p \cdot x  
\left ( \tau - \tau' \right ) - p \cdot \sum_{k=1}^{\infty} \, \left (  
\frac{\sqrt{2 T_{\alpha}}}{k \pi} \sin (k \pi \tau)  a_k^{(\alpha)} - 
\frac{\sqrt{2 T_{\beta}}}{k \pi} \sin (k \pi \tau') a_k^{(\beta)} \right )
\ee
the average of the last exponential in Eq. (\ref{Sint av 1}) is given by 
\be
\exp \left [ \, \frac{i}{2 \kappa_0} \left ( {\cal F}_{\rm int}(\alpha,\beta) 
- {\cal F}_t \right ) \, \right ]
\ee
where 
\bea
{\cal F}_{\rm int}(\alpha,\beta) \EA {\cal F}_M \Bigl [ p \to \tilde \lambda q - 
p (\tau - \tau'), \, A_x \to A_0(1/T_1 + 1/T_2), \, A_k^{(i)}, \, B_k, \non
&& \hspace{1.2cm} f_k^{(i)} = - p \frac{\sqrt{2 T_i}}{k \pi}  \left ( \, 
\delta_{\alpha i} \,
 \sin (k \pi \tau) -  \delta_{\beta i} \, 
\sin (k \pi \tau') \, \right )  \Bigr ] \> .
\eea
With the help of the master integral (\ref{master int}) we obtain
\bea
{\cal F}_{\rm int}(\alpha,\beta) - {\cal F}_t && = \> 
\frac{T_1 T_2}{A_0 (T_1 + T_2)} \, 
\left [ \, p^2 ( \tau - \tau')^2 - 2 \tilde \lambda \, p \cdot q \,  
( \tau - \tau') \, \right ] \non
&& + p^2 \sum_{k=1}^{\infty} \frac{1}{(k \pi)^2}  
\frac{1}{A_k^{(1)} A_k^{(2)}  - B_k^2} \, \Biggl [ \sum_{i=1}^2  2 T_i  
A_k^{(3-i)}   \Bigl (  \delta_{\alpha i} \, \sin (k \pi \tau) \non 
&& - \delta_{\beta i}  \sin (k \pi \tau') \, \Bigr )^2
+ 2 B_k  \prod_{i=1}^2 \sqrt{2 T_i} \, \left (  \delta_{\alpha i} 
\sin (k \pi \tau) -  \delta_{\beta i} \sin (k \pi \tau' ) \, \right )  \Biggr ] 
\> .
\eea
Introducing the modified variational parameter $\lambda$ from Eq. 
(\ref{def lambda}) we therefore have
\bea
\left < S_{\rm int}  \right >_t \EA 
- \frac{1}{8 \kappa_0^2} \, \sum_{\alpha,\beta=1}^2 \,  g_{\alpha}' g_{\beta}' \, 
T_{\alpha} T_{\beta} 
\int_0^1 d\tau \, d\tau' \> \int \frac{d^4 p}{(2 \pi)^4} \> 
\frac{1}{p^2 - m^2 + i0} \non
&& \cdot \exp \left \{ \, \frac{i}{2 \kappa_0} \, 
\left [ \, p^2 \mu_{\alpha \beta}^2(\tau,\tau';T_1,T_2) - 2 \lambda \, 
p \cdot q \, 
\frac{T_1 T_2}{T_1 + T_2} \, (\tau - \tau') \, \right ] \, \right \}
\label{S1 av}
\eea
where
\bea
\mu_{\alpha \beta}^2(\tau,\tau';T_1,T_2) \EA \frac{T_1 T_2}{A_0 (T_1 + T_2)} \, 
( \tau - \tau')^2 + \sum_{k=1}^{\infty} \, \frac{2}{(k \pi)^2} 
\frac{1}{A_k^{(1)} A_k^{(2)}  - B_k^2} \non 
&& \cdot \Biggl \{ \>  \delta_{\alpha,\beta} 
\, T_{\alpha} \, \left ( A_k^{(2)}  \delta_{\alpha 1} + A_k^{(1)}  
\delta_{\alpha 2}
\right ) \, \left [ \, \sin (k \pi \tau) -   \sin (k \pi \tau') \, \right ]^2 \non
&& \hspace{0.5cm} + \left ( 1 - \delta_{\alpha \beta} \right ) 
\, \Bigl  [ \, 
 T_{\alpha} \, A_k^{(\beta)} \, \sin^2 (k \pi \tau) 
+  T_{\beta} \, A_k^{(\alpha)} \, \sin^2 (k \pi \tau') \non
&& \hspace{2.5cm} - 2 B_k \, \sqrt{T_1 T_2} \,  \sin (k \pi \tau)   
\sin (k \pi \tau')\, \Bigr ] \> \Biggr \} \> .
\label{app:mu2(tau,tau')}
\eea
Putting everything together we now have
\bea 
\Pi^{\rm var}(q) \EA - \frac{1}{16 \pi^2} \, \int_0^{\infty} \frac{dT_1 \, dT_2} 
{(T_1 + T_2)^2} \, \exp \Biggl \{ \> \frac{i}{2 \kappa_0} \, 
\Biggl [ \, - (M_1^2 T_1 + M_2^2 T_2) + \frac{T_1 T_2}{T_1 + T_2} \, q^2 \, \left 
( 
2 \lambda - \lambda^2 \right ) \non
&& \hspace{5cm} - (T_1 + T_2) \, \left ( \, \Omega_{12}  +  V
\, \right )(T_1,T_2)  \, \Biggr ]  \, \Biggr \} 
\label{app:Pi var 1}
\eea
where we have defined 
\be
\Omega_{12}(T_1,T_2) \Def \frac{-i d \kappa_0}{T_1 + T_2} 
\left \{  \ln A_0 + \frac{1}{A_0} - 1 + \sum_{k=1}^{\infty} \, 
\left [   \ln \left ( A_k^{(1)} A_k^{(2)} - B_k^2 \right ) + 
\frac{A_k^{(1)}+ A_k^{(2)}}{A_k^{(1)} A_k^{(2)} - B_k^2} - 2 \right ]  
\right \}
\label{def Omega12}
\ee
and 
\be
V(T_1, T_2)  \E - \frac{2\kappa_0}{T_1 + T_2} \, \left < \, S_{\rm int}  \, 
\right >_t  \> .
\label{def V}
\ee
Equation (\ref{app:Pi var 1}) still holds for all $q$, i.e. also away from 
the poles of the polarization propagator.

\section{Asymptotic Behaviour of Profile functions and Pseudotimes}
\label{app: asy}
\setcounter{equation}{0}

Here we derive the limiting behaviour of the solutions of the variational 
equations both for $E, \sigma \to 0$ and for  $E, \sigma \to \infty$ . 
As the pseudotimes 
are Fourier cosine transforms of the (inverse) profile functions we expect, of 
course, that the large $E$-behaviour of the one function is related to the 
small $\sigma$-behaviour of the other and {\it vice versa}.

We first consider
the limit $E \to \infty$ in Eqs. (\ref{var eq for Aminus},\ref{var eq for Aplus}).
Naively one would just replace the squared 
trigonometric functions by their average value $1/2$ and obtain 
$A_{\pm} \to 1 + ({\rm const}_1+ {\rm const}_2)/E^2$ 
where the constants are determined by 
$ \int_0^{\infty} d\sigma \,\delta V_{1j}/\delta \mu^2_{1j}(\sigma) $. 
However, for $j = 1 $ this integral does not exist
as $\mu_{11}^2 \to \sigma$ for $\sigma \to 0$ 
and $ \delta V_{11}/\delta \mu^2_{11}$ diverges as $1/\sigma^2$ in 
that limit (see Eq. (\ref{delta V1j 2})). This is, of course the standard 
behaviour for the
self-energy part and the way to obtain the proper asymptotic behaviour 
is well-known: 
introduce a factor $\sigma^2/\sigma^2$ into the $\sigma-$integral and change to 
the variable $ x = E \sigma$. We then obtain
\bea
A_{\pm}(E) \EA 1 + \frac{2}{\kappa_E E^2} \, \int_0^{\infty} d(x/E) \, 
\left ( \frac{E}{x} \right )^2 \, 
\left ( \sigma^2 \frac{\delta V_{11}}{\delta \mu^2_{11}(\sigma)} 
\right )_{\sigma = x/E} \sin^2 \left ( \frac{x}{2} \right ) \non
&\stackrel{E \to \infty}{\longrightarrow} & \! 
1 + \frac{2}{\kappa_E \, E}  \lim_{\sigma \to 0}  \sigma^2 
\frac{\delta V_{11}}{\delta \mu^2_{11}(\sigma)}  \int_0^{\infty} \!\!  dx \, 
\frac{\sin^2(x/2)}{x^2} =
1 + \frac{\alpha  M^2}{4 \kappa_E \, E} + 
{\cal O} \left ( \frac{\ln^2E}{E^2} \right )
\label{Aplusminus for large E}
\eea
where in the last step the $(\sigma \to 0)$-limit of Eq.  (\ref{delta V1j 2})
has been used. 
It should be emphasized that this is exactly the same asymptotic 
behaviour as in the one-body (self-energy) case (see Eq. (143) in ref. [I]);
the only new information here 
is that both profile functions become identical at asymptotic values of $E$.
However, they have quite a different behaviour at small $E$: whereas $A_-(E)$ 
approaches a constant value
\be
A_-(0) \E 1  + \frac{1}{2 \kappa_E} \,  \int_0^{\infty} d\sigma \> \sigma^2
\sum_{j = 1}^2 \, \frac{\delta V_{1j}}{\delta \mu^2_{1j}(\sigma)} 
\> ,
\ee
$A_+(E)$ diverges for small $E$:
\be
A_+(E) \> \stackrel{E \to 0}{\longrightarrow} \> \frac{\omega_{\rm var}^2}{E^2} + {\rm cons
t.}
\> \> , \hspace{1cm} \omega_{\rm var}^2 \Def  \frac{2}{\kappa_E} \, \int_0^{\infty} 
d\sigma \> \frac{\delta V_{12}}{\delta \mu^2_{1 2}(\sigma)}
\label{def omegavar} 
\ee
as expected from the nonrelativistic limit. Consequently the pseudotimes
also have quite different behaviour at small $\sigma$. First, 
from Eq. (\ref{mu11 by Aplusminus}) it is seen that $\mu^2_{11}$ has the
standard self-energy behaviour
\be
\mu^2_{11}(\sigma) \E
\frac{2}{\pi} \, \sigma \int_0^{\infty} dx \> \frac{\sin^2(x/2)}{x^2} \, \left [ 
\, \frac{1}{A_-(x/\sigma)} + \frac{1}{A_+(x/\sigma)} \, \right ]
\> \stackrel{\sigma \to 0}{\longrightarrow} \> \frac{4}{\pi} \, \sigma 
\int_0^{\infty} dx \> \frac{\sin^2(x/2)}{x^2} \E \sigma \> .
\ee
For the low-$\sigma$ behaviour of $\mu^2_{12}$ we may just set $\sigma = 0$ 
in the definition (\ref{mu12 by Aplusminus}) to obtain
\be
\mu^2_{12}(\sigma) \> \stackrel{\sigma \to 0}{\longrightarrow} \>   
\E \frac{2}{\pi} \, \int_0^{\infty} dE \, \frac{1}{E^2 A_+(E)} \> \equiv \> 
\mu_{12}^2(0) 
\ee
where the integral exists due to the behaviour (\ref{def omegavar}) of $A_+(E)$ 
at small
$E$. Thus for $\sigma \to 0$
the `interaction' pseudotime stays constant and nonzero as assumed. 
Note that this constant is {\it not} the inverse of the $\omega_{\rm var}$ defined 
in Eq. (\ref{def omegavar}) since $ A_+(E) $ deviates 
from $ 1 + \omega_{\rm var}^2/E^2 $ at $E > 0$.

It is also possible to work out higher-order terms by considering
\be
\ddot \mu^2_{12}(\sigma) \E \frac{1}{\pi} \, \int_0^{\infty} dE \> \left [ \, 
\frac{1}{A_-(E)} - \frac{1}{A_+(E)} \, \right ] \, \cos \left ( E \sigma \right )
\> .
\ee
Because both profile functions have the same asymptotic behaviour 
(\ref{Aplusminus for large E}) for large $E$ the integral also exists for 
$\sigma = 0 $. Therefore we obtain after integration with the boundary condition\\
$\dot  \mu^2_{12}(0) = 0 $
\be
\mu^2_{12}(\sigma) 
\> \stackrel{\sigma \to 0}{\longrightarrow} \> \mu_{12}^2(0) + 
\frac{1}{2 \pi} \, \int_0^{\infty} dE \> \left [ \, 
\frac{1}{A_-(E)} - \frac{1}{A_+(E)} \, \right ] \, \cdot \, \sigma^2 + \ldots \> .
\ee
For the self-energy part the situation is different:
\be
\ddot \mu^2_{11}(\sigma) \E \frac{1}{\pi} \, \int_0^{\infty} dE \> \left [ \, 
\frac{1}{A_-(E)} + \frac{1}{A_+(E)} - 2 \, \right ] \, \cos \left ( E \sigma 
\right )
\ee
does not exist for $\sigma = 0 $ as the integrand behaves like $E^{-1}$ 
for large $E$ as can be seen from Eq. (\ref{Aplusminus for large E}). This means
that an expansion only in powers of $\sigma$ is not possible and that logarithmic
terms appear in higher order. The leading term is obtained by realizing that 
the cosine provides an upper limit const.$/\sigma$ for the $E$-integration. 
Therefore
\be
\ddot \mu^2_{11}(\sigma) \> \stackrel{\sigma \to 0}{\longrightarrow} \> 
\frac{1}{ \pi} \, \int^{{\rm const.}/\sigma} dE \> \left [ \, - \frac{\alpha}{2} 
\frac{M^2}{\kappa_E E} \, \right ] \E \frac{\alpha}{2 \pi} \frac{M^2}{\kappa_E} \, 
\ln \frac{\sigma}{\sigma_0}
\ee
where $\sigma_0$ is some undetermined constant. Integration with the appropriate 
boundary conditions gives
\be
\mu^2_{11}(\sigma) 
\> \stackrel{\sigma \to 0}{\longrightarrow} \> \sigma  + 
\frac{\alpha}{4 \pi} \frac{M^2}{\kappa_E} \, \sigma^2 \, \ln \frac{\sigma}{\sigma_1}
+ \ldots 
\ee
where $\sigma_1$ is another constant to be determined at ${\cal O}(\sigma^2)$. 
In the framework of generalized functions one obtains the same result 
from the asymptotic behaviour (\ref{Aplusminus for large E})
of the profile functions in the definition (\ref{mu12 by Aplusminus}) 
of $\mu_{11}^2(\sigma)$ (see Table 1 from ref. \cite{Light} \footnote{Note the 
different definition of the Digamma function in this work compared to 
ref. \cite{Handbook} !}). In turn this
logarithmic behaviour of the pseudotime $\mu_{11}^2(\sigma)$ 
induces the $\ln^2(E)/E^2$-remainder in the asymptotic expansion 
(\ref{Aplusminus for large E}) of the profile functions which is essential for the
convergence of several integrals encountered above.

Finally we consider the pseudotimes for large $\sigma$ 
by adding and subtracting the term $1/A_-(0) $ from the inverse pseudotimes in the
integrand. This gives 
\bea
\mu^2_{11}(\sigma) \EA \frac{2}{\pi}  
\int_0^{\infty} dE \, \frac{\sin^2(E \sigma/2)}{E^2 A_-(0)} + 
\frac{2}{\pi}  
\int_0^{\infty} dE \, \frac{\sin^2(E \sigma/2)}{E^2} \, \left [  \frac{1}{A_-(E)}
- \frac{1}{A_-(0)} + \frac{1}{A_+(E)}  \right ] \non
& \stackrel{\sigma \to \infty}{\longrightarrow} &  \frac{\sigma}{2A_-(0)} + 
\frac{1}{\pi} \, \int_0^{\infty} dE \,\frac{1}{E^2} \, \left [ \, \frac{1}{A_-(E)}
- \frac{1}{A_-(0)} + \frac{1}{A_+(E)} \, \right ] 
\label{mu11 large sigma}
\eea
where the last integral exists due to the proper behaviour of the integrand at 
$E = 0$ and $E = \infty$. Since the last line in  Eq. (\ref{mu11 large sigma}) 
was obtained by replacing $\sin^2(E \sigma/2)$ asymptotically by $1/2$ it also 
holds for $\mu^2_{12}$ where $1/A_+(E)$ is weighted with $\cos^2(E \sigma/2)$. 

\section{Numerical Details}
\label{app: num}
\setcounter{equation}{0}

Here we describe a few modifications and improvements of the numerical methods 
used in ref. (II) in order to achieve stable and reliable results for the 
bound-state problem. As before we have used Gauss-Legendre integration rules 
with $n_g$ gaussian points and $n_e$ extra divisions to evaluate
\be
\int_a^b dy \> f(y) \> \simeq \> \frac{h}{2} \, \sum_{i=1}^{n_e} 
\sum_{j=1}^{n_g} \, w_j \, 
f\left ( y_{i j} \right ) \> \> , \> \> y_{ij} \E a + 
\left ( i - \frac{1}{2} \right ) h +  \frac{h}{2} \, x_j \> \> , \> \> 
h \E \frac{b-a}{n_e}
\ee
where $x_j, w_j$ are the usual abscissae and weights for Gauss-Legendre numerical 
integration in the interval $[-1,+1]$ (see Eq. 25.4.30 in ref. \cite{Handbook}).
A major problem is the calculation of the pseudotimes 
from the profile functions at large $E, \sigma$ since the
integrand $\sin^2 (E \sigma/2) $ leads to huge oscillations 
and loss of accuracy when the integrals are evaluated by
mapping to a finite interval (e.g. by the transformation $ E = E_0 \tan \psi, 
\sigma = E_0^{-1} \tan \phi \> , \> E_0 $ a suitable scale) and 
subsequent gaussian integration. This procedure was adopted 
in ref. (II) and found to be sufficient
for the one-body case. Recently we have shown that one can eliminate the 
profile functions altogether and solve a nonlinear, delay-type equation for
the pseudotime which bears a striking similarity to the classical 
Abraham-Lorentz equation \cite{VALE}. This works perfectly well for 
the one-body case but 
would need a different strategy for the bound-state problem.
Therefore for the present purposes we have decided to stay in 
the proven, conventional scheme but to take {\it finite}
upper limits $E_{\rm max}, \sigma_{\rm max} $ for the integrations. 
The upper limit for $E$ is chosen
such that the asymptotic contribution (\ref{Aplusminus for large E}) at 
$ E = E_{\rm max}$ is less than $5 \times \Delta $ where $ \Delta $ 
is the measure
of deviation which should be reached in the iteration (see below).
Thus for $\kappa_E = 1$ 
\be
E_{\rm max} \E \frac{\alpha}{4} \, \frac{M^2}{ 5 \, \Delta} \> .
\ee
In all $E$-integrals the numerical integration is done up to 
$ E = E_{\rm max} $ and the asymptotic contribution from $ E_{\rm max} $ 
up to $ \> \infty \> $ is evaluated 
analytically and added to the numerical result. Similarly, an upper limit for the 
$\sigma$-integration is determined by requiring that the $u$-integral over the
ubiquitous exponential $e(u,\sigma)$ in Eq. (\ref{def e(u,sigma)})
\be
K(x,y) \E \int_0^1 du \, \exp \left (\, - x \, \frac{1-u}{u} - y \, u \, 
\right ) \> \> , \> x \E \frac{1}{2} \, m^2 \mu^2 (\sigma) \> \> , \> 
y \E   \frac{1}{2} \, \frac{(\lambda q/2)^2 \sigma^2}{ \mu^2(\sigma)} 
\label{def K(x,y)} 
\ee
should be smaller than $e^{-15} = 3.1 \cdot 10^{-7}$. Using the 
large-$\sigma$ limit
of this expression from Eqs. (82) - (84) in ref. \cite{WC3} we therefore 
have to take 
\be 
\sigma_{\rm max} \simeq  \frac{15}{m M} \> .
\ee
Also, in all $\sigma$-integrals the numerical integration is done up to 
$ \sigma  = \sigma_{\rm max} $ and
the asymptotic contribution from $ \sigma_{\rm max} $ till $\infty$ is evaluated 
analytically and added to the numerical result. This strategy works 
very well as we know the large
$(E,\sigma)$-behaviour of profile functions and pseudotimes (see appendix~C).

The number of gaussian points $n_g$ (typically $72$) and subdivisions  $ n_e $ 
(typically $ 3 - 4 $ ) should, of course, be adapted to the values of
$E_{\rm max},  \sigma_{\rm max}$: ideally at least one point should be on each
oscillation of the trigonometric function 
$ \sin^2 (E \sigma/2) = (1 - \cos (E \sigma))/2 $ for {\it all} values of 
$E, \sigma$ but experience showed that 
$ E_{\rm max} \sigma_{\rm max}/(\pi \, n_g \, n_e) \sim 3 - 4$ was sufficient.

In principle, the $u$-integral for $K(x,y)$ in Eq. (\ref{def K(x,y)}) can be 
expressed in terms of a (non-standard) special function, i.e. either by a
special Shkarofsky plasma dispersion function (Eq. (26) in ref. \cite{Rob}) 
or by an incomplete MacDonald function of first order in Bessel form 
(as defined in Eq. (III.1.8) of ref. \cite{inBess}). However, 
by doing so nothing is gained for practical computations. 
Therefore, for simplicity 
and ease of implementation, we have used a direct 
gaussian integration of the defining integral (\ref{def K(x,y)}) with the
same number of gaussian points and subdivisions as for the solution of
the variational equations. Of course, for $m = 0$ this is not necessary as the
$u$-integration can be done analytically.

The problem of numerical computation of integrals with oscillating integrands is also
present when one tries to evaluate the virial theorem in the form of Eq. 
(\ref{virial theorem}). Here the derivatives of the pseudotimes
\be
\frac{\partial \mu^2_{1j}(\sigma)}{\partial \sigma} \E \frac{1}{\pi} \, 
\int_0^{\infty} dE \> \left [ \, \frac{1}{A_-(E)} 
+ (-)^{j+1} \, \frac{1}{A_+(E)} \, \right ] \,  \frac{\sin (E \sigma)}{E}
\label{dot mu2}
\ee
are needed which have oscillating integrands with less damping at large $E$ than 
the pseudotimes themselves. This may lead to a loss of accuracy in the numerical 
evaluation. It can be shown that for theories without derivative interactions -- 
like the WC model but not for QED  -- the derivative term in the virial theorem 
(\ref{virial theorem}) may be eliminated completely.  
However, Table \ref{tab:varbind} and \ref{tab:varbind noself} show that below the 
critical coupling such a form of the virial theorem is not needed as the virial check 
is satisfied within a factor of two of the relative deviation which measures 
how well the variational equations are fulfilled.

\noindent
If one solves the coupled variational equations by iteration there is a crucial 
difference now compared to the one-body case in that one cannot start with 
$A_+(E) = 1$ as this would correspond
to an {\it unbound} system. Instead we have used the nonrelativistic solution
$ \, A_+(E) = 1 + \omega^2/E^2 $ with the Coulombic value 
(\ref{om nonrel d}) for $ d = 4$.
Obviously, the Yukawa value would be a better starting point although 
that doesn't really matter since one has to iterate many times anyway. 
Of course, the singular behaviour of the profile function $  A_+(E) $ 
at $ E = 0 $ has to be treated 
with care; this is done by updating the coefficient $ \omega_{\rm var} $  
from Eq. (\ref{def omegavar}) each time during the iteration. 
To shorten the iteration one can use the $\lambda$-value from the 
one-body case or the previous solution as starting point. 
As a measure of convergence during iteration we have used the relative 
deviation 
\be
\Delta_i(x) \E \left | \frac{x^{(i+1)} - x^{(i)}}{x^{(i)}} \right |
\ee
for each variational parameter $ x = \lambda, \omega_{\rm var} $ 
and the maximal relative deviation of the variational
functions $ x = {\rm max}_n [ A_{\pm}(E_n), \mu^2_{11/12}(\sigma_n)] $ at each 
gaussian point. From these relative deviations a
global measure of convergence 
\be
\Delta_i \E {\rm max} \left [ \,  \Delta_i(\lambda),  \Delta_i(\omega_{\rm var}), 
\Delta_i(A_{\pm}), \Delta_i(\mu^2_{11/12}) \, \right ]
\label{global dev}
\ee
was constructed and monitored during iterations.
The updating after each iteration was performed with
a prescription given by Hauck {\it et al.} \cite{HSA} which suppresses 
oscillating values from one iteration to the next by using an exponential 
weigthing of old and new values:
\be
\left ( 1 - e^{-\Delta_i(x)} \right ) \, x^{(i+1)} +  
e^{-\Delta_i(x)} \, x^{(i)} \> \longrightarrow \> x^{(i+1)} \>.
\ee
This is most effective in the early stages of the iteration when the relative 
deviation $\Delta_i(x)$ is still large.

The iteration was stopped when the global measure $\Delta_i$ was less than a 
prescribed value for which typically $10^{-3}$ was taken. Although this is far 
away from the $10^{-5}$ - accuracy which was easily obtained
in the one-body case, it should be sufficient for the present 
purposes. Given the much more demanding numerical problems of the two-body case
we believe that any improvement would need much more computing power 
(i.e. more integration points) and/or more efficient algorithms \cite{Maas,Mand}
to solve the nonlinear integral equations.

\section{Critical oupling Constant and Width: Analytic Approximation}
\label{app: crit}
\setcounter{equation}{0}

Here we show how one obtains the critical coupling from the simple ansatz
$A_-(E) = 1, A_+(E) = 1 + \omega^2/E^2 $, keeping $\lambda$ and $\omega$ as 
variational parameters. For this choice and $ \kappa_E = 1 $ one easily finds the 
kinetic terms and pseudotimes as
\be
\Omega_- \E 0 \> \> , \hspace{1cm} \Omega_+ \E \omega \> \> , \hspace{1cm} 
\mu_{1j}^2(\sigma) \E \frac{\sigma}{2} + 
\frac{1}{2 \omega} \, \left [ \,  1 + (-)^j e^{-\omega \sigma} \, \right ] \> \> 
( j = 1,2 )\> .
\label{amu supersimp}
\ee
Furthermore we assume weak binding 
\be
\frac{\omega}{(q/2)^2} \> \simeq \> \frac{\omega}{M^2} \> \ll \> 1
\ee
and that $ \omega \sigma  \ll 1 $ in all integrals over the proper time.
Then we may take approximately
\be
\mu_{11}^2(\sigma) \> \simeq \> \sigma \> \> , \hspace{1cm} \mu_{12}^2(\sigma) 
 \> \simeq \> \frac{1}{\omega} 
\ee
and obtain after performing the $\sigma$-integration
\bea
V_{11} & \simeq & \frac{\alpha}{2 \pi} M^2 \, \int_0^{1} du \> \ln \left [ \, 1 + 
\frac{(\lambda q/2)^2}{m^2} \, \frac{u^2}{1-u} \, \right ]
\label{V11 supersimp}\\
V_{12} & \simeq & - \frac{Z \alpha}{\sqrt{2 \pi}} M^2 \, 
\sqrt{ \frac{\omega}{(\lambda q/2)^2}} \, \int_0^1 du \> \frac{1}{2 \sqrt{u}} \, 
\exp \left [ \, - \frac{m^2}{2 \omega} \frac{1-u}{u} \, \right ] \> .
\label{V12 supersimp}
\eea
Equation (\ref{V11 supersimp}) is 
identical with Eq. (46) in ref. [II] if we use the weak binding 
approximation $ q/2 = M $. Mano's equation then approximately reads
\be
\bar M^2 \E \left ( 2 \lambda - \lambda^2 \right ) \left ( \frac{q}{2} \right )^2 
- \omega - 2 \left ( \, V_{11} + V_{12} \, \right ) \> .
\label{Mano approx 1}
\ee
We now vary w.r.t. $\lambda$ and $\omega$ and obtain
(for the case $ m = 0 $ where the $u$-integrals can be performed analytically)
\bea
\frac{1}{\lambda} \EA 1 + \frac{\alpha}{\pi} \frac{M^2}{(\lambda q/2)^2} \, 
\left [ \, 1 + Z \sqrt{\frac{\pi \omega}{2 (\lambda q/2)^2} }\, \right ] 
\label{lambda supersimp 1}\\
\omega \EA \frac{(Z \alpha)^2}{2 \pi} \, \frac{M^4}{(\lambda q/2)^2} \> .
\label{omega supersimp}
\eea
Note that Eq. (\ref{omega supersimp}) reduces to the $d=4$-result in 
Eq. (\ref{om nonrel d}) if we replace $\lambda q/2$ by $M$. Here, however
we keep the $\lambda$-dependence which in the one-body case was essential 
for probing the instability of the WC model.
Inserting Eq. (\ref{omega supersimp}) into the $\lambda$-Eq. 
(\ref{lambda supersimp 1}) and taking $ q/2 \simeq M$ now gives
a {\it quartic} equation for the variational parameter 
\be
\lambda^4 - \lambda^3 + \frac{\alpha}{\pi} \, \lambda^2 + \frac{(Z \alpha)^2}{2 \pi} 
\E 0
\label{lambda supersimp 2}
\ee
which for $Z = 0 $ reduces to the approximate one-body equation (149) in ref. [I]. To 
find the critical value of the coupling constant 
where this equation ceases to have real solutions we do not have to use the 
cumbersome explicit solutions; rather we observe that 
branching into the complex plane occurs when $ \partial \lambda/\partial \alpha = 
\infty$. Therefore we may 
differentiate Eq. (\ref{lambda supersimp 2}) w.r.t. to $\lambda$ and setting 
 $  \partial \alpha/\partial \lambda = 0 $ we obtain 
\be
4 \lambda_{\rm crit}^2 - 3 \lambda_{\rm crit} + 2  \frac{\alpha_{\rm crit}}{\pi} 
\E 0 \> .
\ee
If this is put back into Eq. (\ref{lambda supersimp 2}) we can solve for
\be 
\lambda_{\rm crit} \E \frac{1}{4} \, \frac{1}{1-z} \, \left [ 1 - 3 z + 
\sqrt{1 + 3 z} \right ]
\ee 
where 
\be
z \E 2 \pi Z^2 
\label{def z}
\ee
and obtain the critical coupling as
\be
\alpha_{\rm crit} \E \frac{\pi}{8} \, \frac{1 - 9 z + (1 + 3 z )^{3/2}}{(1 - z)^2} \> .
\ee
By simple algebraic 
manipulations this can be brought into the form (\ref{alpha crit}).

To obtain an analytic estimate for the width of the ground state above the critical 
coupling one allows the product $ \zeta = \lambda q/2 \deF \zeta_0 \, 
\exp(- i \chi) $ to become complex.
The quartic equation (\ref{lambda supersimp 2}) then reads
\be
\frac{1}{\lambda} \E 1 + \frac{\alpha}{\pi} \, \frac{M^2}{\zeta^2} + 
\frac{(Z \alpha)^2}{2 \pi} \, \frac{M^4}{\zeta^4} \>. 
\label{lambda supersimp 3}
\ee
If this and the variational solution (\ref{omega supersimp}) are inserted 
into the approximation (\ref{Mano approx 1}) to Mano's equation
we obtain
\be
\zeta^2 \E \bar M^2 - \frac{2 \alpha}{\pi} M^2 - \frac{3 (Z \alpha)^2}{2 \pi} \, 
\frac{M^4}{\zeta^2} + \frac{\alpha}{\pi} M^2 \, \int_0^{1} du \> \ln \left [ \, 1 + 
\frac{\zeta^2}{m^2} \, \frac{u^2}{1-u} \, \right ] 
\label{zeta eq 1}
\ee
which is the generalization of Eq. (48) in ref. [II]. For the imaginary part of 
this equation
it is possible to set immediately the meson mass to zero 
\footnote{This can be also done 
for the real part if we eliminate $\bar M$ from Mano's one-body equation, e.g.
by setting $ Z = 0 $ in Eq. (\ref{zeta eq 1}).} with the 
result
\be
\zeta_0^2 \sin 2 \chi \, \left [ \, 1 - \frac{3 (Z \alpha)^2}{2 \pi} \, 
\frac{M^4}{\zeta_0^4} \, \right ] \E \frac{2 \alpha}{\pi} M^2 \, \chi 
\label{Im zeta eq}
\ee
(compare with Eq. (49) in ref. [II]).
Equation (\ref{Im zeta eq}) allows the modulus $\zeta_0$ to be determined 
in terms of the phase $\chi$:
\be
\zeta_0^2 \E \frac{\alpha}{\pi} M^2 \, \frac{1}{2 S(\chi)} \, \left [ \, 
1 + \sqrt{1 + 6 \pi Z^2 \, S^2(\chi)} \, \right ] \> \> , \hspace{0.5cm} 
S(\chi) \Def \frac{\sin 2 \chi}{2 \chi}  \> 
\stackrel{\chi \to 0}{\longrightarrow} \> 
1 - \frac{2}{3} \chi^2 + \ldots \> .
\label{zeta0 by chi}
\ee
For $Z = 0 $ this reduces to Eq. (49) in ref. [II].
The width is obtained from Eqs. (\ref{def zeta}) and (\ref{lambda supersimp 3})
\be
M - i \frac{\Gamma}{4} \E \zeta + \frac{(Z \alpha)^2}{2 \pi} \, 
\frac{M^4}{\zeta^3} + \frac{\alpha}{\pi} \, \frac{M^2}{\zeta}
\ee
by taking real and imaginary parts:
\bea
M \EA \zeta_0 \, \cos \chi \, \left [ \, 1 + 
\frac{\alpha}{\pi} \, \frac{M^2}{\zeta_0^2} 
\, \right ] + \frac{(Z \alpha)^2}{2 \pi} \, 
\frac{M^4}{\zeta_0^3} \, \cos 3 \chi  
\label{zeta eq 2}\\
\Gamma \EA 4 \zeta_0 \, \sin \chi \, \left [ \, 1  - \frac{\alpha}{\pi} \, 
\frac{M^2}{\zeta_0^2} \, \right ] - \frac{2 (Z \alpha)^2}{\pi} \, 
\frac{M^4}{\zeta_0^3} \, \sin 3 \chi  \> .
\label{Gamma eq 1}
\eea
With the help of Eq. (\ref{Im zeta eq}) the last equation can 
be brought into the form
\be
\Gamma \E  4 \zeta_0 \, \sin \chi \, \left [ \, 1  - S(\chi) \, \right ] + 
\frac{2 (Z \alpha)^2}{\pi} \, 
\frac{M^4}{\zeta_0^3} \, \left [ \, 3 \sin \chi \, S(\chi) - \sin 3 \chi \, 
\right ]  
\label{Gamma eq 2}
\ee
which shows that close to the critical coupling ($ \chi \to 0 $) the width behaves as
\be
\Gamma \> \stackrel{\alpha \to \alpha_{\rm crit}}{\longrightarrow} \> 
\frac{8}{3} \zeta_{\rm crit} \, 
\left [ \, 1 + \frac{3 (Z \alpha_{\rm crit})^2}{2 \pi} \, 
\frac{M^4}{\zeta_{\rm crit}^4} \, \right ] \, \chi^3 \> .
\label{Gamma near crit 1}
\ee
Here 
\be 
\zeta_{\rm crit}^2 \E \frac{\alpha_{\rm crit}}{\pi} M^2 \, \frac{1}{2} \, \left [ \, 
1 + \sqrt{1 + 3 z} \, \right ]
\ee
is the value of the modulus at the critical coupling obtained from Eq. 
(\ref{zeta0 by chi}) by setting $\chi = 0$. The parameter $z$ has been defined in 
Eq. (\ref{def z}).

It remains to determine $\chi(\alpha)$. As in ref. (II) one obtains the inverse
relation by combining Eq. (\ref{zeta eq 2}) with Eq. (\ref{zeta0 by chi}). After some
algebra this gives 
\bea
\alpha \EA 2 \pi \, S(\chi) \, [ w(z,\chi)]^3 \non
&& \cdot \, \left \{ \, \cos \chi \left [ 1 + S(\chi) \right ] \, 
[ w(z,\chi)]^2 + z \, S^2(\chi) \, \left [ \cos 3 \chi - 3 \, 
S(\chi) \cos \chi \right ] \, \right \}^{-2}
\label{app:alpha by chi}
\eea
where $ w(z,\chi) = 1 + \sqrt{1 + 3 z S(\chi)} $. This result 
is the generalization of Eq. (53) in ref. [II] to which it exactly reduces for 
$Z = z = 0$.
Expanding $\alpha$ around the critical value 
in powers of $\chi$ yields 
\be
\alpha \E \alpha_{\rm crit} \, \left [ \, 1 + \alpha_2 \chi^2 + \ldots \, \right ]
\ee
where 
\be
\alpha_2 \E \frac{1 + 2 z + \sqrt{1 + 3 z}}{1 + z + \sqrt{1 + 3 z}} \> .
\label{alpha2}
\ee
Finally instead of Eq. (57) in ref. [II] one gets from  Eq. (\ref{Gamma near crit 1})
the following width for the ground state
\be
\Gamma  \simeq \frac{2}{3} \, 2 M \, 
 \left ( \frac{\alpha - \alpha_{\rm crit}}{\alpha_{\rm crit}} \right )^{3/2} \, 
f_{\rm corr} (Z) + 
\ldots
\ee
where the explicit form of the correction factor is given in Eq. (\ref{fcorr}).

\section{Weak-Binding Limit}

\setcounter{equation}{0}

Here we derive the weak-coupling limit for the binding energy in the case of massless
pions (which is the original Wick-Cutkosky model). Setting $ m = 0 $ allows us to perform 
the calculations analytically. First this is done for the 
variational approximation and then in the potential version of effective field theory 
appropriate for the present model \cite{PiSo}.

\subsection{Variational Calculation}
\label{app: weak var bind}
For the variational approach a convenient starting point is the Feynman-Hellmann theorem 
(\ref{Hell-Fey}) for which one only has to evaluate
\be
V_{12} \E - \frac{\alpha}{\pi} \, \frac{M^2}{(\lambda q/2)^2} \, \int_0^{\infty} 
d\sigma \> \frac{1}{\sigma^2} \, \left \{ \, 1 - \exp \left [ - 
\frac{(\lambda q/2)^2 \sigma^2}{2 \mu_{12}^2(\sigma)} \right ] \, \right \} \> .
\label{V12 m=0}
\ee
Again, we have choosen to work in the reparametrization ``gauge'' $\kappa_E = 1$.
From Eq. (\ref{amu supersimp}) we infer that approximately 
$ \mu_{12}^2(\sigma) \simeq 1/\omega + \omega \sigma^2/4 + \ldots $ with
$ \omega = {\cal O}\left ( (Z\alpha)^2 \right)$ (see  Eq. (\ref{omega supersimp})).
Therefore we insert the expansion
\be 
\mu_{12}^2(\sigma) \E u_0 + u_2 \sigma^2 + \ldots \> \> , \> \> u_2 \ll u_0
\ee
into Eq. (\ref{V12 m=0}). A simple calculation then gives
\be
V_{12} \E - \frac{Z \alpha}{\sqrt{2 \pi}} \, \frac{M^2}{\lambda q/2} \, 
\frac{1}{\sqrt{u_0}} \, 
\left [ 1 - \frac{1}{2} \frac{u_2}{(\lambda q/2)^2} + \ldots \right ] \> .
\ee
We have to determine the coefficients $u_0, u_2$ from the variational equations. 
This we can do perturbatively except that the low-$E$ behaviour 
$A_+(E) \to \omega_{\rm var}^2/E^2 $ needs to be included to all orders to generate
a bound state. Thus from Eq. (\ref{mu12(0)}) and Eq. (\ref{var eq for Aplus}) we have
\bea
u_0 \EA \frac{2}{\pi} \, \int_0^{\infty} dE \> \left \{ \, E^2 + \omega_{\rm var}^2 + 
2 \int_0^{\infty} d\sigma \left [ \delta V_{11} - \delta V_{12} \right ] \, 
\sin^2 (E\sigma/2) \, \right \}^{-1} \non
&\simeq& \frac{1}{\omega_{\rm var}} -  \int_0^{\infty} d\sigma \> \left [ \delta V_{11} - 
\delta V_{12} \right ] \, \frac{4}{\pi} \int_0^{\infty} dE \> 
\frac{\sin^2 (E\sigma/2)}{(\omega_{\rm var}^2 + E^2)^2} + \ldots \non
&\simeq& \frac{1}{\omega_{\rm var}} \left [ \, 1 - \frac{1}{4}  \int_0^{\infty} d\sigma \> 
\sigma^2 \, \left ( \delta V_{11} - 
\delta V_{12} \right ) \, \right ] \> .
\label{u0 omega}
\eea 
Here we have used $ \omega_{\rm var} \sigma \ll 1 $ and the abbreviation
$\delta V_{1j} \Def \delta V_{1j}/\delta \mu_{1j}^2(\sigma) $. If the result 
(\ref{u0 omega}) is inserted into the equation defining $\omega_{\rm var}^2$
\be
\omega_{\rm var}^2 \> \equiv \> 2 \int_0^{\infty} d\sigma \> \delta V_{12} \E 
2 \frac{\partial V_{12}}{\partial u_0} \E 
\frac{Z \alpha}{\sqrt{2 \pi}} \, \frac{M^2}{\lambda q/2} \, \frac{1}{u_0^{3/2}} \, 
\left [ 1 - \frac{1}{2} \frac{u_2}{(\lambda q/2)^2} + \ldots \right ] 
\label{u0}
\ee
we obtain 
\be 
\omega_{\rm var}^{1/2} \E \frac{Z \alpha}{\sqrt{2 \pi}} \, \frac{M^2}{\lambda q/2} \, 
\left [ \, 1 - \frac{1}{4}  \int_0^{\infty} d\sigma \> 
\sigma^2 \, \left ( \delta V_{11} - 
\delta V_{12} \right ) \, \right ]^{-3/2} \, \left [ 1- 
\frac{1}{2} \frac{u_2}{(\lambda q/2)^2} \right ] \> .
\ee
The correction terms may be safely expanded and the approximations
$ \lambda \simeq 1 , \> q/2 \simeq M , \> \mu_{11}^2 \simeq \sigma , \> 
\mu_{12}^2 \simeq 1/\omega , \> 
u_2 \simeq \omega/4 , \> \omega \simeq (Z \alpha)^2/(2 \pi) \> $ used without 
impunity. Thus we obtain
\be
\omega_{\rm var} \E \frac{(Z \alpha)^2}{2 \pi} \, \frac{M^4}{(\lambda q/2)^2} \, 
\left [ \, 1 + \frac{3 \alpha}{4 \pi} - \frac{5}{16} \frac{(Z \alpha)^2}{\pi} + \ldots 
\, \right ] \> .
\ee
Note that the RHS of Eq. (\ref{u0}) is just $ -V_{12}/u_0$. Hence
\be 
V_{12} \> \simeq \> - u_0 \, \omega_{\rm var}^2 \E - \frac{(Z \alpha)^2}{2 \pi} \, 
\frac{M^4}{(\lambda q/2)^2} \, 
\left [ \, 1 + \frac{\alpha}{2 \pi} - \frac{(Z \alpha)^2}{4\pi} + \ldots 
\, \right ] \> .
\ee
It remains to determine the variational parameter $\lambda$ up to order 
$ \alpha, (Z \alpha)^2$.

This can be done by using the variational Eq. (\ref{var eq for lambda 2}) with the
zeroth-order approximations for the pseudotimes or simply by expanding Eq. 
(\ref{lambda supersimp 1}) which is correct to that  order. In both cases one obtains
\be
\lambda \E 1 - \left [ \, \frac{\alpha}{\pi} + \frac{(Z \alpha)^2}{2 \pi} \, \right ] 
+ \ldots \> ,
\ee
which now allows application of the Feynman-Hellmann theorem (\ref{Hell-Fey})
\be
\left ( \frac{q}{2} \right )^2 \, \frac{\partial}{\partial Z} \left ( \frac{q}{2} 
\right )^2 \E - \frac{Z \alpha^2}{\pi} M^4 \, \left [ \, 1 + \frac{7}{2} 
\frac{\alpha}{\pi} + \frac{5}{4} \frac{Z^2 \alpha^2}{\pi^2} + \ldots \right ] \> .
\ee
Integrating on both sides and using $ q/2 = M + \epsilon_0 $ gives
\be
\left (  M + \epsilon_0 \right )^4 \E {\rm const.} - \frac{(Z \alpha)^2}{\pi} M^4 \, 
\left [ \, 1 + \frac{7}{2} \frac{\alpha}{\pi}\, \right ] - 
\frac{5}{8} \frac{(Z \alpha)^4}{\pi^2} M^4 + \ldots  \> .
\ee
The integration constant must equal $M^4$ because the two particles are unbound 
for $Z = 0$. Therefore
the ground-state binding energy has the weak-coupling expansion
\be
\epsilon_0 \E - \frac{M}{2} \, \left [ \, \frac{(Z \alpha)^2}{\pi} 
\left ( \, 1 + \frac{7}{2} \frac{\alpha}{\pi}\, \right ) + 
\frac{(Z \alpha)^4}{\pi^2}  + \ldots  \, \right ] \> .
\ee

\subsection{Effective Field-Theory Calculation}
\label{app: perturb bind}

\noindent
For simplicity, we consider here not the Lagrangian (\ref{L2 WC}) but a system of 
identical nucleons described by the Lagrangian
\be
{\cal L} \E \Phi^{\dagger} \, \left ( - \partial^2 - M^2 + g' \chi \right ) \Phi 
+ {\cal L}_0(\chi) 
\label{L3 WC}
\ee
where the last term is the free Lagrangian for the mesons.
In the nonrelativistic limit ( $ M \to \infty$) we make the ansatz
\be
\Phi(x) \E \frac{1}{\sqrt{2 M}} \, e^{-i M x_0} \, \phi(\bfx,x_0=t)
\ee
to describe particles (anti-particles would have a different sign in the phase 
factor and are omitted as explicit degrees of freedom in the following). This leads
to
\be
{\cal L} \E \phi^{\dagger} \, \left ( \, i \partial_t + \frac{\Delta}{2 M} + 
\frac{g'}{2 M} \chi - \frac{1}{2 M}  \partial_t^2 \, \right ) \, \phi
+ {\cal L}_0(\chi) \> .
\ee
The action is the 4-dimensional integral over the Lagrangian (density) so that
an integration by parts brings the last term in the brackets 
into the form $ \> \dot \phi^{\dagger} \dot \phi/(2 M) \> $. 
Using the equation of motion for the nonrelativistic field $\phi$ 
\be
\left ( \, - 2 i M \partial_t - \Delta + \partial_t^2 \, \right ) \phi(\bfx,t) 
\E g' \chi \, \phi(\bfx,t)
\ee
one obtains
\be
\dot \phi(\bfx,t) \>  \simeq \> \frac{i}{2 M} \, \left ( \Delta + g' \chi \right ) 
\, \phi(\bfx,t) \> .
\label{phi dot}
\ee
Therefore 
\be
{\cal L}_3 \E - \frac{1}{2M} \, \phi^{\dagger}  \partial_t^2 \,  \phi
\> \longrightarrow \> \frac{1}{(2 M)^3} \, \phi^{\dagger} \, \left ( \, \Delta + 
g' \chi \right )^2 \, \phi 
\ee
is the $1/M^3$-correction to the leading nonrelativistic Lagrangian. Its form
is in agreement with Eq. (3.6) or Eq. (4.12) in ref. \cite{AGGR} and 
can also be easily obtained from the Foldy-Wouthusen-Tani Hamiltonian
$ \> \phi^{\dagger} \, \sqrt{M^2 - \Delta - g' \chi} \, \phi \> $ by an expansion 
in inverse powers of the heavy mass $M$. Note that a ``seagull'' term 
$ \> \phi^{\dagger} \chi^2 \phi $ appears in the nonrelativistic field theory.

Since one has changed the high-energy behaviour of the theory the correct 
effective Lagrangian (for particles) is 
\bea
{\cal L}_{\rm eff} \EA   \phi^{\dagger} \, \left ( \, i \partial_t + 
\frac{\Delta}{2 M} + \frac{\Delta^2}{8 M^3} + \frac{g'}{2 M} c_1 \chi \, \right ) 
\phi \non
&& + \frac{1}{(2 M)^3} \, \phi^{\dagger}  \left [ \, g' c_2  \left ( 
\Deltaleft \chi
+ \chi \Deltaright \right ) +  g'^2 c_3 \chi^2 + g' \, d_1 \, (\Delta \chi)  
\,  \right ] \phi + \ldots + {\cal L}_0 (\chi) \> .
\label{L eff}
\eea
This should be accurate up to order $M v^4$ in the (velocity) counting rules 
\footnote{Since $ v \equiv v/c $ the counting is equivalent to the expansion in 
appendix \ref{app: nonrel} but the procedure is a systematic one which also allows
to include loop effects.}
in which
\bea 
\partial_t \EA {\cal O} \left ( M v^2 \right ) \hspace{4.1cm} 
(\mbox{nonrelativistic energy}) \non
\Delta \EA  {\cal O} \left ( M^2 v^2 \right ) \> \Longrightarrow \> 
\frac{\Delta}{2 M} \E {\cal O} \left ( M v^2 \right ) \hspace{0.5cm}
(\mbox{nonrelativistic kinetic energy}) \non
\frac{g'}{2 M} \EA \sqrt{4 \pi \alpha} \E  {\cal O} \left ( v^{1/2} \right ) 
\hspace{2.6cm} (\mbox{velocity in Bohr orbit} = \alpha) \non
\chi \EA {\cal O} \left ( M v^{3/2} \right ) \> \Longrightarrow \> 
\frac{g'}{2 M} \chi =  {\cal O} \left ( M v^2 \right )  \> \> \> (\mbox{potential 
energy}) \> .
\label{count rules} 
\eea
In Eq. (\ref{L eff}) the ``Wilson-coefficients'' $ c_i = 1 + {\cal O}(g'^2) $ 
encode the missing high-energy (short-distance) information. They are obtained 
by ``matching'', i.e. by comparing physical amplitudes in the full and 
the effective theory. The coefficient $d_1$ is expected to be  $ {\cal O} (g'^2) $ 
since it is not present in the tree-level calculations and this is confirmed by
the explicit calculation below. Note that the kinetic terms 
$ \> \Delta/(2M), \Delta^2/(8 M^3) $ are not
renormalized since they reflect the exact Lorentz symmetry of the underlying 
relativistic theory.

We determine the Wilson coefficients $ c_1, c_2, d_1 $ by evaluating
the meson-nucleon scattering amplitude in both theories including 
the lowest-order radiative corrections.
To do that we also need the residue $Z_r$ of the 
nucleon 2-point function
\be
\frac{1}{p^2 - M_0^2 + \Sigma(p^2)} \To \frac{Z_r}{p^2 - M^2}
\ee
where $M^2 = M_0^2 + \Sigma(M^2)$ and 
\be
Z_r^{-1}\E  1 + \frac{\partial \Sigma(p^2)}{\partial p^2} \Bigr |_{p^2 = M^2} \> .
\ee
Of course, in lowest order we have $ Z_r^{(0)} = 1 $ .
Using Muta's conventions \cite{Muta} we obtain for the 1-loop self-energy 
(depicted in fig. \ref{fig: one_loop}a ) 
\bea
\Sigma^{(1)}(p^2) \EA g'^2 \, \nu^{4-d} \, \int \frac{d^d k}{(2 \pi)^d i} \> 
\frac{1}{M_0^2 - (p-k)^2} \, \frac{1}{m^2 - k^2} \non
\EA g'^2 \, \frac{\Gamma(2 - d/2)}{(4 \pi)^{d/2}} \, \int_0^1 dx \>  
\frac{1}{[ M_0^2 x + m^2 (1-x) - p^2 x (1-x) ]^{2 - d/2}} \> .
\eea
Here we have introduced standard Feynman parameters and 
performed the diverging momentum integral
in $d$ dimensions. In this order we may replace the bare mass $M_0$ by the 
physical mass $M$, differentiate w.r.t. $p^2$, expand and find 
\bea
Z_r^{(1)} \EA - g'^2 \nu^{4-d} \, \frac{\Gamma(3 - d/2)}{(4 \pi)^{d/2}}  \, 
\int_0^1  dx \, 
\frac{x (1-x) }{[ M^2 x^2 + m^2 (1-x) ]^{3 - d/2}}  \non
&& \stackrel{d=4}{=} \>  - \frac{g'^2}{16 \pi^2} \, 
\int_0^1  dx \, \frac{x (1-x) }{M^2 x^2 + m^2 (1-x)}
\eea
which is UV-finite but IR-divergent for $ m = 0 $. 

The tree-level truncated nucleon-meson (2,1)-point function (multiplied by $-i$)
is just the coupling constant $g'$. The one-loop correction to this vertex 
function is shown in fig. \ref{fig: one_loop}b and leads to  
\be
\Gamma^{(1)} (p,q) \E g'^3 \int \frac{d^4 k}{(2 \pi)^4 i} \> 
\frac{1}{M^2 - (p - k)^2} \,  \frac{1}{M^2 - (p - k + q)^2} \, \frac{1}{m^2 - k^
2} \> .
\ee
For $q \to 0 $ we obtain by standard techniques
\be
\Gamma^{(1)} (p,q \to 0) \E \frac{g'^3}{16 \pi^2} \, \int_0^1 dx \> 
\frac{x}{ M^2 x + m^2 (1-x) - p^2 x (1-x)} + {\cal O}(g'^3 \, q^2) \> .
\ee
The physical amplitude for meson-nucleon scattering  
has two external nucleon legs and therefore requires
twice a wavefunction renormalization constant $\sqrt{Z_r}$ applied to the 
truncated (2,1)-point function. Hence
\bea
T(q \to 0) \EA Z_r \, \Gamma(p,q \to 0) \Bigr |_{p^2 = M^2}  \deF 
g'_{\rm eff} \, \left [ \, 1 + \frac{1}{6} q^2 \left < r^2 \right >  + 
{\cal O} (q^4) \, \right ] 
\label{phys amplit}\\
g'_{\rm eff} \EA 
g' \, \left [ \, 1 + \frac{g'^2}{16 \pi^2} \, \int_0^1 dx \> \frac{x^2}{ M^2 x^2 + 
m^2(1-x)} + \ldots \, \right ]  
\eea
Here $ \left < r^2 \right > $ is the root-mean square radius of the nucleon
due to radiative corrections.

The meson-nucleon scattering amplitude 
in the nonrelativistic theory described by the Lagrangian
(\ref{L eff}) reads
\be
T^{\rm nonrel.}(\bfp,\bfq) \E \left \{ \, c_1 \frac{g'}{2 M} + 
\frac{g'}{(2M)^3} \, \left [
c_2 \left (- \bfp^2 - ( \bfp + \bfq )^2 \right )  - d_1 \bfq^2 \right ] \, 
\right \} \, \sqrt{2 E_{\bfp} \, 2 E_{\bfp + \bfq}}
\ee
where the square-root factor is due to the different normalization of
relativistic and nonrelativistic single-particle states \cite{Gall}.
Expanding $\> E_{\bfp} = \sqrt{M^2 + \bfp^2} \> $ and $\> E_{\bfp+\bfq} \> $ for 
low three-momenta we obtain
\be
T^{\rm nonrel.}(\bfp \to {\bf 0},\bfq \to {\bf 0}) \E c_1 g' + \left ( c_1 - c_2 \right) 
\, g' \, \frac{\bfp^2 + (\bfp+\bfq)^2}{4 M^2} - d_2 \, g' \, \frac{\bfq^2}{4 M^2}
+ \ldots \> .
\ee
Comparing with Eq. (\ref{phys amplit}) we find for massless mesons
\be
c_1 \E c_2 \E \frac{g'_{\rm eff}}{g'} \E 1 + \frac{\alpha}{\pi} + \ldots \> .
\label{c1}
\ee
and ($ q^2 \to - \bfq^2 $ when all momenta are small)
\be
d_1 \E \frac{2 M^2}{3} \, \frac{g'_{\rm eff}}{g'} \, \left < r^2 \right > \E 
{\cal O}(g'^2) \> .
\label{d1}
\ee
The coefficients $c_1, c_2$ 
account for an enhancement of the effective meson-nucleon 
coupling constant through loop effects similar as in Schwinger's famous 
determination of the anomalous magnetic moment of the electron in QED.
The coefficient $d_1$ describes the finite extension of nucleons due to the 
meson cloud and therefore is infrared divergent. This 
can also be seen from its explicit lowest-order 
perturbative expression as given in Eq. (60) of ref. \cite{WC3} but we 
do not need it in the following.

In the next step one has to integrate out the mesons in order to reduce the
relativistic bound-state problem to a quantum-mechanical one.
This we do by solving the equation of motion for the meson field (operator)
obtained from varying Eq. (\ref{L eff})
\be
\left ( \partial + m^2 \right ) \chi_0 \E \frac{Z g'}{2 M} \,  
c_1 \, \phi^{\dagger} \phi + \ldots
\label{meson eom}
\ee
Since  in Eq. (\ref{c1}) we have determined $c_1$ only to one-loop order, i.e. to order
$v$ in the counting rules, we have retained only the leading term and for consistency
have neglected the ${\cal O}(v^2)$ corrections in the effective Lagrangian. 
Furthermore, we replace $ g' \to Z g'$  to indicate that the mesons are 
from the other particle (self-energy effects vanish in the effective
nonrelativistic theory if dimensional regularisation is used \cite{Man}).
For the case that no external mesons are present the solution of 
Eq. (\ref{meson eom})
is
\bea
\chi_0(\bfx,t) \EA   \frac{Z g'}{2 M} c_1 \,
\int d^4y \, \left < x \left | \frac{1}{ \partial^2 + m^2} \right | y \right > \, 
\phi^{\dagger}(y) \phi(y)  \non
&& \simeq  \> \frac{Z g'}{2 M}  c_1  \int d^3 y \, 
 \left < \bfx \left |  \frac{1}{-\Delta + m^2}  
 \right | \bfy \right > \, \phi^{\dagger}(\bfy,t) \phi(\bfy,t) \> .
\label{chi0}
\eea
According to the counting rules,
the time derivative in the d'Alembertian is also 
suppressed by a factor ${\cal O}(v^2)$ compared to the Laplacian. Equation (\ref{chi0}) 
is therefore correct up to order $M v^{5/2}$ and after substitution into 
Eq. (\ref{L eff}) gives the following effective (potential) Lagrangian 
\be
{\cal L}_{\rm eff}^{\rm pot} \E \phi^{\dagger}(\bfx,t) \, \left \{  \, i 
\partial_t + \frac{\Delta}{2 M} 
- \frac{1}{2} \int d^3 y \> \phi^{\dagger} (\bfy,t)
V_{\rm eff} (\bfx - \bfy) \phi(\bfy,t) + {\cal O} ( M v^4) \, \right \} \, 
\phi(\bfx,t) 
\ee
where
\be
V_{\rm eff}\left (\bfx - \bfy \right ) \E - c_1^2  \, \left ( 
\frac{Z g'}{2 M} \right )^2 \, \left < \bfx \left | \, \frac{1}{-\Delta + m^2} \, 
\, \right | \bfy \right > \E - c_1^2 \, \frac{Z \alpha}{|\bfx - \bfy|} 
e^{-m |\bfx-\bfy|}
\ee
is the usual Yukawa potential {\it enhanced} by a factor $c_1^2$ from radiative 
corrections. In the Coulombic case ($m = 0$) this means that the leading term for 
the binding energy is
\be
\epsilon_n \E  - \frac{M}{2} \, \frac{(Z \alpha)^2}{2 (n+1)^2} \, c_1^4 + \ldots
\E  - \frac{M}{2} \, \frac{(Z \alpha)^2}{2 (n+1)^2} \, \left [ \, 1 + 
4 \, \frac{\alpha}{\pi}  \, \right ]  + {\cal O} \left ( (Z \alpha)^2 \alpha^2, 
(Z \alpha)^4 \right ) 
\> \> , \> n = 0, 1, 2 \ldots \> .
\ee


\vspace{1.5cm}


\begin{thebibliography}{99}


\bibitem{SaBe} 
Salpeter, E.~E, Bethe, H.~A:
Phys.\ Rev.\  {\bf 84}, 1232 (1951)

\bibitem{BSreview}  Nakanishi, N. (ed.): Progr. Theor. Phys. Suppl. {\bf 95} (1988)

\bibitem{Saul} \v{S}auli, V., Adam, J.:  Phys. Rev. {\bf D 67}, 085007 (2003)
$\>$  arXiv:hep-ph/0111433

\bibitem{crossedBS}
Karmanov, V.~A., Carbonell, J.:
$\>$  arXiv:nucl-th/0510051; $\>$  arXiv:hep-th/0505262

\bibitem{LoTa}
Logunov, A.~A., Tavkhelidze, A.~N.: Nuovo Cim.\  {\bf 29}, 380 (1963)  

\bibitem{BlSu} 
Blankenbecler, R., Sugar R.: Phys.\ Rev.\  {\bf 142}, 1051 (1966) 

\bibitem{Gross} 
Gross, F.: Phys. Rev. {\bf 186}, 1448 (1969); Phys. Rev. {\bf D 10}, 223 (1974);
Phys. Rev. {\bf C 26}, 2203 (1982) 

\bibitem{light}
Mangin-Brinet, M., Carbonell, J., Karmanov, V.~A.:
Phys.\ Rev.\  {\bf C 68}, 055203 (2003) $\>$ [arXiv:hep-th/0308179];\\ 
Bernard, D., Cousin, T., Karmanov, V.~A., Mathiot, J.~F.:
Phys.\ Rev.\  {\bf D 65}, 025016 (2002) $\>$ [arXiv:hep-th/0109208];\\
Mangin-Brinet, M., Carbonell, J., Karmanov, V.~A.:
Phys.\ Rev.\  {\bf D 64}, 125005 (2001) $\>$ [arXiv:hep-th/0107235];\\
Sales, J.~H.~O., Frederico, T., Carlson, B.~V., Sauer,  P.~U.: 
Phys.\ Rev.\  {\bf C 61}, 044003 (2000) $\>$ [arXiv:nucl-th/9909029]; 
Phys.\ Rev.\ {\bf C 63}, 064003 (2001);\\ 
Carbonell, J., Desplanques, B., Karmanov, V.~A., Mathiot, J.~F.:
Phys.\ Rept.\  {\bf 300}, 215 (1998) $\>$ [arXiv:nucl-th/9804029]; \\
Brodsky, S.~J., Pauli, H.~C., Pinsky, S.~S.:
Phys.\ Rept.\  {\bf 301}, 299 (1998)
$\>$ [arXiv:hep-ph/9705477]

\bibitem{point} 
Desplanques, B.: Nucl.\ Phys.\  {\bf A 748}, 139 (2005) 
$\>$ [arXiv:nucl-th/0405059]; \\
Desplanques, B., Theussl, L.: Eur.\ Phys.\ J.\  {\bf A 13}, 461 (2002)
$\>$ [arXiv:nucl-th/0102060];\\
Klink, W.~H.: Phys. Rev. {\bf C 58}, 3587 (1998) 

\bibitem{FSR}
\c{S}avkl{\i}, \c{C}., Gross, F., Tjon, J.: Phys.\ Atom.\ Nucl.\  {\bf 68}, 
842 (2005) $\>$ [Yad.\ Fiz.\  {\bf 68}, 874 (2005)]
$\>$   [arXiv:nucl-th/0404068];\\
Simonov, Y.~A., Tjon, J.~A.: Ann. Phys. (N. Y.)  {\bf 300}, 54 (2002)
$\>$ [arXiv:hep-ph/0205165];\\
\c{S}avkl{\i}, \c{C}.: Comput.\ Phys.\ Commun.\  {\bf 135}, 312 (2001)
$\>$ [arXiv:hep-ph/9910502];\\
\c{S}avkl{\i}, \c{C}., Tjon, J., Gross, F.:
Phys.\ Rev.\  {\bf C 60}, 055210   (1999)  $\>$ [arXiv:hep-ph/9906211],
[Erratum: ibid.\  {\bf C 61}, 069901  (2000)]


\bibitem{boundQED} 
Eides, M.~I., Grotch, H., Shelyuto, V.~A.:
Phys.\ Rept.\  {\bf 342}, 63 (2001)
$\>$ [arXiv:hep-ph/0002158];\\
Kinoshita, T. (ed.): Quantum Electrodynamics, World Scientific (1990)

\bibitem{emdeut} 
Gilman, R., Gross, F.: J. Phys. {\bf G 28}, R37 (2002) 
$\>$ [arXiv:nucl-th/0111015]; \\
Gross, F.: Eur.\ Phys.\ J.\  {\bf A 17}, 407 (2003)
$\>$ [arXiv:nucl-th/0209088]

\bibitem{PWD} 
Phillips, D.~R., Wallace, S.~J., Devine, N.~K.: Phys.\ Rev.\  {\bf C 58}, 
2261 (1998) $\>$ [arXiv:nucl-th/9802067]

\bibitem{MoMue} 
Montvay, I., M\"unster, G. : Quantum Fields on a Lattice. Cambridge:  
Cambridge University Press 1994

\bibitem{AlSm}
Alkover, R., von Smekal, L.:
Phys.\ Rept.\  {\bf 353}, 281 (2001)
$\>$ [arXiv:hep-ph/0007355]

\bibitem{DS}
Maris, P., Roberts, C. D.: 
Int. J. Mod. Phys. {\bf E 12}, 297 (2003)
$\>$ [arXiv:nucl-th/0301049];\\
Roberts, C. D., Schmidt, S. M.:
Progr. Part. Nucl. Phys. {\bf 45}, S1 (2000) 
$\>$ [arXiv:nucl-th/0005064]

\bibitem{NRQCD} 
Brambilla, N., Pineda, A., Soto. J., Vairo, A.: 
Rev.\ Mod.\ Phys.\  {\bf 77}, 1423 (2005) 
$\>$  [arXiv:hep-ph/0410047]

\bibitem{varfield} 
Polley, L., Pottinger, D. E. L. (eds.): Variational Calculations in Quantum 
Field Theory. Singapore: World Scientific 1988

\bibitem{Dar} 
Darewych, J. W., Terekidi, A. G.:  
J.\ Math.\ Phys.\  {\bf 46}, 032302  (2005)
$\>$ [arXiv:hep-ph/0311132]; \\
Terekidi, A. G., Darewych, J. W.: 
arXiv:hep-th/0303250; \\
Darewych, J. W., Shpytko, V.:
Phys.\ Rev.\  {\bf D 64}, 045012 (2001)
$\>$ [arXiv:nucl-th/0012080]

\bibitem{DarWC}
Darewych, J. W., Ding, B.:
Nucl.\ Phys.\ Proc.\ Suppl.\  {\bf 90}, 136 (2000);\\ 
Ding, B.~f., Darewych, J. W.:
J.\ Phys.\  {\bf G 26}, 907 (2000)
$\>$ [arXiv:nucl-th/9908022];\\
Darewych, J. W.:
Can.\ J.\ Phys.\  {\bf 76}, 523 (1998)
$\>$ [arXiv:nucl-th/9807006];\\
Darewych, J. W., Barham, M.:
J.\ Phys.\  {\bf A 31}, 3481 (1998); \\
Darewych, J. W., Shapoval, D. V., Simoneg, I. V., Sitenko, A. G.:
J.\ Math.\ Phys.\  {\bf 38}, 3908 (1997);\\
Darewych, J. W., Polozov, A. D., Di Leo, L.:
J.\ Phys.\  {\bf G 21}, 1167 (1995)

\bibitem{Fey} 
Feynman, R. P.: Phys. Rev. {\bf 97}, 660 (1955)

\bibitem{Mano} 
Mano, K.: Progr. Theor. Phys. {\bf 14}, 435 (1955)

\bibitem{worldline} 
Strassler, M. J.: Nucl.\ Phys.\  {\bf B 385}, 145 (1992)
$\>$ [arXiv:hep-ph/9205205];\\ 
Reuter, M., Schmidt, M. G., Schubert, C.:
Ann. Phys. (N.Y.)  {\bf 259}, 313 (1997)
$\>$ [arXiv:hep-th/9610191];\\ 
Schubert, C.: Phys.\ Rept.\  {\bf 355}, 73 (2001)
$\>$ [arXiv:hep-th/0101036]

\bibitem{WC1} 
Rosenfelder, R., Schreiber, A. W.: Phys. Rev. {\bf D 53}, 3337 
(1996) $\>$ [arXiv:nucl-th/9504002],
denoted by (I) in the following

\bibitem{WC2} 
Rosenfelder, R., Schreiber, A. W.: Phys. Rev. {\bf D 53}, 3354 
(1996) $\>$ [arXiv:nucl-th/9504005], denoted by (II) in the following

\bibitem{WC3} 
Schreiber, A. W., Rosenfelder, R., Alexandrou, C.:  
Int. J. Mod. Phys. {\bf E 5}, 681 (1996)  $\>$ [arXiv:nucl-th/9504023]

\bibitem{WC4} 
Schreiber, A. W., Rosenfelder, R.: Nucl. Phys. {\bf A 601}, 397 (1996) 
$\>$ [arXiv:nucl-th/9510032]

\bibitem{WC56} 
Alexandrou, C., Rosenfelder, R., Schreiber, A. W.: 
Nucl. Phys. {\bf A 628}, 427 (1998) $\>$  [arXiv:nucl-th/9701036]; \\
Fettes, N.,  Rosenfelder, R.: Few-Body Syst. {\bf 24}, 1 (1998)

\bibitem{WC7} 
Rosenfelder, R., Schreiber, A. W.: Eur. Phys. J. {\bf C 25}, 139 (2002) 
$\>$ [arXiv:hep-th/0112212]

\bibitem{WiCu} 
Wick, G. C.: Phys. Rev. {\bf 96}, 1124 (1954);\\
Cutkosky, R. E.: Phys. Rev. {\bf 96}, 1135 (1954) 

\bibitem{QED1} 
Alexandrou, C., Rosenfelder, R., Schreiber, A. W.: 
Phys. Rev. {\bf A 59}, 1762 (1999)
$\>$ [arXiv:hep-th/9809101]

\bibitem{VALE} 
Rosenfelder, R., Schreiber, A. W.: Eur. Phys. J. {\bf C 37}, 161 (2004)
$\>$ [arXiv:hep-th/0406062]

\bibitem{unstable} 
Baym, G.: Phys. Rev. {\bf 117}, 886 (1960);\\
Rosenfelder, R., Schreiber, A. W.: arXiv:hep-ph/9911484;\\
Gross, F., \c{S}avkl{\i}, \c{C}., Tjon, J.:
Phys.\ Rev.\ D {\bf 64}, 076008 (2001)
$\>$ [arXiv:nucl-th/0102041]

\bibitem{QED2} 
Alexandrou, C., Rosenfelder, R., Schreiber, A. W.: 
Phys. Rev. {\bf D 62}, 085009 (2000)
$\>$ [arXiv:hep-th/0003253]

\bibitem{ItZu} 
Itzykson, C., Zuber, J. - B.: Quantum Field Theory. New York: McGraw-Hill 1980

\bibitem{BRS} 
Barro-Bergfl\"odt, K., Rosenfelder, R., Stingl, M.:
Mod. Phys. Lett. {\bf A 20}, 2533 (2005) $\>$ [arXiv:hep-ph/040330]

\bibitem{bipol}
Verbist, G., Peeters, F. M.,  Devreese, J. T.: 
Phys. Rev. {\bf B 43}, 2712 (1991);\\
Adamowski, J.: Phys. Rev. {\bf B 39}, 3649 (1989)

\bibitem{VSPD}
Verbist, G., Smondyrev, M. A., Peeters, F. M.,  Devreese, J. T.: 
Phys. Rev. {\bf B 45}, 5262 (1992) 

\bibitem{Pol} 
Poliatzky, N.: J. Phys. {\bf A 25}, 3649 (1992); see Eq. (37) and Table 1.

\bibitem{FeWa} 
Fetter, A. L., Walecka, J. D.: Quantum Theory of Many-Particle Systems. New York: 
McGraw-Hill 1971

\bibitem{SVZ} 
Shifman, M. A., Vainshtein, A. I., Zakharov, V. I.: 
Nucl. Phys. B {\bf 147}, 385, 448, 519 (1979) 

\bibitem{Handbook} 
Abramowitz, M., Stegun, I. (eds.): Handbook of Mathematical Functions. New York:  
Dover 1965

\bibitem{Ynd} 
Yndur\'ain, F. J.: Quantum Chromodynamics. An Introduction to the Theory of Quarks 
and Gluons. Berlin:  Springer 1983

\bibitem{Fey-Hell} 
Thirring, W.: Quantum Mathematical Physics (2nd edition), p. 527. Berlin:  
Springer 2002; \\
Robinett, R. W.: Quantum Mechanics, p. 421. Oxford: Oxford University Press 1997

\bibitem{mesh} 
Baye, D., Heenen, P.-H.: J. Phys. A {\bf 19}, 2041 (1986); \\
Semay, C., Baye, D., Hesse, M., Silvestre-Brac, B.: Phys. Rev. {\bf E 64}, 016703 
(2001) \\ 
Buisseret, F., Semay, C.: Phys.\ Rev.\  {\bf E 71}, 026705 (2005) 
$\>$  [arXiv:hep-ph/0409033]

\bibitem{AhAl} 
Ahlig, S., Alkofer, R.: Ann. Phys. (N.Y.)  {\bf 275}, 113 (1999)

\bibitem{induced} 
Affleck, I. K., De Luccia, F.: Phys. Rev.  {\bf D 20}, 3168 (1979) ;\\
Voloshin, M.: Phys. Rev. {\bf D 49}, 2014 (1994);\\
Calzetta, E., Roura, A., Verdaguer, E.: Phys.\ Rev.\ {\bf D 64}, 105008 (2001)
$\>$ [arXiv:hep-ph/0106091]\\
Gorsky, A., Voloshin, M. B.: arXiv:hep-th/0511095

\bibitem{Efi1} 
Efimov, G. V.: Few Body Syst. {\bf 33}, 199 (2003) 
$\>$ [arXiv:hep-ph/0304194]

\bibitem{NiTj2} 
Nieuwenhuis, T., Tjon, J. A.: Few Body Syst. {\bf 21}, 167 (1996) 
$\>$ [arXiv:nucl-th/9607041]

\bibitem{KaCa}
Karmanov, V. A., Carbonell, J.: Eur. Phys. J. {\bf A 27}, 1 (2006)
$\>$ [arXiv:hep-th/0505261]

\bibitem{Efi2} 
Efimov, G. V.: private communication (February 2004).

\bibitem{NiTj1} 
Nieuwenhuis, T., Tjon, J. A.: Phys. Rev. Lett. {\bf 77}, 814 (1996)
$\>$ [arXiv:hep-ph/9606403]

\bibitem{EFT}
Caswell, W. E., Lepage, G. P.: Phys.\ Lett.\  {\bf B 167}, 437 (1986);\\
Pineda, A., Soto, J.: Phys.\ Rev.\  {\bf D 59}, 016005 (1999)
$\>$ [arXiv:hep-ph/9805424]

\bibitem{Gries} 
Griesshammer, H. W.: Phys.\ Rev.\  {\bf D 58}, 094027 (1998) 
$\>$ [arXiv:hep-ph/9712467]

\bibitem{Ji} 
Ji, C. R.: Phys.\ Lett.\  {\bf B 322}, 389 (1994)

\bibitem{Math} 
Mathiot, J. F., Karmanov, V. A., Smirnov, A.:
Few Body Syst. {\bf 36}, 173 (2005) $\>$ [arXiv:hep-th/0412199]

\bibitem{Pil} 
Pilkuhn, H. M.: Relativistic Quantum Mechanics. Berlin: Springer 2003

\bibitem{Tod}
Todorov, I. T.: Phys.\ Rev.\  {\bf D 3}, 2351 (1971) 

\bibitem{BIZ}
Brezin, E., Itzykson, C., Zinn-Justin, J.: Phys.\ Rev.\  {\bf D 1}, 2349 (1970)

\bibitem{Jen} 
Jentschura, U. D.: Phys. Lett.  {\bf B 564}, 225 (2003) 
$\>$ [arXiv:hep-ph/0305072];\\
Jentschura, U. D., Mohr, P. J., Soff, G.: Phys. Rev. Lett.  {\bf 82}, 53 (1999)  
$\>$ [arXiv:physics/0001068]

\bibitem{cum2} 
Marshall, J. T., Mills, L. R.: Phys. Rev. {\bf B 2}, 3143 (1970); \\
Rosenfelder, R., Lu, Y.: Phys. Rev. {\bf B 46}, 5211 (1992)  

\bibitem{LuLu}
Luttinger, J. M., Lu, C.-Y.: Phys. Rev. {\bf B 21}, 4251 (1980); \\
Rosenfelder, R.: J. Phys. {\bf A 27}, 3523 (1994)
    
\bibitem{AlDi}
Alexandrou, C., Diakonos, F. K.: Z.\ Phys.\  {\bf A 353}, 149 (1995)
$\>$ [arXiv:nucl-th/9503010]

\bibitem{RoPANIC}
Rosenfelder, R.: arXiv:hep-ph/0601088

\bibitem{HoGa}
D'Hoker, E., Gagne, D. G.: Nucl.\ Phys.\  {\bf B 467}, 297 (1996)
$\>$ [arXiv:hep-th/9512080]

\bibitem{Kondo} Kondo, K.-I.: arXiv:hep-th/0307270, hep-th/0311033

\bibitem{Rophi4} Rosenfelder, R.: to be published

\bibitem{Jaxo}
Binosi, D., Theussl, L.: Comput.\ Phys.\ Commun.\  {\bf 161}, 76 (2004)
$\>$  [arXiv:hep-ph/0309015]

\bibitem{ErWe} 
Ericson, T., Weise, W.: Pions and Nuclei. Oxford:  Clarendon Press 1988

\bibitem{Efi3} 
Efimov, G. V.: arXiv:hep-ph/9907483; arXiv:hep-ph/9607425

\bibitem{BeFu}
B\`eg, M.~A.~B., Furlong, R. C.:  Phys.\ Rev.\ {\bf D 31}, 1370 (1985)

\bibitem{Light} 
Lighthill, M. J.: Introduction to Fourier Analysis and 
Generalized Functions. Cambridge: Cambridge University Press 1958

\bibitem{Rob} 
Robinson, P. A.: J. Math. Phys. {\bf 27}, 1206 (1986); 
J. Math. Phys. {\bf 30}, 2484 (1989)

\bibitem{inBess} 
Agrest, M. M., Maksimov, M. S.: Theory of Incomplete 
Cylindrical Functions and their Application (Die Grundlehren der mathematischen 
Wissenschaften, Band 160). Berlin: Springer 1971

\bibitem{HSA}
Hauck, A., von Smekal, L., Alkofer, R.: Comput.\ Phys.\ Commun.\  {\bf 112}, 166 
(1998) $\>$ [arXiv:hep-ph/9804376]

\bibitem{Maas} 
Maas, A.: arXiv:hep-ph/0504110

\bibitem{Mand}
Mandelzweig, V. B.: Phys.\ Atom.\ Nucl.\  {\bf 68}, 1227 (2005)
[Yad.\ Fiz.\  {\bf 68}, 1277 (2005)]

\bibitem{PiSo}
Pineda, A., Soto, J.: Nucl.\ Phys.\ Proc.\ Suppl.\  {\bf 64}, 428 (1998) 
$\>$ [arXiv:hep-ph/9707481]; Phys.\ Rev.\ {\bf D 59}, 016005 (1999) 016005 $\>$
[arXiv:hep-ph/9805424]

\bibitem{AGGR}
Antonelli, A., Gall, A. Gasser, J., Rusetsky, A.: Ann. Phys. (N.Y.) 
{\bf 286}, 108 (2001)
$\>$  [arXiv:hep-ph/0003118]

\bibitem{Muta}
Muta, T.: Foundations of Quantum Chromodynamics. An Introduction to Perturbative 
Methods in Gauge Theories, Chap. 2.3.2. Singapore:  World Scientific 1987

\bibitem{Gall}
Gall, A.: arXiv:hep-ph/9910364

\bibitem{Man}
Manohar, A. V.: Phys.\ Rev.\ {\bf D 56}, 230 (1997)
$\>$ [arXiv:hep-ph/9701294]


\end{thebibliography}
\end{document}